\documentclass[fleqn,usenatbib]{mnras}

\usepackage{newtxtext,newtxmath}
\usepackage{tikz}
\usetikzlibrary{arrows.meta,calc,patterns,shapes.geometric,fit,backgrounds,positioning}
\usepackage{bbm}

\usepackage[T1]{fontenc}

\DeclareRobustCommand{\VAN}[3]{#2}
\let\VANthebibliography\thebibliography
\def\thebibliography{\DeclareRobustCommand{\VAN}[3]{##3}\VANthebibliography}

\usepackage{graphicx}
\usepackage{amsmath}
\usepackage{bm}
\usepackage{xspace}

\usepackage{dblfloatfix}
\usepackage{placeins}
\makeatletter
\renewcommand\figure{\def\baselinestretch{1}%
  \let\@makecaption\SFB@makefigurecaption \@float{figure}}
\let\endfigure\end@float
\@namedef{figure*}{\def\baselinestretch{1}\st@rredfloattrue
  \let\@makecaption\SFB@makefigurecaption \@dblfloat{figure}}
\@namedef{endfigure*}{\end@dblfloat\st@rredfloatfalse}
\makeatother

\newcommand{\ceri}{\textsc{Ceridwen}}
\newcommand{\pros}{\textsc{Prospector}}
\newcommand{\jwst}{\textit{JWST}\xspace}
\newcommand{\fob}{\ensuremath{f_{\rm ob}}}
\newcommand{\fesc}{\ensuremath{f_{\rm esc}}}

\title[Stellar Population Inference with \ceri]{\ceri: Fast and Flexible GPU-Accelerated Stellar Population Inference}

\author[A. Stoffers et al]{
Amanda Stoffers,$^{1, 2}$\thanks{E-mail: aas208@cam.ac.uk}
Sandro Tacchella,$^{1,2}$
Benjamin D. Johnson$^{3}$
\\
$^{1}$The Kavli Institute for Cosmology (KICC), University of Cambridge, Madingley Road, Cambridge, CB3 0HA, UK\\
$^{2}$Cavendish Laboratory, University of Cambridge, 19 JJ Thomson Avenue, Cambridge, CB3 0HE, UK\\
$^{3}$Center for Astrophysics $|$ Harvard \& Smithsonian, 60 Garden St., Cambridge, MA 02138, USA
}

\date{Accepted XXX. Received YYY; in original form ZZZ}

\pubyear{\the\year{}}

\begin{document}
\label{firstpage}
\pagerange{\pageref{firstpage}--\pageref{lastpage}}
\maketitle

\begin{abstract}
\jwst has increased both the number of high-redshift galaxies with high-quality spectral energy distributions (SEDs) and their information content. In parallel, wide-area surveys from \textit{Euclid}, Rubin’s \textit{LSST}, and \textit{Roman} will increase galaxy samples by orders of magnitude. Analysing these rapidly growing datasets requires stellar-population models that are both flexible and computationally efficient. We present \ceri, a GPU-native SED fitting framework written in \textsc{JAX}, with an end-to-end differentiable forward model spanning stellar populations, nebular emission, dust attenuation and emission, and cosmological projection into the observer frame. Its vectorised, compiled architecture lets nested sampling replace a batch of live points in parallel on the GPU, which makes flexible stellar-population models tractable under full Bayesian inference. Automatic differentiation also provides exact gradients for the gradient-based samplers included in the package. We jointly infer time-dependent chemical-enrichment histories instead of a single stellar metallicity, and demonstrate non-parametric star-formation histories (SFHs) with ${\sim}120$ age bins. Using $\alpha$-enhanced stellar libraries from \textsc{FSPS}, \ceri\ can sample stellar $[\alpha/\mathrm{Fe}]$ jointly with $[\mathrm{Fe/H}]$, mass, and SFH, so that the joint posterior represents the $[\mathrm{Fe/H}]$--$[\alpha/\mathrm{Fe}]$ degeneracy explicitly. In controlled mocks, \ceri\ recovers parameters with well-calibrated posterior uncertainties, while fits to real \jwst observations reproduce posteriors from the established \pros\ framework: on a single GPU, \ceri\ completes a fit in a median sampling time of ${\sim}4$\,min, ${\sim}134\times$ faster per fit than equivalent CPU-based \pros\ runs. \ceri\ therefore makes full Bayesian inference practical for larger galaxy samples and more flexible stellar-population models, reducing computational constraints on the physical complexity explored in SED fitting.
\end{abstract}

\section{Introduction}
\label{sec:introduction}

The physical properties and evolutionary histories of galaxies are not directly observable, but must be inferred from their emitted light. Spectral energy distribution (SED) modelling provides the bridge between the two, comparing observations with models of stellar populations to reconstruct the underlying physical properties. Over the past several decades, this inverse problem has developed into a mature field and become one of the principal tools for studying galaxy formation and evolution \citep[see, e.g.,][]{Tinsley:1980aa,walcher_fitting_2011,conroy_modeling_2013}.

Great observational advances have the habit of exposing apparent cracks in our physical picture of the Universe. Often, however, the tension lies in the comprehensiveness of the models used to interpret increasingly informative data, rather than in a failure of the prevailing physical picture.
This is particularly important in the early Universe, where inferred stellar masses, star-formation histories (SFHs), and star-formation efficiencies directly shape our conclusions about baryonic physics and cosmology.  Early \jwst observations provide several examples. Massive galaxy candidates at $z\gtrsim7$ initially appeared to require stellar-mass densities approaching the available baryonic reservoir of their dark-matter haloes \citep{Labbe:2023aa, Boylan-Kolchin:2023aa}; for some sources, allowing an obscured active galactic nucleus (AGN) to contribute to the observed continuum substantially reduced the inferred stellar mass \citep{Kocevski:2023aa, Wang:2024aa}. Similarly, the interpretation of apparently massive quiescent galaxies depends sensitively on separating genuinely old stellar populations from alternative SFHs \citep{Glazebrook:2024aa, Turner:2025aa}, while heavily obscured star-forming galaxies can yield very different stellar masses and star-formation rates depending on the adopted treatment of dust \citep{Xiao:2024aa, Lapasia:2026aa}. These examples illustrate a common problem: as observations become more constraining, limitations in the forward model can become limitations on the astrophysics we infer.
This problem is neither new nor unique to \jwst. Balmer breaks inferred from \textit{Spitzer}/IRAC photometry at $z\simeq6$ were initially interpreted as evidence for highly evolved stellar populations \citep{Eyles:2007aa}, before the inclusion of nebular continuum and line emission substantially reduced the inferred stellar ages \citep{Schaerer:2009aa}. The broader lesson is that more informative data make assumptions in the forward model more important. Robust constraints on galaxy formation, including quantities such as the efficiency with which dark-matter haloes convert baryons into stars, therefore require models flexible enough to represent the range of physical processes permitted by the data.

This need for greater model flexibility grows as the observational datasets themselves are becoming both richer and larger. Deep NIRCam imaging and NIRSpec spectroscopy have pushed the number of high-redshift ($z>2$) galaxies with high-quality SEDs into the many thousands, while simultaneously increasing the information content available for each of them. Galaxies during the Epoch of Reionisation observed with \jwst may now be characterised by a dozen or more photometric bands and a rest-frame optical spectrum, whereas a decade ago the same galaxy might have been measured with only a handful of broad-band fluxes and non-detections in the remaining bands. The next generation of large-scale surveys, including \textit{Euclid} \citep{Laureijs:2011aa}, the Vera C.\ Rubin Observatory’s Legacy Survey of Space and Time (LSST; \citealt{Ivezic:2019aa}), and the \textit{Nancy Grace Roman Space Telescope} \citep{Spergel:2015aa}, will add further orders of magnitude of galaxies with moderate-quality, few-band SEDs, for which the cost per fit becomes the limiting factor. These two developments create a fundamental computational tension: richer data justify increasingly flexible physical models, while larger samples require each individual inference to become substantially cheaper.

Flexibility is also needed in the construction of the forward model itself. An SED model combines prescriptions for stellar evolution, dust, and nebular photoionisation, all of which remain uncertain at a level that increasingly informative data can resolve. Differences in the treatment of post-main-sequence evolution, low-metallicity and non-solar-abundance stellar libraries, attenuation geometry, and ionising spectra can therefore propagate directly into the inferred galaxy properties. A fitting framework cannot remove these uncertainties, but it should make the underlying ingredients easy to vary and compare.

\ceri\ is designed around both requirements: computationally efficient inference at the scale of current and forthcoming surveys, and a modular forward model in which the relevant physical ingredients can be varied and compared.

Allowing the necessary level of physical freedom, however, is exactly where the computational difficulty begins. An SED fit has three ingredients: the data, a forward model, and an inference method. Most forward models begin with single stellar populations (SSPs), generated by stellar population synthesis codes such as \textsc{GALAXEV} \citep{Bruzual:2003aa}, \textsc{FSPS} \citep{conroy_propagation_2009, Conroy:2010aa} or \textsc{BPASS} \citep{stanway_re-evaluating_2018}. These are combined according to the parameters of the model, such as its SFH and chemical-enrichment history (ZH), into a composite stellar population (CSP), which is then altered by dust, nebular continuum and line emission and other sources of radiation such as AGN, and finally projected into the observer frame. The inference method then asks which parameters of that model the data support. Maximising the likelihood, which quantifies the probability of the data given the model parameters, is cheap, but rarely sufficient: SED posteriors are non-Gaussian, strongly degenerate, often multimodal, and sensitive to the priors, so the object of interest is the full posterior distribution rather than just its mode.

Every physical degree of freedom added to the forward model is therefore paid for twice. It raises the dimensionality of the parameter space, and it can introduce new degeneracies, which make that space harder to explore. Both effects increase the number of likelihood evaluations needed for a converged posterior, while a more complex forward model can also make each evaluation more expensive. This is the origin of the speed--flexibility trade-off that every SED-fitting code negotiates: codes are fast because they restrict the model, or flexible because they accept the cost, and neither choice scales comfortably to surveys of $10^5$ galaxies with excellent quality data. 

Existing codes sit at varying points along this speed--flexibility trade-off. Some of the earliest and fastest approaches rely on precomputed template libraries or model grids. Codes such as EAZY \citep{Brammer:2008aa}, MAGPHYS \citep{da-Cunha:2008aa} and CIGALE \citep{Boquien:2019aa} evaluate the expensive physics in advance, reducing fitting to interpolation over a fixed set of models.
The cost is paid at construction, where the grid grows exponentially with the number of physical parameters, and the physics is frozen from the moment the grid is built. More flexible Bayesian forward-modelling codes such as BEAGLE \citep{chevallard_modelling_2016}, BAGPIPES \citep{Carnall:2018aa} and \pros\ \citep{Johnson:2021aa} instead assemble the SED from SSPs at every likelihood call and sample the resulting posterior with MCMC or nested sampling, which enables continuous parameters, flexible SFH priors, and uncertainties that reflect the degeneracies. It is this class of code that made non-parametric SFHs standard practice \citep{leja_how_2019}, and it is also this class that can be very expensive depending on its complexity. At the other end of the flexibility axis, full-spectrum fitting codes such as \textsc{alf} \citep{Conroy:2018aa} vary the abundances of individual elements and the initial mass function, but do so for a restricted family of star-formation histories.

A third strand attacks the cost of the model evaluation directly. Neural emulators, such as \textsc{SPECULATOR} \citep{Alsing:2020aa}, replace the SPS calculation with a network trained to reproduce it, reducing the per-evaluation cost by orders of magnitude (see also \citealt{Mathews:2023aa} for how far such emulators can be simplified before they break). Simulation-based inference (SBI) sidesteps the likelihood entirely and learns the posterior from simulated observations \citep{Hahn:2022aa, Khullar:2022aa}. Both are highly effective within the model and prior volume they were trained on. The cost is not that every galaxy needs a bespoke model, but that most changes of the model require a new training set and a retrained network. Model exploration, which should be routine, therefore becomes more expensive.

A fourth approach retains the physical model but changes how it is evaluated. DSPS \citep{Hearin:2023aa} implements the construction of composite stellar populations in \textsc{JAX} \citep{jax2018github}, including integration over the star-formation history and metallicity distribution, nebular emission, dust attenuation, and photometry. This makes the predicted photometry automatically differentiable with respect to the population parameters and allows the calculations to run natively on graphics processing units (GPUs), which are well suited to large matrix operations. \ceri\ shares this design for its CSP assembly (\S\ref{sec:maths}) and extends it to the remaining stages of the SED fit, including the projection onto spectra and emission-line catalogues and the likelihood with its nuisance terms. Together, the forward model and likelihood form a single compiled, vectorised, differentiable computation built from interchangeable physical components, which is evaluated at each sampler step.

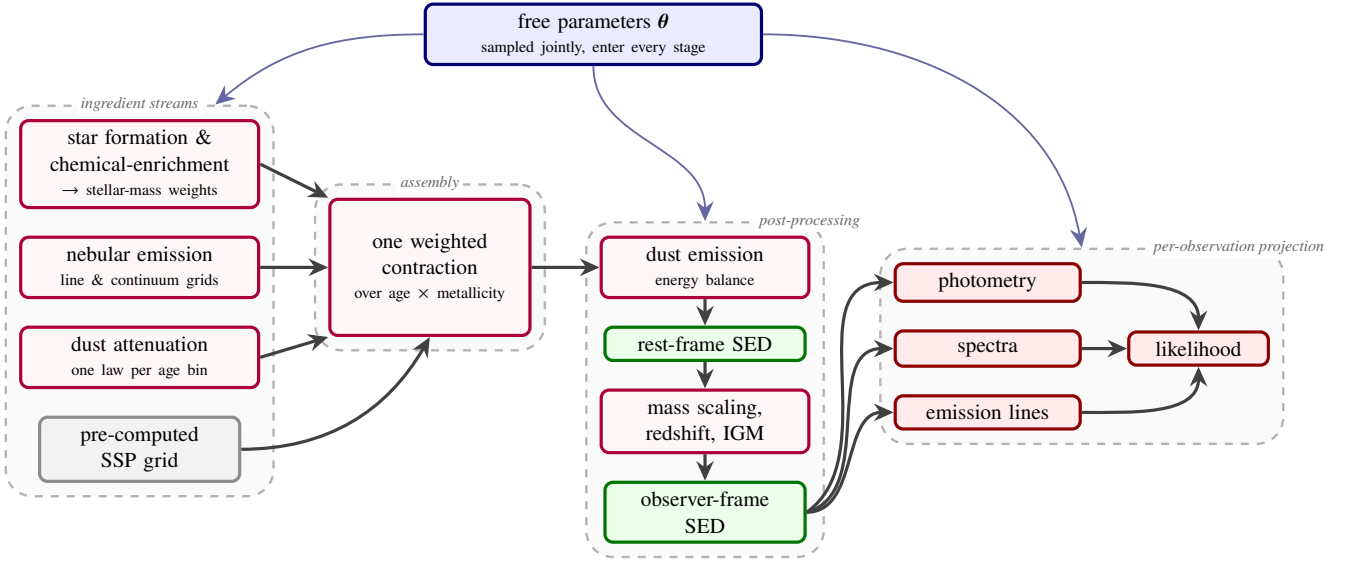
\begin{figure*}[!tb]
\centering
\begin{tikzpicture}[
    >={Stealth[length=2.6mm,width=2.3mm]},
    font=\footnotesize,
    node distance=3.5mm,
    box/.style ={rectangle, rounded corners=3pt, very thick, align=center, inner sep=3.5pt},
    par/.style ={box, draw=blue!45!black,   fill=blue!8,  text width=26mm},
    hyphenpenalty/.code={},
    sta/.style ={box, draw=black!45,        fill=black!5, text width=24mm},
    opp/.style ={box, draw=purple!90!black, fill=pink!12, text width=29mm},
    qty/.style ={box, draw=green!45!black,  fill=green!8, text width=24mm},
    obs/.style ={box, draw=red!55!black,    fill=red!8,   text width=22mm},
    blob/.style={rectangle, rounded corners=10pt, draw=black!30, dashed,
                 line width=0.9pt, fill=black!2, inner sep=5pt},
    blab/.style={font=\scriptsize\itshape, text=black!60, inner sep=1.5pt, fill=white},
    flow/.style={->, line width=1.2pt, draw=black!75},
    thin/.style={->, line width=0.7pt, draw=blue!40!black!60}
]
\node[opp]                    (sfh) {star formation \& chemical-enrichment\\[-1pt]{\scriptsize $\rightarrow$ stellar-mass weights}};
\node[opp, below=of sfh]      (neb) {nebular emission\\[-1pt]{\scriptsize line \& continuum grids}};
\node[opp, below=of neb]      (dust){dust attenuation\\[-1pt]{\scriptsize one law per age bin}};
\node[sta, below=of dust]     (ssp) {pre-computed\\SSP grid};
\node[opp, right=9mm of neb, text width=24mm, minimum height=18mm]
                              (sum) {one weighted\\contraction\\[-1pt]{\scriptsize over age $\times$ metallicity}};
\node[opp, right=9mm of sum, text width=25mm] (dem) {dust emission\\[-1pt]{\scriptsize energy balance}};
\node[qty, below=of dem]      (rest){rest-frame SED};
\node[opp, below=of rest, text width=25mm] (glob){mass scaling,\\redshift, IGM};
\node[qty, below=of glob]     (osed){observer-frame\\SED};
\node[obs, right=11mm of dem, yshift=-2mm]  (phot) {photometry};
\node[obs, below=of phot]     (spec) {spectra\\[-1pt]};
\node[obs, below=of spec]     (line) {emission lines};
\node[box, draw=red!55!black, fill=red!8, right=6mm of spec, text width=16mm] (like) {likelihood};
\node[par, above=7mm of sfh, xshift=60mm, text width=42mm]
                              (theta) {free parameters $\boldsymbol{\theta}$\\[-1pt]{\scriptsize sampled jointly, enter every stage}};
\begin{scope}[on background layer]
\node[blob, fit=(sfh)(neb)(dust)(ssp)] (BA) {};
\node[blob, fit=(sum)]                 (BB) {};
\node[blob, fit=(dem)(rest)(glob)(osed)] (BC) {};
\node[blob, fit=(phot)(spec)(line)(like)] (BD) {};
\end{scope}
\node[blab] at (BA.north) {ingredient streams};
\node[blab] at (BB.north) {assembly};
\node[blab] at ([xshift=-2mm]BC.north east) {post-processing};
\node[blab] at ([xshift=-6mm]BD.north east) {per-observation projection};
\draw[flow] (sfh.east)  -- (sum.north west);
\draw[flow] (neb.east)  -- (sum.west);
\draw[flow] (dust.east) -- (sum.south west);
\draw[flow] (ssp.east)  to[out=0, in=-120] (sum.south);
\draw[flow] (sum.east)  to[out=0, in=180] (dem.west);
\draw[flow] (dem) -- (rest);
\draw[flow] (rest) -- (glob);
\draw[flow] (glob) -- (osed);
\draw[flow] (osed.east) to[out=35, in=180] (phot.west);
\draw[flow] (osed.east)       to[out=15, in=180] (spec.west);
\draw[flow] (osed.east) to[out=0, in=180] (line.west);
\draw[flow] (phot.east) to[out=0, in=90]  (like.north);
\draw[flow] (spec.east) -- (like.west);
\draw[flow] (line.east) to[out=0, in=-90] (like.south);
\draw[thin] (theta.west) to[out=180, in=40] ([xshift=10mm]BA.north);
\draw[thin] (theta.south) to[out=-90, in=90] (BC.north);
\draw[thin] (theta.east) to[out=0, in=100] (BD.north);
\end{tikzpicture}
\caption{The \ceri\ forward model at a glance. Three parallel ingredient streams, the star-formation and chemical-enrichment history, the nebular line and continuum grids, and the per-age-bin dust attenuation, are assembled with the pre-computed SSP grid in a single weighted contraction over age and metallicity. The resulting rest-frame SED is post-processed (dust emission by energy balance; mass scaling, redshift, and IGM transmission) and the observer-frame SED is projected onto each observation: photometry, spectra, and emission lines, which share one likelihood. The galaxy's stellar and gas velocity dispersions are set once on the model and enter at this projection stage, where they are combined with the instrumental line-spread function of each spectrum. All free parameters are sampled jointly, and the whole graph is differentiable end to end, so likelihood gradients come for free. The complete data-flow diagram is given in Appendix~\ref{app:flow} (Fig.~\ref{fig:ceridwen_flow}).}
\label{fig:flow_cartoon}
\end{figure*}

\ceri\ adopts the modular modelling philosophy of \pros\ and translates many of its physical and observational treatments directly to \textsc{JAX}, allowing components to be added, exchanged, or omitted without redesigning the inference layer. It extends this framework with time-dependent chemical enrichment, variable $[\alpha/\mathrm{Fe}]$, and more flexible dust and escape geometries. Evolving metallicity histories are not new in themselves: \textsc{ProSpect} ties the metallicity to the mass growth of the SFH \citep[e.g.][]{Thorne:2022aa}. \ceri\ instead samples a free, non-parametric metallicity history at the resolution of the SFH. Its default vectorised nested slice sampler \citep[NSS;][]{Yallup:2025aa} evaluates populations of parameter vectors in parallel on the GPU while retaining the robustness to multimodality and Bayesian-evidence calculation of nested sampling \citep{Skilling2006}. \ceri\ is aimed at large samples of galaxies and at models too flexible for serial samplers, and it is designed for GPUs.

This paper documents the code and demonstrates which new parameter spaces the architecture allows us to explore. \S\ref{sec:forward_model} describes the forward model and its ingredients, \S\ref{sec:observations} its projection onto photometry, spectra and emission lines, and \S\ref{sec:inference} the likelihood, priors and samplers. On mock data (\S\ref{sec:demo_mock}), \ceri\ first reproduces known results on the information in photometry, spectroscopy and wavelength coverage (\S\ref{sec:demo_phot_spec_bands}). It then fits a time-varying chemical-enrichment history jointly with the SFH (\S\ref{sec:demo_zh}), and SFHs with about one hundred age bins (\S\ref{sec:high_detail_sfh}); the GPU throughput makes both regimes practical. With the $\alpha$-enhanced aMIST/C3K libraries of \textsc{FSPS}, it also samples $[\alpha/\mathrm{Fe}]$ jointly with $[\mathrm{Fe/H}]$, stellar mass, SFH and dust (\S\ref{subsec:alpha}). A mock suite matched to the \jwst Advanced Deep Extragalactic Survey (JADES) tests the calibration of the posteriors (\S\ref{subsec:mock_jades_data}). On real data (\S\ref{sec:demo_real}), we refit JADES galaxies with matched data and priors. \ceri\ returns posteriors consistent with \pros\ and, per fit, obtains them a median ${\sim}134\times$ faster on one GPU than \pros\ on CPU (\S\ref{sec:demo_jades}; \S\ref{sec:performance}). \S\ref{sec:future} outlines the extensions the modular structure invites next: an update to the nebular emission handling, $\alpha$-enhanced photoionisation grids, a time-varying $[\alpha/\mathrm{Fe}]$, and an AGN component.

\section{Building a forward model}
\label{sec:forward_model}

The forward model in stellar population inference should be tailored to the data and scientific question, balancing sufficient physical flexibility against the information content of the observations. The adopted parametrisation and priors form an integral part of the model.

In \ceri, the forward model is assembled from modular components, including pre-computed stellar population grids, prescriptions for the star-formation and chemical-enrichment histories, and optional treatments of dust attenuation, nebular continuum and line emission, and observational effects. Components can be selected according to the data and scientific application, while additional physical prescriptions and parameters can be incorporated as needed. Figure~\ref{fig:flow_cartoon} provides an overview of this framework, with a more detailed schematic shown in Fig.~\ref{fig:ceridwen_flow}.

\begin{figure*}[!tb]
    \centering
    \includegraphics[width=\linewidth]{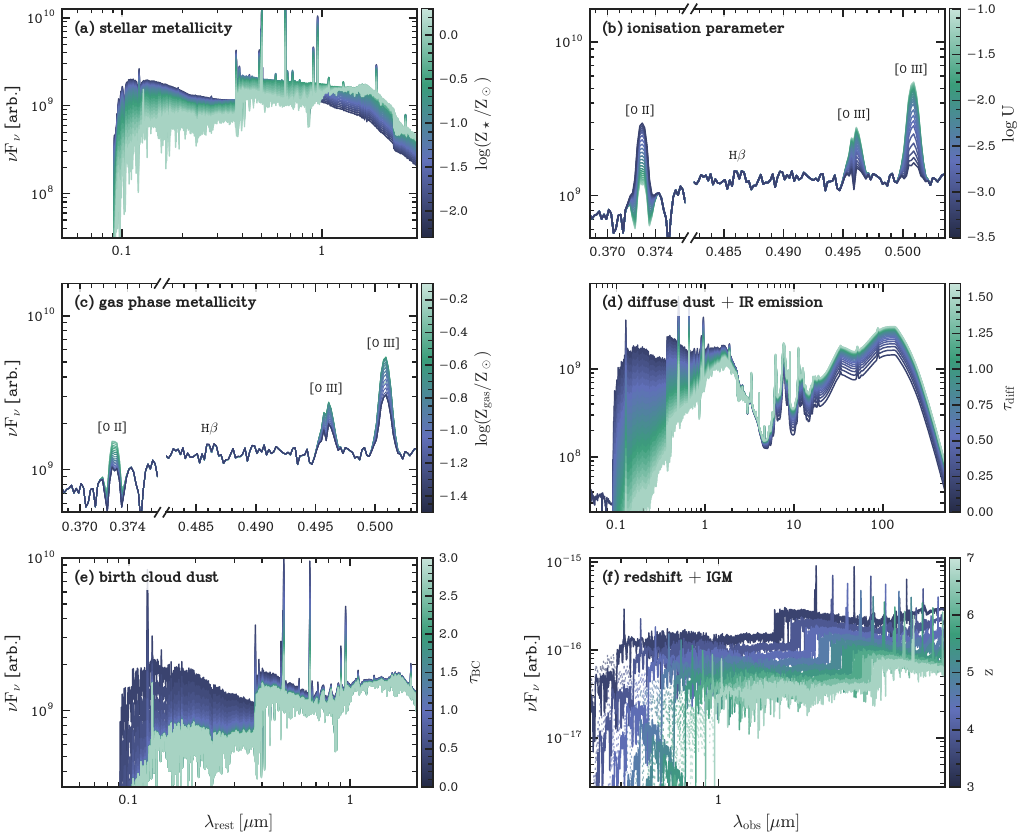}
    \caption{Examples of the modular components of the \ceri\ forward model. Starting from a fiducial composite stellar population, each panel varies one parameter group while keeping all others fixed. \textbf{(a)} The effect of stellar metallicity; \textbf{(b)} the ionisation parameter; \textbf{(c)} the gas-phase metallicity; \textbf{(d)} the diffuse dust attenuation optical depth and associated dust emission; \textbf{(e)} the birth-cloud dust attenuation optical depth, which affects only stars younger than $10$~Myr; and \textbf{(f)} redshift and intergalactic-medium (IGM) absorption, with the unabsorbed spectrum shown as dotted lines and the absorbed spectrum as solid lines. Panels \textbf{(b)} and \textbf{(c)} show the same two rest-frame wavelength regions around [\ion{O}{ii}]\,$\lambda\lambda3726,3729$ and H$\beta$+[\ion{O}{iii}]\,$\lambda\lambda4959,5007$ using a broken wavelength axis. The ionisation parameter primarily changes the relative strengths of [\ion{O}{ii}] and [\ion{O}{iii}] (O32 $=$ [\ion{O}{iii}]/[\ion{O}{ii}]), whereas the gas-phase metallicity changes their strengths relative to H$\beta$ (R23 $=$ ([\ion{O}{ii}]$+$[\ion{O}{iii}])/H$\beta$).}
    \label{fig:param_effects}
\end{figure*}

This section describes the main building blocks used by \ceri\ to generate model galaxy SEDs and their role in defining the inferred physical parameters. Throughout, $\boldsymbol{\theta}$ denotes the model-parameter dictionary, with $\theta[\texttt{key}]$ referring to the entry stored under \texttt{key}. Entries may be scalars (e.g. \texttt{logzsol}, \texttt{logmass}, or \texttt{gas\_logu}) or vectors. Parameters may be sampled directly under their own priors or derived from sampled parameters through user-defined transforms (\S\ref{sec:priors}); fixed parameters are implemented as constant transforms and therefore do not enter the sampled parameter set. Figure~\ref{fig:param_effects} illustrates how a subset of these parameters shapes the predicted SED.

\subsection{Single stellar populations}
\label{sec:ssp}
The foundation of the \ceri\ forward model is a grid of SSPs. Each SSP describes the spectrum of a coeval stellar population with a fixed stellar metallicity, age, initial mass function, and set of stellar-evolution assumptions. In practice, the SSP grid is a 3D object---spectra evaluated as a function of wavelength, with grid points in stellar age and stellar metallicity---or, when $\alpha$-enhancement is included, a 4D object with an additional [$\alpha$/Fe] axis. The rest of the model is built by interpolating within this grid and combining SSPs according to the assumed star-formation and chemical-enrichment histories. At present, \ceri\ does not generate SSP grids internally. Instead, we use a thin wrapper around \texttt{python-fsps} to generate the SSP spectra and store them in the \ceri-compatible \texttt{SSPData} format. This design keeps the expensive and currently non-differentiable SPS step separate from the fitting step: the SSP grid is generated once, saved to disk, and then passed to the forward model. However, this means that stellar-population properties, such as the slope of a parametric initial mass function (IMF), are not yet accessible as fit parameters. Through \textsc{FSPS}, \ceri\ can use a range of stellar libraries, including \textsc{MILES}~\citep{falcon-barroso_updated_2011} and \textsc{C3K} \citep{Park:2025aa}, as well as several commonly used isochrone sets, including \textsc{BaSTI}~\citep{Pietrinferni:2006aa}, \textsc{Geneva} \citep{Schaller:1992aa}, \textsc{MIST} \citep{choi_mesa_2016, Dotter:2016aa}, \textsc{Padova} \citep{Marigo:2008aa}, \textsc{PARSEC} \citep{Bressan:2012aa}, and the \textsc{BPASS} binary-population models \citep{stanway_re-evaluating_2018}, which come with their own stellar library.

The \texttt{SSPData} interface is independent of \textsc{FSPS}: any externally generated SSP library can be used once converted to the \ceri\ format. This includes the $\alpha$-enhanced grids of \citet{Park:2025aa} distributed with \ceri. The main limitation is nebular emission. The current implementation uses the \textsc{FSPS}-based photoionisation grids of \citet{Byler:2017aa} and is therefore self-consistent only for the stellar-population models for which matching grids are available: \textsc{BPASS}, \textsc{MIST}, \textsc{Padova}, and \textsc{PARSEC}; no equivalent grids are available for \textsc{BaSTI} or \textsc{Geneva}. \ceri\ selects the appropriate grid from the isochrone set recorded in the SSP file. No corresponding $\alpha$-enhanced photoionisation grids are currently available. The $\alpha$-enhanced fits in \S\ref{subsec:alpha} therefore omit nebular emission and are restricted to continuum-dominated, quiescent systems, where [$\alpha$/Fe] is most cleanly constrained. Emission lines can still be fitted with free fluxes (\S\ref{sec:likelihood}), but no nebular continuum can be included. Future versions of \ceri\ will support more flexible nebular-emission models, including on-the-fly photoionisation calculations, as well as stellar-population parameters such as the IMF slope (\S\ref{sec:future}).

\subsection{Star-formation history and chemical evolution}
\label{sec:maths} 

This subsection describes how the SFH and ZH set the weights of the SSP grid, expanding on the treatments of \citet{Johnson:2021aa} and \citet{Hearin:2023aa}. In particular, \texttt{sfh} and \texttt{logzsol\_hist} contain one value per node of the lookback-time grid, with $N_T$ elements in total; no parameters are stored per node of the SSP age–metallicity grid. The lookback-time grid increases from the present day ($T_0=0$, index 0) to the oldest node, so element 0 of each vector is the most recent.  The full derivation of the SFH/ZH-to-weights step, for both the step and the linear scheme, is given in Appendix~\ref{app:sfh_interp}.

\subsubsection{Star-formation history}
\label{sec:sfh}

\ceri\ adopts a flexible (non-parametric) parametrisation of the SFH: no analytic form is imposed, and the star-formation rate (SFR) is specified piecewise on a user-defined time grid. The free parameters are the tabulated SFRs, one value per node of the lookback-time grid or one per bin. Analytic SFHs can still be implemented by transforming a small set of sampled parameters into tabulated SFRs (\S\ref{sec:priors}). The CSP spectrum is assembled as 
\begin{equation}
    F_{\lambda,\,\mathrm{intrinsic}}(\boldsymbol{\theta}) \;=\;
    \sum_{m=1}^{N_{\rm Z}}\sum_{a=1}^{N_{\rm age}}
    W_{ma}(\boldsymbol{\theta})\,f^{\rm SSP}_{ma}(\lambda),
    \label{eq:csp_sum_main}
\end{equation}
where $f^{\rm SSP}_{ma}(\lambda)$ is the pre-computed SSP spectrum at metallicity node $m$ and log-age node $a$, and the weight $W_{ma}$ is the stellar mass formed at that grid point (Fig.~\ref{fig:ceridwen_flow}, Tier A).

\ceri\ supports both \pros\ schemes for mapping a tabulated SFH onto the SSP grid. The default \textit{step} scheme (\pros’s \texttt{FastStepBasis}) assumes a constant SFR within each bin, equal to the mean of its two node values when one SFR per node is given, while the \textit{linear} scheme of \citet[their appendix~B]{Johnson:2021aa} assumes a piecewise-linear SFR and integrates it analytically over the log-age-interpolated SSP grid. Both schemes conserve stellar mass; their derivations are given in Appendix~\ref{app:sfh_interp} and compared in Fig.~\ref{fig:sfh_interp}. In both schemes the weights sum to the time integral of the tabulated SFR; the forward model does not renormalise them. When the lookback-time grid is rescaled to a sampled redshift (\mbox{\S\ref{sec:global}}), the SFR is rescaled with it, so this integral is unchanged. In the parametrisation used in this paper, a transform generates the SFRs from sampled log-SFR ratios and normalises them to one solar mass formed (\S\ref{sec:priors}). The SFRs then set only the shape of the SFH. The overall normalisation is set separately by the stellar-mass parameter, which rescales the rest-frame spectrum at the end of the forward model (Fig.~\ref{fig:ceridwen_flow}). This separation allows independent priors on the SFH shape and total formed stellar mass. As in \pros, the latter is the time integral of the SFH; the surviving mass in stars and remnants is obtained in post-processing from the \textsc{FSPS} surviving-mass table, when available, using the same $W_{ma}$ weights as for the spectrum.

\subsubsection{Chemical evolution history}
\label{sec:zh}
The stellar metallicity of a galaxy is a running record of its baryon cycle: successive stellar generations enrich the gas that fuels later star formation. Stars born at different epochs can therefore have substantially different metallicities, and recovering this evolution jointly with the SFH is one route to constraining the gas flows that regulate galaxy growth. Star formation that continues at constant or declining metallicity can indicate dilution by inflowing metal-poor gas \citep{Lilly:2013aa, Cresci:2010aa}, while the metallicity reached before quenching can help distinguish starvation from rapid gas removal \citep{Peng:2015aa, Trussler:2020aa}. Modelling this evolution is also important for the SFH itself, since forcing all stellar populations to share a single metallicity distorts the inferred distribution of star formation \citep{Thorne:2022aa}.

The SFH interpolation of \S\ref{sec:sfh} gives weights along the SSP age axis; these must then be distributed across metallicity to obtain the full $W_{ma}$ matrix in Eq.~\eqref{eq:csp_sum_main}. \ceri\ supports either a single stellar metallicity or a time-varying chemical-enrichment history. In the \textit{constant-metallicity} mode, all stellar populations share one fitted metallicity, $\theta[\texttt{logzsol}]$ in $\log_{10}(Z/Z_\odot)$, where $Z_\odot$ is defined by the SSP grid in use. At each age, the stellar mass is distributed between the two bracketing metallicity grid points by linear interpolation in $\log_{10}(Z/Z_\odot)$. This reproduces the standard single-metallicity assumption used in many SED-fitting models.

In the \textit{time-varying} mode, the metallicity is instead described by a vector $\boldsymbol{\zeta}=\theta[\texttt{logzsol\_hist}]$ sampled on the same lookback-time grid as the SFH. Each SFH bin is assigned a metallicity from the neighbouring values of $\boldsymbol{\zeta}$, and its mass is then interpolated across the SSP metallicity grid in the same way as above. The SFH and enrichment history are therefore recovered at the same temporal resolution, at the cost of $N_T$ additional parameters. The elements of $\boldsymbol{\zeta}$ need not be sampled independently. As with other model parameters, the full enrichment history may instead be generated through a transform (\S\ref{sec:priors}), for example from a lower-dimensional chemical-evolution model. 

\subsection{Nebular emission: continuum and lines}
\label{sec:nebular}

Nebular lines and continuum can dominate the broad-band fluxes of young, high-redshift galaxies (\S\ref{sec:introduction}). Nebular emission in \ceri\ currently uses the grids of \citet{Byler:2017aa}, computed with \textsc{Cloudy} \citep{ferland_2013_2013}. The model is controlled by the gas-phase metallicity $\theta[\texttt{gas\_logz}] = \log_{10}(Z_{\rm gas}/Z_\odot)$, defined relative to the solar metallicity of the photoionisation grid, and the ionisation parameter $\theta[\texttt{gas\_logu}]$. The SSP age sets the ionising-photon budget and therefore the overall strength of the nebular emission. By default \mbox{\ceri} uses the grids computed without dust grains in the ionised gas, as \mbox{\textsc{FSPS}} does. It is possible to select the grids with dust instead. The demonstrations of \mbox{\S\ref{sec:demo_mock}} and the \mbox{\ceri} fits of \mbox{\S\ref{sec:demo_jades}} used the grids with dust.

\subsubsection{Grid evaluation and line insertion}
\label{sec:nebular_grids}

As in \textsc{FSPS}, the continuum and line luminosities per ionising photon are interpolated over gas metallicity, ionisation parameter, and SSP age. They are then scaled by the ionising-photon rate of each SSP node and added to the stellar SED before the contraction in Eq.~\eqref{eq:csp_sum_main} (Appendix~\ref{app:nebular}). Unlike \textsc{FSPS}, which inserts emission lines directly into the spectrum at the resolution of the stellar library, \ceri\ keeps the line luminosities separate from the continuum. On the model wavelength grid, lines are represented with the minimum two-pixel width required to conserve their flux. Their physical and instrumental broadening is applied only when projecting the model onto a specific observation (\S\ref{sec:spectra}): spectra receive analytic line profiles evaluated on their native pixels, while emission-line catalogues receive the line luminosities directly (\S\ref{sec:lines}). The interpolation formulas are given in Appendix~\ref{app:nebular}, and the line-profile treatment in Appendix~\ref{app:spec_smoothing}.
 
\subsubsection{Ionising-photon escape}
\label{sec:frac_obrun}
 
The escape of ionising photons is modelled by a single free parameter, $\theta[\texttt{frac\_obrun}] \equiv \fob \in [0,1]$, which follows the \textsc{FSPS} OB-runaway convention \citep{Conroy:2010aa}: \fob\ is the fraction of the young stellar populations that is not embedded in birth clouds, which are ``runaway OB stars or escaping ionizing radiation''. The runaway fraction modifies both the nebular emission and the escaping ionising continuum of the stellar populations that contribute to nebular emission. These populations correspond to the SSP ages covered by the chosen photoionisation library. Of their ionising emission, only the covered fraction $(1-\fob)$ contributes to the nebular line and continuum emission, while the remaining fraction \fob\ is allowed to escape. In the default $\fob=0$ case, the ionising continuum is fully absorbed and reprocessed. Priors on \fob\ and their influence on the inferred escape fractions are discussed in \citet{Stoffers:2026aa}.
 
The sampled parameter \fob\ is distinct from the emergent escape fraction relevant for reionisation,
\begin{equation}
    \fesc \;=\; \frac{\dot N_{\rm ion,esc}}{\dot N_{\rm ion,int}}.
    \label{eq:fesc_def}
\end{equation}
The emergent escape fraction is the ratio of the ionising-photon rate leaving the galaxy to the intrinsic stellar production rate. It equals \fob\ only for particular assumptions about the dust along the escape path. In \ceri, these assumptions are specified by the \texttt{fesc\_geometry} switch, which selects between two geometries:\\

\texttt{fesc\_geometry = 'runaway\_bc'} (default) reproduces the \textsc{FSPS} treatment. The per-age birth-cloud transmission of the young light is replaced by the runaway mixture
\begin{equation}
    T_{\rm BC}(\lambda) \;\longrightarrow\; (1-\fob)\,e^{-\tau_{\rm BC}(\lambda)} + \fob,
    \label{eq:runaway_mix}
\end{equation}
applied to the combined young array, which consists of the stellar continuum \emph{and} the nebular continuum, and equivalently to the emission-line luminosities; the full spectrum then crosses the diffuse screen $e^{-\tau_{\rm diff}(\lambda)}$ (\S\ref{sec:dust}). The emergent escape fraction in this geometry is therefore
\begin{equation}
    \fesc \;=\; \fob\, e^{-\tau_{\rm diff}(912\,\mbox{\scriptsize\AA})} \;\leq\; \fob,
    \label{eq:fesc_runaway}
\end{equation}
with equality only in the limit of vanishing diffuse dust.\\

\texttt{fesc\_geometry = 'picket'} implements a picket-fence geometry. A fraction \fob\ of the young stellar light, including the ionising continuum, escapes through sightlines free of both birth-cloud and diffuse dust. The remaining fraction $(1-\fob)$ is fully covered, powers the nebular continuum and lines, and is attenuated by both dust components. Older stars are unaffected by the birth-cloud geometry and experience only diffuse attenuation. In this case, the emergent escape fraction is simply the sampled parameter, $\fesc=\fob$. 

\subsection{Dust attenuation and emission}
\label{sec:dust}

Dust is one of the most uncertain, yet most consequential, components of the forward model. SED fitting constrains an \emph{attenuation} curve rather than an extinction curve along a single sight line: attenuation combines absorption, scattering, and the relative geometry of stars and dust within the aperture \citep{calzetti_dust_2000,salim_dust_2020}. Its shape therefore varies between galaxies, in both slope and normalisation \citep{kriek_dust_2013,Chevallard:2013aa}. Assuming a fixed attenuation law can consequently bias inferred stellar masses, star-formation rates, and ultraviolet slopes \citep{leja_deriving_2017,salim_dust_2020,Lower:2020aa,narayanan_ultraviolet_2024}. The attenuation is also age dependent: stars younger than ${\sim}10\,$Myr are still embedded in their birth clouds and more heavily obscured than the populations that have drifted out of them \citep{charlot_simple_2000}, as the systematically larger attenuation of the nebular lines relative to the stellar continuum confirms \citep{price_direct_2014,reddy_connection_2016}. Because those young populations also dominate the light, their obscuration is degenerate with the recent SFHs and can bias stellar masses substantially at high redshift \citep{Narayanan:2024aa}. A part of the absorbed light is re-emitted in the infrared, so panchromatic data can constrain the total attenuation through energy balance \citep{Draine:2007aa}.

Dust attenuation in \ceri\ is built to be as flexible as the data warrant. The stellar SSP grid can be divided along the age axis into any number of user-defined, potentially overlapping bins, each with its own attenuation law and free parameters. An optional diffuse component can additionally apply a single attenuation law to all stellar populations, independent of age. The default is the two-component model of \citet{charlot_simple_2000}, with a power law applied to stars younger than ${\approx}11$~Myr (the default birth-cloud bin edge $\log_{10}(\mathrm{age}/\mathrm{Gyr})=-1.97$) and a diffuse component following the attenuation law of \citet{kriek_dust_2013}. In $\boldsymbol{\theta}$, the birth-cloud power law has optical depth $\tau_{\rm BC}=\theta[\texttt{tau\_pow}]$ at $5500$\,\AA\ and slope $\theta[\texttt{alpha\_pow}]$; the diffuse component has $\tau_{\rm diff}=\theta[\texttt{diffuse\_tau\_kc}]$ and slope $\delta_{\rm dust}=\theta[\texttt{diffuse\_dust\_index}]$. This reproduces the \textsc{FSPS} \texttt{dust1}/\texttt{dust2} treatment used by \pros. The generalisation to an arbitrary number of age bins, each with the same or different attenuation laws, is new to \ceri. The available attenuation curves are stored in a registry distributed with the code. This currently contains twelve laws: nine of the curves in \texttt{sedpy} \citep{johnson_2021_4582723}, ported to \textsc{JAX}, together with the \citet{kriek_dust_2013} curve used by \textsc{FSPS} and \pros, the SMC-bar curve of \citet{Gordon:2003aa} and the curve of \citet{Reddy:2015aa}, as implemented in \textsc{FSPS} and \pros. Newly registered laws are immediately available to every age bin. The total attenuation enters as a multiplicative factor in the same contraction as the SSP grid and SFH weights (Fig.~\ref{fig:ceridwen_flow}, Tier~B). The detailed mapping between attenuation bins and the SSP age grid is given in Appendix~\ref{app:dust}.

\begin{figure}[!tb]
    \centering
    \includegraphics[width=\linewidth]{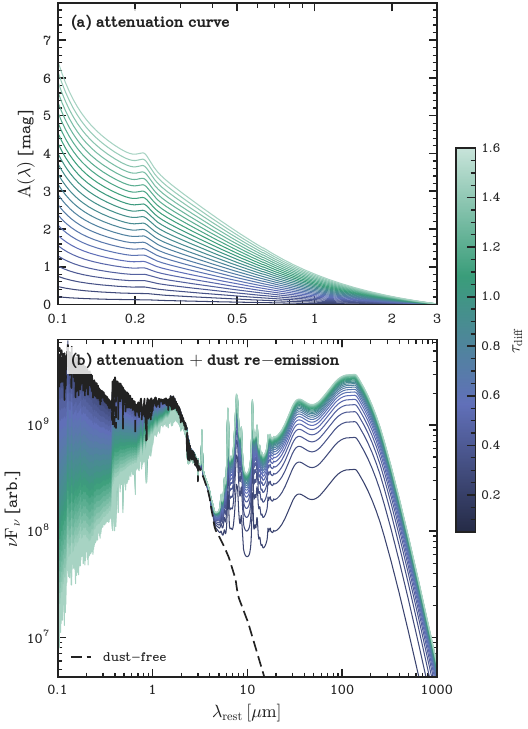}
    \caption{Dust attenuation and emission in \ceri. \textbf{(a)} The effective attenuation $A_\lambda$ (mag) of the default diffuse attenuation curve \citep{kriek_dust_2013}, swept over the optical depth $\tau_{\rm diff}$ (colour); age-binned birth-cloud laws enter the model in the same way (\S\ref{sec:dust}). \textbf{(b)} The same populations with energy-balance dust emission. Each curve is the attenuated stellar spectrum plus the infrared emission for the matching $\tau_{\rm diff}$ in panel~\textbf{(a)}, from an interpolated DL07 template with a single self-absorption correction. The dashed line is the dust-free spectrum.}
    \label{fig:dust}
\end{figure}

Optionally, the energy absorbed by dust is re-emitted in the infrared under energy balance. \ceri\ supports either the \citet{Draine:2007aa} (DL07) templates used by \textsc{FSPS} and \pros, or the \citet{Jones:2017aa} (THEMIS) templates, selected at model construction. Both are parametrised by the PAH mass fraction $\theta[\texttt{duste\_qpah}]$, the minimum starlight intensity $\theta[\texttt{duste\_umin}]$, and the fraction $\theta[\texttt{duste\_gamma}]$ of dust heated by stronger radiation fields, although the $q_{\rm PAH}$ parameter has a different numerical scale in the two models. The emitted infrared spectrum is then attenuated once by the diffuse dust to account for self-absorption (Appendix~\ref{app:dust_emission}; Fig.~\ref{fig:ceridwen_flow}, Tier~C). Figure~\ref{fig:dust} shows the default diffuse attenuation and the resulting dust emission.

\subsection{Kinematics}
\label{sec:kinematics}

The line-of-sight motions of stars and ionised gas broaden their spectral features. \ceri\ models these effects with two source-level parameters stored in a \texttt{Kinematics} object: the stellar velocity dispersion $\sigma_{\rm gal}$, which broadens the continuum, and the gas velocity dispersion $\sigma_{\rm gas}$, which sets the widths of the emission lines. The continuum is everything except the emission lines: the stars, the nebular continuum and the dust emission. Each dispersion is applied once, when the model is projected onto an observation. $\sigma_{\rm gal}$ acts only on the continuum, and $\sigma_{\rm gas}$ only on the emission lines. They are modelled as Gaussian widths in velocity, equivalently in $\ln\lambda$, and therefore combine in quadrature with other broadening terms (\S\ref{sec:spectra}, Eq.~\eqref{eq:one_kernel}). By default, the gas dispersion is tied to the stellar dispersion, but either can be specified independently. Each may be fixed or associated with an entry of $\boldsymbol{\theta}$ and inferred from the data under its own prior. By default both dispersions are $300$\,km\,s$^{-1}$.

The source broadening is applied when the model is projected onto the observations (\S\ref{sec:observations}), including both fitted spectra and, by default, the spectrum integrated through the photometric filters. Instrumental broadening is treated separately: the line-spread function belongs to each individual spectrum and is specified by the corresponding \texttt{Instrument} attached to the \texttt{Spectrum} observation (\S\ref{sec:spectra}).

\subsection{Observer-frame conversion}
\label{sec:global}

In the final stage of the forward model, the rest-frame SED from Eq.~\eqref{eq:csp_sum_main} is converted into an observer-frame spectrum (Fig.~\ref{fig:ceridwen_flow}, Tier~C).

The spectrum is first scaled by the total formed stellar mass, $10^{\theta[\texttt{logmass}]}$, which sets the luminosity normalisation independently of the SFH shape (\S\ref{sec:sfh}). It is then converted from the rest frame at $10$~pc to the observer-frame flux density using the luminosity distance $D_L(z)$ and the standard cosmological flux factor $(1+z)(d_{10,\mathrm{pc}}/D_L)^2$. The SSP spectra are tabulated per unit solar luminosity, with $L_\odot=3.839\times10^{33}$\,erg\,s$^{-1}$, the value adopted by \textsc{FSPS}. The resulting spectrum is expressed in cgs $F_\nu$; conversion to maggies is performed only during the photometric projection (\S\ref{sec:photometry}).

\ceri\ evaluates $D_L(z)$ and the age of the Universe with a differentiable flat-$\Lambda$CDM cosmology implemented in \textsc{JAX}. This allows redshift to be sampled jointly with the stellar-population parameters, with its effect propagated through both the flux normalisation and the age-dependent SFH grid. The cosmology is chosen at model construction: a predefined model (Planck 2018, Planck 2015 or WMAP9) or user-supplied parameters. At fixed redshift, the oldest SFH node must not predate the Universe. With a sampled redshift, the lookback-time grid can be rescaled to the age of the Universe at each sampled $z$. The rescaling preserves the formed mass of the SFH.

Finally, an optional intergalactic-medium (IGM) transmission curve attenuates the spectrum, using by default the \citet{madau_radiative_1995} prescription with an optional free scaling $\theta[\texttt{igm\_factor}]$. \ceri\ can additionally include a damping wing from a partly neutral IGM \citep{Miralda-Escude:1998aa, Totani:2006aa} and a damped Ly$\alpha$ absorber (DLA) \citep{Tepper-Garcia:2006aa}. The neutral fraction $\theta[\texttt{x\_HI}]$, DLA column density $\theta[\texttt{logN\_HI}]$, and absorber redshift $\theta[\texttt{z\_dla}]$ may each be fixed or fitted. For a spectrum, and for photometry with velocity broadening, the IGM transmission is applied after the galaxy's velocity broadening and before the instrumental line-spread function. The model remains on the rest-frame wavelength grid; conversion to the observer frame is performed separately for each observation during the projection step (Fig.~\ref{fig:ceridwen_flow}, Tier~D).

\section{Observation projections}
\label{sec:observations}

The forward model described in the previous section produces the observer-frame SED of a galaxy, but this is not what our instruments measure. Instead, observations sample the spectrum through the instrumental response and the galaxy's kinematics (\S\ref{sec:kinematics}): broadband photometry integrates the spectrum through non-tophat filter transmission curves, and spectroscopy is further broadened by the instrument's line-spread function. The \texttt{Observation} classes in \ceri\ project the model SED into each of these observables.
Only a subset of the projection parameters is specific to the observations themselves. For spectra, these include the spectrophotometric calibration parameters $\theta[\texttt{spectrum\_scaling}]$ and $\theta[\texttt{spectrum\_calib}]$, together with the instrumental-width scale; for emission-line catalogues, the aperture factor is $\theta[\texttt{eline\_scaling}]$. Each observation has its own noise parameters: an error-bar scale, an additive jitter, model- and data-anchored fractional floors and an outlier fraction (\S\ref{sec:likelihood}; Appendix~\ref{app:sys_floor}). Their names end in the kind of observation, \texttt{phot}, \texttt{spec} or \texttt{lines}, for example $\theta[\texttt{log\_jitter\_phot}]$. With several observations of one kind, the observation's name is appended as well. All remaining quantities in the projection are fixed at model construction and folded into precomputed operators. We describe the photometric projection in \S\ref{sec:photometry}, the spectral projection and broadening in \S\ref{sec:spectra}, and the emission-line fluxes in \S\ref{sec:lines}.

\subsection{Photometry}
\label{sec:photometry}

Photometry in \ceri\ follows \citet{Johnson:2021aa}. A \texttt{Photometry} observation stores the measured AB fluxes, uncertainties, and filter set, with optional masks and upper-limit flags. Fluxes and uncertainties are in maggies, the linear AB unit (1 maggy $=3631$\,Jy, so $m_{\rm AB}=-2.5\log_{10}f$). Filter curves are drawn from the \textsc{sedpy} library or supplied by the user.

Model photometry is obtained by integrating the observer-frame spectrum through each filter in the photon-counting AB convention. At fixed redshift, this projection is precomputed as a linear operator; when $\theta[\texttt{zred}]$ is sampled, the spectrum is interpolated onto the filter grid at each evaluation so that the predicted fluxes remain differentiable in redshift. Implementation details are given in Appendix~\ref{app:phot}, while upper limits and additional photometric uncertainties are treated in the likelihood (\S\ref{sec:likelihood}).

\subsection{Spectra}
\label{sec:spectra}

A \texttt{Spectrum} observation holds the observer-frame flux density $F_\nu$, in erg\,s$^{-1}$\,cm$^{-2}$\,Hz$^{-1}$, on a fixed pixel grid with its per-pixel $1\sigma$ uncertainty and an inclusion mask, plus optional vectors for an unsubtracted sky background, a multiplicative spectrophotometric calibration, and \texttt{Instrument}, the instrumental line-spread function (LSF). Its role is to project the rest-frame model spectrum (\S\ref{sec:global}) onto the detector grid.

\begin{figure*}[!tb]
    \centering
    \includegraphics[width=\linewidth]{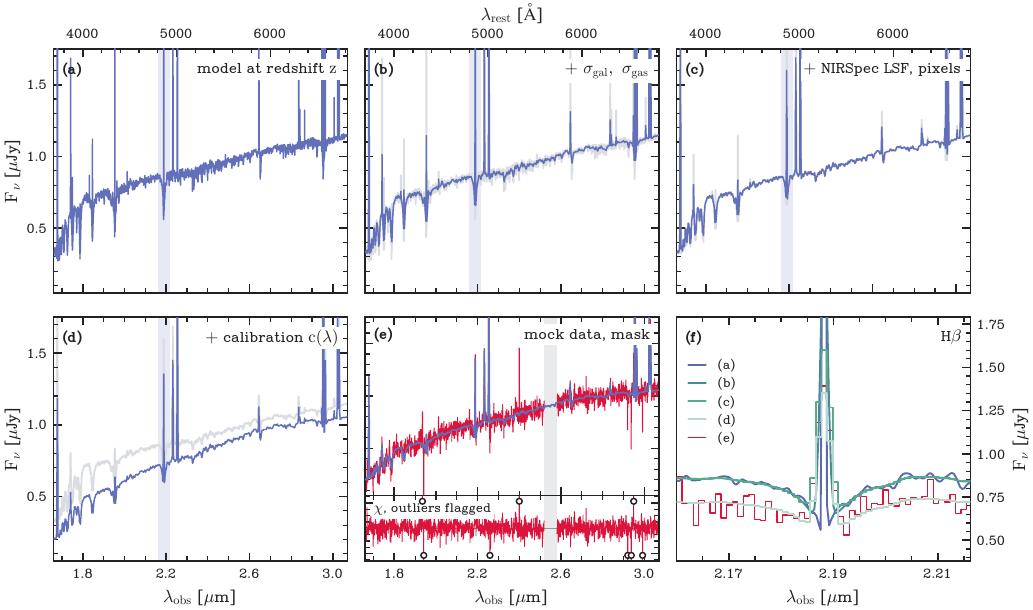}
    \caption{Spectral projection in \ceri\ for a mock galaxy at $z=3.5$ observed with \jwst/NIRSpec G235M/F170LP ($1.66$--$3.07\,\mu$m). Each panel adds one operation to the previous one, shown in grey. \textbf{(a)} Intrinsic model spectrum at the SSP-grid resolution, placed at the sampled redshift. \textbf{(b)} Velocity broadening: $\sigma_{\rm gal}$ broadens the continuum and $\sigma_{\rm gas}$ the emission lines. \textbf{(c)} Convolution with the G235M line-spread function, with the library resolution removed, and resampling onto the detector pixels. \textbf{(d)} Spectrophotometric calibration through \texttt{spectrum\_scaling} and \texttt{spectrum\_calib}. \textbf{(e)} Mock data (red) with Gaussian noise, injected outliers and a masked detector region (grey band), with the \ceri\ prediction (blue); the lower strip shows the normalised residuals $\chi$, and open circles mark the pixels flagged by the outlier mixture. \textbf{(f)} The H$\beta$ region (blue band in panels a--d) with the model of every stage (a)--(d) and the mock data (e). All broadening, calibration and noise parameters can be sampled.}
    \label{fig:spec_smoothing}
\end{figure*}

The spectral projection accounts jointly for the intrinsic velocity broadening of the source, the instrumental LSF, and the finite resolution of the stellar library. Assuming Gaussian broadening in velocity space, the effective widths are

\begin{align}
    \sigma_{\rm cont}^{2}(\lambda_{i}) &\;=\; \sigma_{\rm gal}^{2} + \sigma_{\rm inst}^{2}(\lambda_{i}) - \sigma_{\rm lib}^{2}(\lambda_{i}),\nonumber\\
    \sigma_{\rm line}^{2}(\lambda_{i}) &\;=\; \sigma_{\rm gas}^{2} + \sigma_{\rm inst}^{2}(\lambda_{i}),
    \label{eq:one_kernel}
\end{align}

for the continuum (stars, nebular continuum and dust emission) and the emission lines, respectively. Here $\sigma_{\rm gal}$ and $\sigma_{\rm gas}$ are the stellar and gas velocity dispersions introduced in \S\ref{sec:kinematics}, $\sigma_{\rm inst}$ is the instrumental LSF, and $\sigma_{\rm lib}$ is the spectral resolution already present in the SSP library. The library term is removed only from the continuum because the emission lines are generated independently of the stellar-library resolution. The emission lines are not part of the spectrum that is convolved with $\sigma_{\rm cont}$. They are added afterwards, on the detector pixels, as Gaussians of width $\sigma_{\rm line}$ (Appendix~\ref{app:spec_smoothing}). Both source dispersions may be fixed or sampled, and an optional scale factor can likewise be applied to the instrumental width. Details of the smoothing and resampling are given in Appendix~\ref{app:spec_smoothing}.
For a free redshift, the rest-frame spectrum is shifted continuously in $\ln\lambda$, while emission lines are placed at their corresponding observed wavelengths. This keeps the spectral projection differentiable with respect to redshift. The same machinery supports both constant and wavelength-dependent LSFs. Fig.~\mbox{\ref{fig:spec_smoothing}} follows one model galaxy through each step of the projection onto the wavelength-dependent LSF and the detector pixels of \mbox{\jwst}/NIRSpec G235M.

Spectrophotometric calibration can be supplied as a fixed correction or fitted jointly with the physical model. The sampled form consists of a grey scaling $\theta[\texttt{spectrum\_scaling}]$ and wavelength-dependent polynomial coefficients $\theta[\texttt{spectrum\_calib}]$. Alternatively, a calibration polynomial may be optimised at each likelihood evaluation following \citet[][their \S2.3]{Johnson:2021aa}. It may also be marginalised analytically. The likelihood is linear in the polynomial coefficients, so with Gaussian priors on them the coefficients integrate out in closed form (Appendix~\ref{app:spec_polymarg}). The calibration uncertainty then enters the posterior and the evidence, at the same cost per likelihood evaluation as the optimised polynomial. The post-processing reports the coefficients conditional on each posterior draw. A supplied sky spectrum is subtracted from the data before evaluating the residuals. Noise terms and correlated residuals are treated in \S\ref{sec:likelihood}.

\subsection{Lines}
\label{sec:lines}

The \texttt{Lines} observation handles catalogued emission-line fluxes that have been collapsed to a single number per line. It stores the \textsc{FSPS} line indices to predict, the vacuum rest-frame wavelengths, the observed integrated fluxes (typically in erg\,s$^{-1}$\,cm$^{-2}$) with their per-line $1\sigma$ uncertainties and an inclusion mask, and optionally a list of human-readable line names and a boolean array flagging upper limits.

The corresponding model prediction is the observed-frame integrated luminosity of the line itself, read from the nebular grid at the model's ionising budget and carried through dust, mass, distance and IGM exactly like the continuum (Appendix~\ref{app:line_proj}); unresolved blends are the sums of their component grid lines. No profile and no broadening enter, since an integrated flux carries no line-shape information. This is the same quantity \pros\ compares with a line catalogue, and a \texttt{Lines} observation therefore requires a nebular component. 

\section{Inference}
\label{sec:inference}

The forward model described above maps a parameter vector $\boldsymbol{\theta}$ to predicted observables; inference uses the data to constrain the parameters that produced them. In \ceri, this is expressed through the posterior $p(\boldsymbol{\theta}\mid\boldsymbol{d}) \propto \mathcal{L}(\boldsymbol{\theta})\,\pi(\boldsymbol{\theta})$, where the likelihood compares the model predictions with the data (\S\ref{sec:likelihood}) and $\pi(\boldsymbol{\theta})$ specifies the prior (\S\ref{sec:priors}). The posterior is then explored with one of the samplers described in \S\ref{sec:sampling}. Because the forward model is compiled, vectorised, and differentiable, posterior evaluations can be batched for nested sampling and differentiated directly for gradient-based inference.

\subsection{Likelihood}
\label{sec:likelihood}

The observations, such as photometry, emission lines and one or more spectra, are modelled as a set of conditionally independent arms (one term, or `arm', per observation), so that the total log-likelihood is the sum of the per-arm contributions assembled by the model's \texttt{predict} dispatch (\S\ref{sec:observations}). Each arm uses a diagonal Gaussian likelihood,
\begin{equation}
    \ln\mathcal{L}(\boldsymbol{\theta})
    \;=\;
    \sum_{\rm arms}\;\sum_{i\in\mathcal{M}}
    \left[
        -\frac{1}{2}\frac{\bigl(d_{i}-m_{i}(\boldsymbol{\theta})\bigr)^{2}}
                          {\sigma_{{\rm eff},i}^{2}}
        -\frac{1}{2}\ln \bigl(2\pi\,\sigma_{{\rm eff},i}^{2}\bigr)
    \right],
    \label{eq:total_lnlike}
\end{equation}

where $d_{i}$ and $m_{i}$ are the observed and predicted data, $\mathcal{M}$ the inclusion mask, and $\sigma_{{\rm eff},i}$ the effective per-datum uncertainty. By default, each datum uses its quoted uncertainty, but \ceri\ can augment this with several additional noise terms. These include a fractional uncertainty on the data to represent calibration errors, a fractional uncertainty on the model to capture model imperfections, a constant jitter term, and a global scale factor $s$ that rescales the quoted uncertainties when they are believed to be systematically under- or over-estimated (Appendix~\ref{app:sys_floor}). The model-dependent term is particularly useful for emission lines, where residual errors in the photoionisation-grid line ratios can dominate the uncertainty of the brightest features. Because some of these terms depend on $\boldsymbol{\theta}$, \ceri\ retains the normalisation term of the Gaussian log-likelihood (Appendix~\ref{app:sys_floor}). Upper limits are treated with the corresponding one-sided likelihood (Appendix~\ref{app:line_lik}).
Each observation can additionally use the pixel-outlier mixture model of \citet[their Eq.~D11]{Johnson:2021aa}, following \citet{Hogg:2010aa}, with a fixed or sampled outlier fraction and inflation factor; this option is disabled by default. Correlated residuals between spectral pixels can be modelled with a Gaussian process (GP) in the likelihood used by all samplers, with its amplitude and correlation length fixed or sampled (Appendix~\ref{app:spec_GP}).
Emission-line amplitudes may also be treated as linear nuisance parameters and marginalised analytically, following \citet[their appendix~E]{Johnson:2021aa}. For either a flat prior or a Gaussian prior centred on the photoionisation-grid prediction, the marginal likelihood reduces to a weighted least-squares solve at each evaluation, while retaining the information from the stellar absorption beneath the lines. In \ceri, the emission-line amplitudes are marginalised for a given spectrum and shared consistently with the corresponding photometry and emission-line catalogue, making the marginalisation joint across all three observables. In this mode, the emission lines no longer constrain the photoionisation model, so the nebular parameters are held fixed. With a flat prior, no photoionisation grid is required, which also makes this treatment available for the $\alpha$-enhanced fits of \S\ref{subsec:alpha}. The line profiles use the gas velocity dispersion of \S\ref{sec:kinematics} together with the instrumental LSF; Ly$\alpha$ is always assigned a flat prior. The marginal likelihood is given in Appendix~\ref{app:line_marg}.

The resulting log-likelihood combines with the log-prior to form the log-posterior, which remains differentiable with respect to all sampled parameters.

\subsection{Priors}
\label{sec:priors}

Priors are an integral part of the inference, particularly when the data do not fully constrain the parameters of a flexible model. In that regime, the posterior can depend strongly on the adopted prior. The SFH is a particularly important example: restricting it to a parametric family can impose strong constraints on inferred ages and stellar masses \citep{carnall_how_2019, Lower:2020aa}, while in non-parametric models the prior controls how strongly neighbouring time bins may vary and therefore the degree of smoothness or burstiness that can be recovered \citep{ocvirk_stecmap_2006, leja_how_2019}. Recent work has moved towards physically motivated priors rather than choices made primarily for analytical convenience. For example, the stochastic prior of \citet{wan_stochastic_2024} is calibrated to the SFH power spectra of cosmological simulations \citep{Iyer:2020aa}. \ceri\ is designed to make such priors straightforward to implement, exchange, and test.

Each sampled parameter is assigned either a user-defined prior or one from a small built-in library: uniform (\texttt{TopHat}), normal, truncated normal (\texttt{ClippedNormal}), log-normal, log-uniform, or Student-$t$. These priors wrap the corresponding \textsc{TensorFlow Probability} distributions \citep{Dillon:2017aa} on the \textsc{JAX} backend and provide the operations required by the samplers: evaluation of the log-density, random sampling, and the inverse of the cumulative distribution function, which maps the unit interval onto the prior.

The joint prior factorises over the sampled coordinates, $\pi(\boldsymbol{\theta}_{\rm samp})=\prod_p \pi_p(\theta_p),$ but this does not restrict the physical parameters to independent priors. Correlations are introduced through differentiable transforms, analogous to \pros\ \texttt{depends\_on}. These transforms map sampled coordinates onto the parameters used by the forward model, so derived quantities are removed from the sampled set. This mechanism can represent, for example, parametric SFHs, chemical-enrichment models that generate $\theta[\texttt{logzsol\_hist}]$ (\S\ref{sec:zh}), or redshift-dependent lookback-time grids. As an example, the fits presented later in this paper use the non-parametric SFH transform shipped with \ceri, which samples the logarithmic ratios between adjacent time bins under a Student-$t$ continuity prior. A transform then maps these ratios to the unit-mass SFH vector $\theta[\texttt{sfh}]$. Independent priors on the sampled ratios therefore induce a correlated prior on the SFH, with the scale $\sigma$ and degrees of freedom $\nu$ controlling the allowed variation between neighbouring bins.

Nested sampling draws directly from the specified priors, whereas the gradient-based samplers operate in an unconstrained parameter space (\S\ref{sec:sampling}). Under nested sampling, every sampled parameter must therefore have an explicit prior; for gradient-based sampling, a missing prior is treated as improper flat.

After a fit, \ceri\ reports how strongly the data, rather than the prior, constrain each parameter. It quantifies this in post-processing using the information gain $D_{\rm KL}\!\left[p(\theta_p \mid \boldsymbol{d})\,\|\,\pi(\theta_p)\right]$, the Kullback--Leibler divergence between the marginal posterior and its prior, measured in nats. The divergence is zero when posterior and prior are identical and increases as the data shift or concentrate the posterior relative to the prior. As a guide, for a Gaussian posterior centred in a Gaussian prior, $D_{\rm KL}=\ln r+1/(2r^{2})-1/2$, with $r$ the ratio of prior to posterior width. Then $0.1$\,nat corresponds to $r\approx1.4$, $1$\,nat to $r\approx4.4$ and $2$\,nats to $r\approx12$. We call a parameter data-constrained when $D_{\rm KL}\gtrsim1$\,nat. It therefore provides a common measure of how informative the data are for each parameter. Comparing $D_{\rm KL}$ with properties of the observations can further indicate which measurements carry the information that constrains a given parameter.

\begin{figure*}[!tb]
    \centering
    \includegraphics[width=\linewidth]{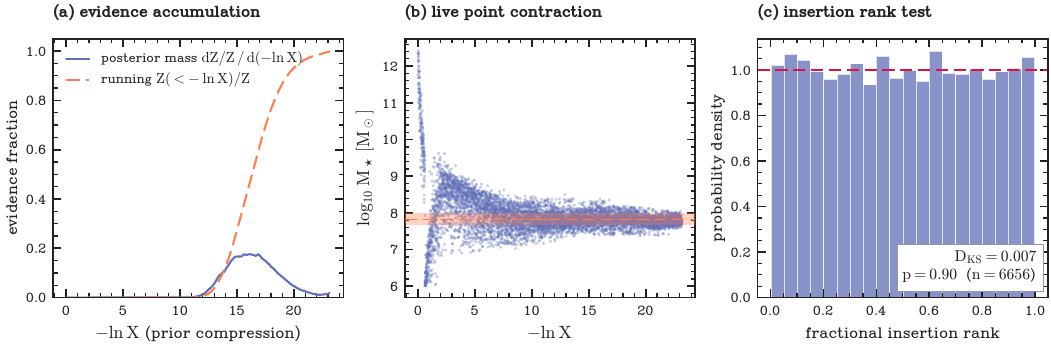}
        \caption{Convergence diagnostics of the nested slice sampler for the JADES 170891 fit (\S\ref{sec:demo_jades}). \textbf{(a)} The normalised posterior mass per unit prior compression, $\mathrm{d}Z/Z/\mathrm{d}(-\ln X)$, where $X$ is the prior volume enclosed by the current likelihood contour (blue solid), and the running evidence fraction (coral, dashed): the run accumulates the bulk of the evidence before termination, the residual mass at the right edge being carried by the final live population. \textbf{(b)} Values at the dead (discarded) points of the stellar mass against compression: the live population fills its prior early in the run and contracts onto the final posterior (coral band and dashed line: posterior $68\%$ interval and median). \textbf{(c)} The fractional insertion ranks of new live points \citep{Fowlie:2020aa}, uniform for a correctly mixing inner kernel, with the Kolmogorov--Smirnov distance and $p$-value annotated; the crimson dashed line is the uniform expectation.}
    \label{fig:nss_diag}
\end{figure*}

\subsection{Sampling algorithms}
\label{sec:sampling}
We adopt nested sampling \citep{Skilling2006} as the default inference engine behind every posterior in this paper, for three reasons:
\begin{itemize}
    \item it is robust to the multimodal posteriors that flexible SFHs induce
    \item it returns the Bayesian evidence alongside the posterior, enabling per-galaxy model comparison 
    \item it replaces a batch of live points per iteration in parallel, which makes sampling on a GPU efficient
\end{itemize}
Specifically, \ceri\ uses NSS \citep{Yallup:2025aa} implemented in \textsc{BlackJAX} \citep{Cabezas:2024aa}. Nested sampling progressively restricts the prior to regions of increasing likelihood; discarded live points are replaced by short Markov chains constrained to the current likelihood contour. NSS uses a gradient-free hit-and-run slice sampler for these replacement chains, with proposal directions adapted to the covariance of the live population. Nested sampling in \ceri\ therefore does not require derivatives of the forward model.

The replacement chains are vectorised over the same GPU batch dimension as the forward model, so several live points can be evolved in parallel. The main numerical settings are the number of live points, the number $k$ of live points replaced per iteration, and the number of slice steps per replacement. We scale the latter with problem dimensionality and report the settings used for each demonstration in \S\ref{sec:demo_mock} and \S\ref{sec:demo_jades}. \citet{Yallup:2025aa} demonstrate NSS for problems with up to ${\sim}100$ dimensions and argue that it remains suitable for $\mathcal{O}(100)$-dimensional inference. The $125$-dimensional fits of \S\ref{sec:high_detail_sfh} therefore probe the upper end of this regime; their computational cost is examined in \S\ref{sec:performance}.

Runs can be checkpointed periodically, when a checkpoint directory is set, and resumed from the complete sampler state. Sampling terminates when the evidence remaining in the live population satisfies $\ln(Z_{\rm live}/Z)<-5$, corresponding to less than $0.7$\% of the total evidence remaining unaccumulated.

We assess convergence using the evolution of the accumulated evidence with prior compression and the insertion-rank test of \citet{Fowlie:2020aa}. The former checks that the posterior bulk has been traversed before termination, while the latter tests whether newly inserted live points are consistent with correct sampling of the likelihood-constrained prior. Figure~\ref{fig:nss_diag} shows these diagnostics for the JADES~170891 fit of \S\ref{sec:demo_jades}. Its insertion ranks are consistent with uniformity ($D_{\rm KS}=0.007$, $p=0.90$ for $6656$ insertions), and the evidence is accumulated almost entirely before termination. The uncertainty in $\ln Z$ is estimated by resampling the stochastic prior-volume compression with \textsc{anesthetic} \citep{anesthetic}; it is $0.22$ for this fit and remains $\lesssim0.25$ across the JADES sample. In the $44$-dimensional posteriors of \S\ref{sec:demo_zh}, the seed-to-seed scatter of $\ln Z$ ($0.6$--$1.0$) is about twice the internal estimate. Differences in $\ln Z$ below ${\approx}2$ should therefore not be over-interpreted.
\ceri\ also provides two optional gradient-based samplers through the same compiled log-posterior interface. The first is the No-U-Turn sampler \citep[NUTS;][]{Hoffman:2011aa}, an adaptive form of Hamiltonian Monte Carlo, using the \textsc{BlackJAX} implementation. During warm-up, NUTS adapts the step size and a dense mass matrix. It samples in an unconstrained parameter space, with bounded parameters transformed through their prior support and the appropriate Jacobian included. Multiple chains can be run in parallel across available devices. NUTS can scale favourably to higher-dimensional problems, but it does not provide the Bayesian evidence and can be sensitive to multimodality and non-smooth regions of the forward model.

\ceri\ supports variational preconditioning with NeuTra \citep{Hoffman:2019aa}. A full-rank Gaussian or inverse-autoregressive flow is first trained to approximate the posterior, after which NUTS samples in the transformed coordinates. The variational approximation can also be used on its own as a fast approximate posterior. Alternatively, the gradient-based samplers may be initialised from a maximum-a-posteriori solution found with L-BFGS; this is used only for initialisation, not as the final inference result.

Neither gradient-based approach is used for the results presented in this paper. Across the parameter spaces considered here, from modest dimensionality to the ${\sim}125$-parameter SFH fits of \S\ref{sec:high_detail_sfh}, the vectorised nested slice sampler converged faster in wall-clock time on the same GPU while also providing the Bayesian evidence. We have therefore not yet reached the regime in which the gradient-based alternatives become preferable for our applications.

\begin{figure*}[!tb]
    \centering
    \includegraphics[width=\linewidth]{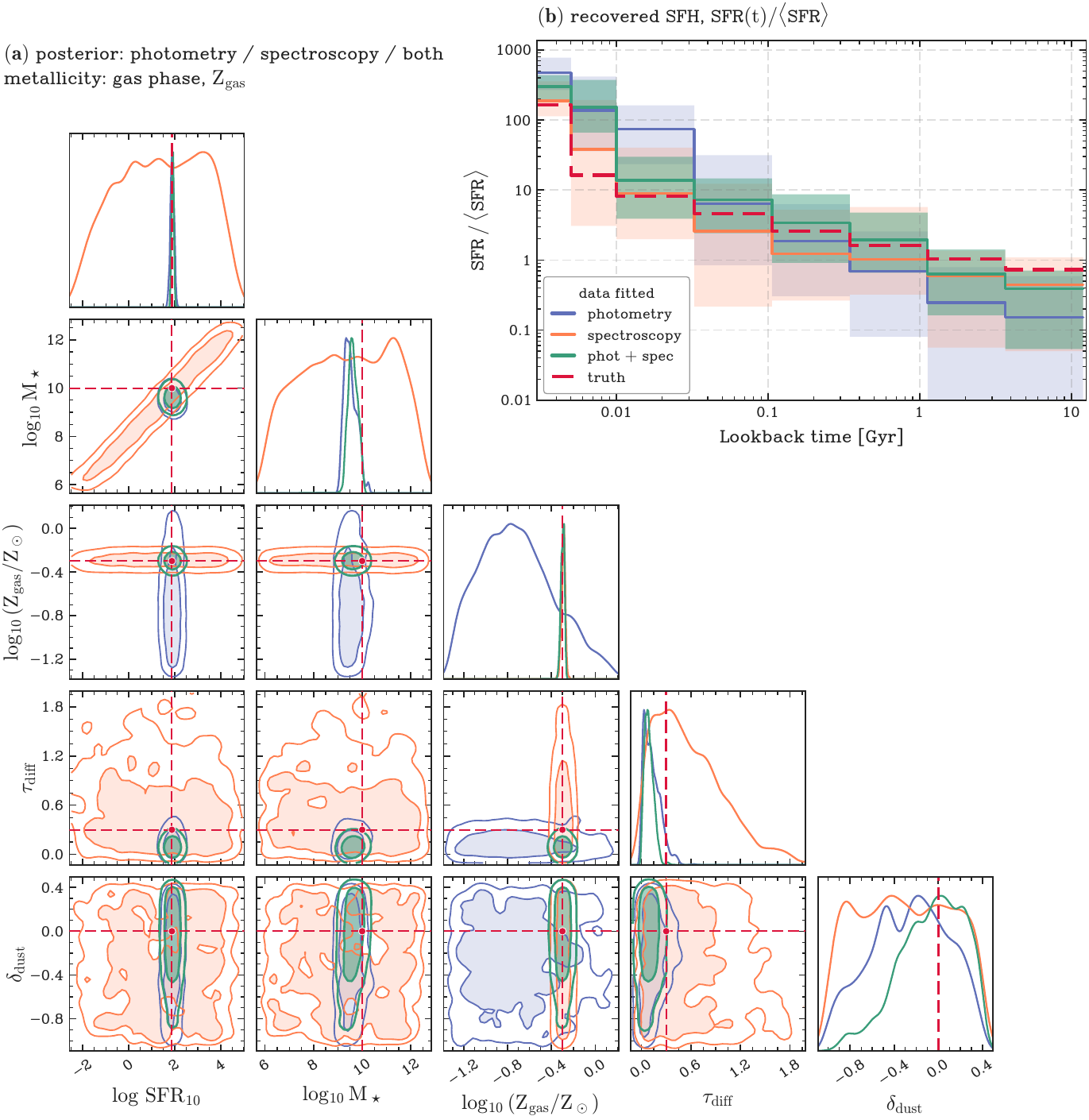}
    \caption{Information content of photometry versus spectroscopy. A mock $z=0.1$ galaxy is fit from broadband photometry alone (blue), the optical spectrum alone (coral), and both jointly (green). The spectrum is fitted through a 12th-order Chebyshev calibration polynomial, optimised at every likelihood call, which removes its normalisation and continuum shape. \textbf{(a)} The marginalised posterior for five key parameters, including the gas-phase metallicity $\log_{10}(Z_{\rm gas}/Z_\odot)$; filled regions and contours mark the $1$ and $2\sigma$ credible regions for each dataset and the crimson lines/point the input truth. \textbf{(b)} The recovered SFHs, each divided by its own time-averaged SFR (median and $68\%$ band), against the truth (crimson dashed). The emission lines in the spectrum pin the gas-phase metallicity, but with its continuum shape removed the spectrum leaves the stellar mass and the dust unconstrained; these come from the photometry.}
    \label{fig:demo_phot_spec}
\end{figure*}

\section{Demonstrations with mock data}
\label{sec:demo_mock}
\begin{figure*}[!tb]
    \centering
    \includegraphics[width=\linewidth]{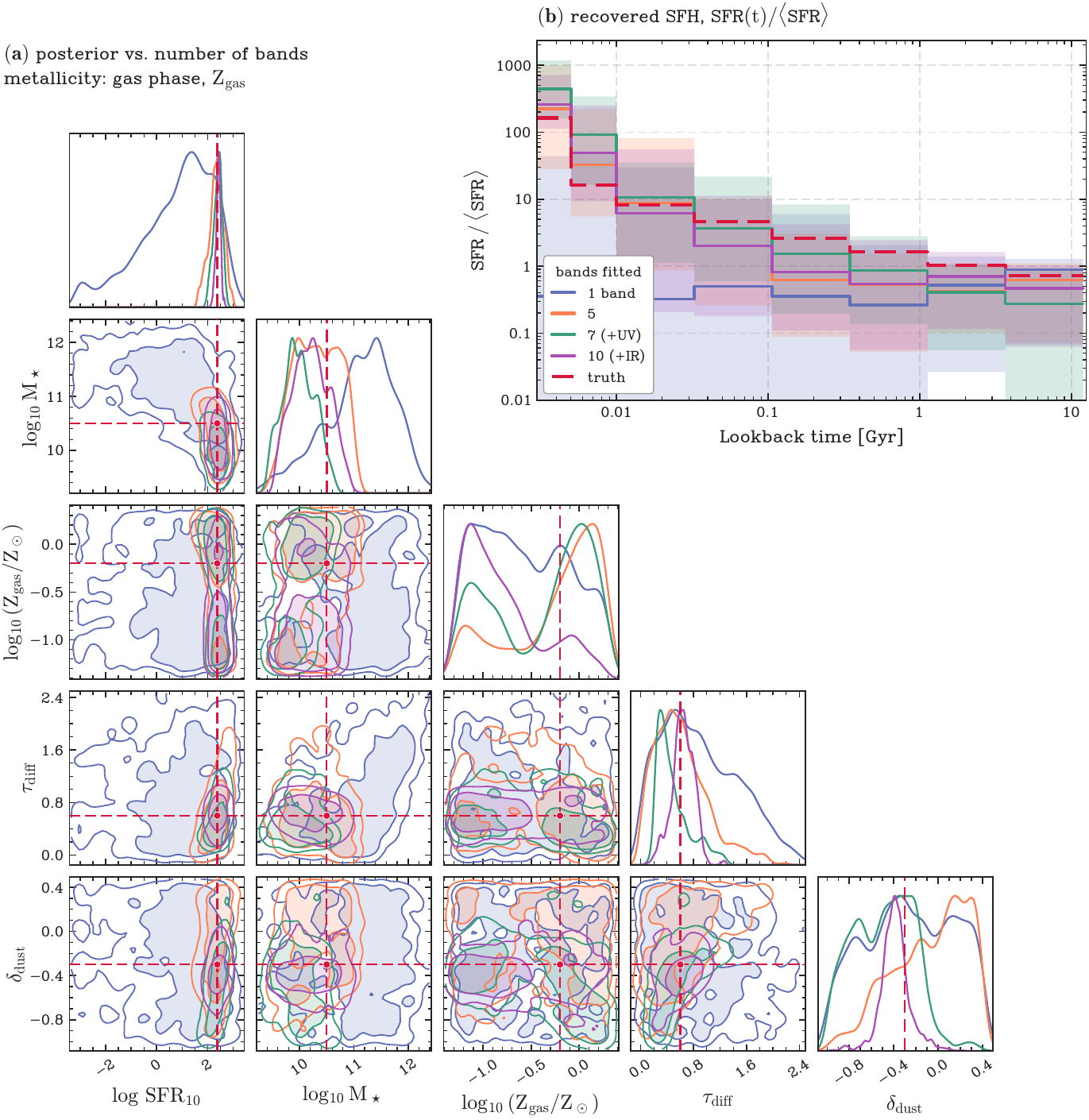}
    \caption{Effect of photometric wavelength coverage. A mock $z=0.1$ galaxy described in \S\ref{sec:demo_phot_spec_bands} is fit from progressively richer photometry: one optical band, the full SDSS optical, $+$GALEX UV, and $+$2MASS near-IR (blue, orange, green, purple), with the fiducial eight-bin model; the input galaxy is the dustier, more massive variant described in the text. \textbf{(a)} The marginalised posterior for five key parameters; filled regions and contours mark the $1$ and $2\sigma$ credible regions for each band set, and the crimson lines/point the input truth. \textbf{(b)} The recovered SFHs, each divided by its own time-averaged SFR (median and $68\%$ band), for each band set against the truth (crimson dashed). The photometry leaves the gas-phase metallicity close to its prior, which spans the nebular grid.}
    \label{fig:demo_more_bands}
\end{figure*}

Having specified the forward model (\S\ref{sec:forward_model}--\S\ref{sec:observations}) and inference framework (\S\ref{sec:inference}), we now test \ceri\ on controlled mock data, for which the input parameters are known exactly. Such recovery tests are an essential part of any stellar-population inference analysis: the information that can be extracted depends not only on the inference method, but also on the adopted physical model and on the quality and type of the data, including their wavelength coverage and signal-to-noise. We begin in \S\ref{sec:demo_phot_spec_bands} by examining how the recovered constraints change between photometry and spectroscopy and with increasing wavelength coverage (Figs~\ref{fig:demo_phot_spec} and \ref{fig:demo_more_bands}). We then turn to capabilities enabled by \ceri's GPU-native architecture, including the joint inference of time-varying star-formation and chemical-enrichment histories (\S\ref{sec:demo_zh}), high-resolution, many-parameter SFHs (\S\ref{sec:high_detail_sfh}), and a sampled $[\alpha/\mathrm{Fe}]$ (\S\ref{subsec:alpha}), before closing with a survey-realistic recovery suite that mirrors the \jwst data of \S\ref{sec:demo_real} (\S\ref{subsec:mock_jades_data}). Several additional components are available but not exercised by these demonstrations: dust with more than two age bins, the picket-fence escape geometry, THEMIS dust emission, the IGM damping wing and DLA, the outlier mixture, the analytic line marginalisation and the GP likelihood.

The mock tests are based on a common fiducial forward model which we briefly introduce here.
The assumed mock is a $z=0.1$ galaxy of $\log_{10}(M_\star/\mathrm{M}_\odot)=10$ with a rising SFH ending in a recent burst, generated from the same forward model used in the fit and perturbed by Gaussian noise at the quoted uncertainties. We built the model on a \textsc{BPASS} SSP grid \citep{stanway_re-evaluating_2018}, whose solar metallicity is $Z_\odot=0.020$. The SFH is piecewise constant in eight lookback-time bins. The two youngest bins span $0$--$5$ and $5$--$10$\,Myr; the other six are spaced uniformly in $\log$ lookback time out to an assumed formation redshift $z_{\rm f}=10$ ($11.97$\,Gyr at $z=0.1$), following \citet{leja_how_2019}. The model has a single stellar metallicity and two-component dust attenuation: a power law for stars younger than ${\sim}10$\,Myr and a diffuse \citet{kriek_dust_2013} component with free optical depth and slope. Nebular continuum and line emission are interpolated from the \citet{Byler:2017aa} grids, with free gas-phase metallicity and ionisation parameter. IGM absorption follows \citet{madau_radiative_1995}. Throughout, $M_\star$ is the total stellar mass formed, $10^{\theta[\texttt{logmass}]}$; the surviving mass is smaller and is available in post-processing (\S\ref{sec:sfh}). The redshift is held fixed at its true value. The input galaxy has $\tau_{\rm BC}=\tau_{\rm diff}=0.3$, a flat attenuation slope ($\delta_{\rm dust}=0$), $\log_{10}(Z_{\rm gas}/Z_\odot)=-0.3$, $\log_{10}U=-2$.

We sampled fourteen parameters: the seven logarithmic ratios of adjacent star-formation-history bins under a Student-$t$ continuity prior ($\nu=2$, $\sigma=0.5$); the stellar mass, $\log_{10}M_\star\sim\mathcal{U}[6,12.5]$; the stellar metallicity, $\log_{10}(Z_\star/Z_\odot)\sim\mathcal{U}[-2.30,0.30]$, the full metallicity range of the \mbox{\textsc{BPASS}} grid; the birth-cloud and diffuse dust optical depths, $\mathcal{N}_{[0,4]}(0.3,0.5)$ and $\mathcal{N}_{[0,4]}(0.3,1.0)$; the attenuation slope, $\delta_{\rm dust}\sim\mathcal{U}[-1,0.4]$; and the gas-phase metallicity and ionisation parameter, $\log_{10}(Z_{\rm gas}/Z_\odot)\sim\mathcal{U}[-1.3,0.3]$ and $\log_{10}U\sim\mathcal{U}[-4,-1]$, the metallicity and ionisation-parameter ranges of the nebular grid. Every posterior in this section was sampled with the nested slice sampler of \S\ref{sec:sampling} on a single A100 GPU, with $300$ live points, $25$ replaced per iteration and $40$ slice steps per replacement unless stated otherwise. The fits of \mbox{\S\ref{sec:demo_phot_spec_bands}} terminated at $\ln(Z_{\rm live}/Z)=-5$.

\subsection{Photometry, spectroscopy, and wavelength coverage}
\label{sec:demo_phot_spec_bands}
Figure~\ref{fig:demo_phot_spec} revisits the canonical comparison of the information carried by photometry and by spectroscopy \citep[cf.][their Fig.~2]{Johnson:2021aa}. The fiducial mock of \S\ref{sec:demo_mock} is observed in two complementary ways. The photometric arm is a ten-band ultraviolet-to-near-infrared set, GALEX \textit{FUV} and \textit{NUV}, SDSS \textit{ugriz}, and 2MASS \textit{JHK}$_{\rm s}$, with a constant $5$\% uncertainty on every band, i.e. a signal-to-noise ratio $\mathrm{S/N}=20$. The spectroscopic arm spans $3500$--$7500$~\AA\ in the observer frame on $2000$ pixels ($2$~\AA\ sampling), covering the $4000$~\AA\ break and the Balmer series at $z=0.1$, and carries photon-counting-shaped noise, $\sigma_{i}\propto\sqrt{|y_{i}|}$, normalised so that the brightest emission line reaches a per-pixel $\mathrm{S/N}$ of $50$. With a line-to-continuum contrast of ${\sim}50$ this leaves the continuum at a per-pixel $\mathrm{S/N}$ of ${\sim}7$, so the spectrum is line-dominated in the way a real survey spectrum designed against a target line would be. We fitted the same galaxy three times---using photometry alone, the spectrum alone, and both jointly---with identical model, priors, and sampler settings. The runs therefore differ only in the data entering the likelihood. As in \citet{Johnson:2021aa}, the photometry primarily constrains the broad continuum shape, while the spectrum carries information from absorption features and the $4000$~\AA\ break; the joint fit gives the strongest overall constraints on the SFH. As in \mbox{\citet{Johnson:2021aa}}, the model spectrum is multiplied by a 12th-order Chebyshev polynomial in wavelength whose coefficients are not sampled but optimised at every likelihood call, a weighted linear least-squares solve (\mbox{\S\ref{sec:spectra}}). This mimics the treatment of real spectra, whose absolute normalisation and broad continuum shape are uncertain because of slit losses, aperture corrections and flux-calibration errors; the mock spectrum itself is perfectly calibrated. The polynomial absorbs the normalisation and the broad continuum shape, so the spectrum constrains the model only through features narrower than it can follow: the emission lines and the absorption features. Unlike \mbox{\citet{Johnson:2021aa}}, the SFH is non-parametric.

\begin{figure*}[!tb]
    \centering
    \includegraphics[width=\linewidth]{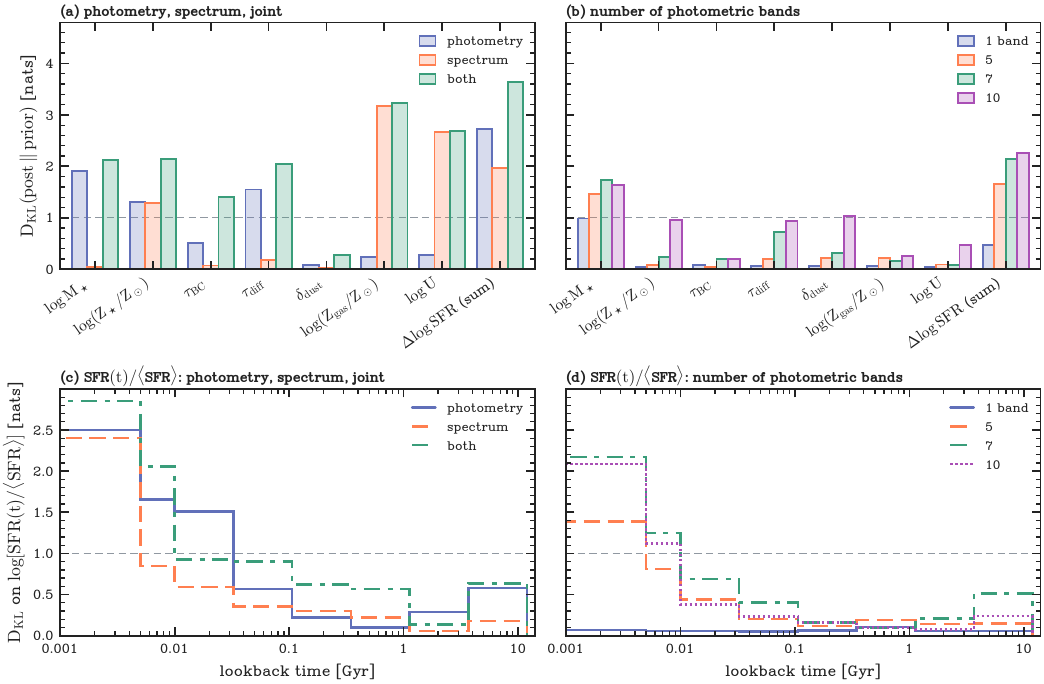}
    \caption{Information gain per parameter, $D_{\rm KL}[p(\theta_p\mid\boldsymbol{d})\,\|\,\pi(\theta_p)]$ in nats (\S\ref{sec:priors}), for the data-content experiments of \S\ref{sec:demo_phot_spec_bands}. \textbf{(a)} Fiducial mock fitted with photometry, spectroscopy, and both jointly (Fig.~\ref{fig:demo_phot_spec}). \textbf{(b)} Band-ladder mock fitted with one, five, seven, and ten photometric bands (Fig.~\ref{fig:demo_more_bands}). The SFH bar sums the information gain over the seven adjacent-bin $\log\mathrm{SFR}$ ratios; $Z_\star$ and $Z_{\rm gas}$ are the stellar and gas-phase metallicities. In \mbox{\textbf{(a)}} the spectrum enters through the optimised 12th-order calibration polynomial of Fig.~\mbox{\ref{fig:demo_phot_spec}}, which removes its continuum shape, so it informs the emission-line parameters and the stellar metallicity but hardly the stellar mass or the dust. \textbf{(c)} and \textbf{(d)} Information gain in $\log[\mathrm{SFR}(t)/\langle\mathrm{SFR}\rangle]$, the SFH divided by its time average, from which the stellar mass drops out, for the same runs as \textbf{(a)} and \textbf{(b)}, respectively, shown across each SFH bin. The prior for this derived quantity is obtained by propagating prior draws through the same SFH transform used for the posterior. The youngest bin ($<5$~Myr) is the best constrained in every fit with more than one band, unsurprisingly. All runs use the same priors and are therefore directly comparable.}
    \label{fig:kl_phot_spec}
\end{figure*}

Figure~\ref{fig:kl_phot_spec} quantifies the contribution of each dataset through the information gain $D_{\rm KL}[p(\theta_p\mid\boldsymbol{d})\,\|\,\pi(\theta_p)]$ defined in \S\ref{sec:priors}. Because the calibration polynomial removes the continuum shape, the two datasets inform different parameters. The spectrum carries almost all of the information on the nebular parameters, $3.2$ and $2.7$\,nats for $\log_{10}(Z_{\rm gas}/Z_\odot)$ and $\log_{10}U$, against $0.2$ and $0.3$\,nats from the photometry, because its emission lines survive the polynomial. It carries almost none on the stellar mass ($0.04$\,nats, against $1.9$\,nats from the photometry) or on the dust ($0.2$\,nats or less for each of $\tau_{\rm diff}$, $\tau_{\rm BC}$ and $\delta_{\rm dust}$), whose imprint is the normalisation and continuum shape that the polynomial absorbs. The stellar metallicity gains $1.3$\,nats from either dataset alone, through the absorption features of the spectrum and the continuum colours of the photometry, and $2.1$\,nats jointly. Relative to photometry alone, the joint fit also gains $0.9$, $0.5$ and $0.2$\,nats on $\tau_{\rm BC}$, $\tau_{\rm diff}$ and $\delta_{\rm dust}$.

For the SFH ratios, the summed information gain is $2.7$, $2.0$ and $3.6$\,nats for the photometry-only, spectrum-only and joint fits. Five of the seven ratio posteriors are narrower jointly than from the spectrum alone, the youngest by $0.46$\,dex in its $68\%$ half-width; the other two are broader by $0.02$\,dex. Within one run, the bootstrap uncertainty on the summed gain is $0.02$--$0.03$\,nats, but repeating each fit with three further sampler seeds on the same data gives $2.0$--$2.7$, $1.8$--$2.0$ and $2.5$--$3.6$\,nats. The run-to-run sampling variance therefore dominates, and the same repeats move the joint-fit median of $\log_{10}M_\star$ between $9.6$ and $10.0$.

Figure~\ref{fig:demo_more_bands} isolates the effect of photometric wavelength coverage. The mock galaxy has the same SFH shape, stellar metallicity, redshift, and priors as the fiducial case, but is more massive and dustier: $\log_{10}(M_\star/\mathrm{M}_{\odot})=10.5$, $\tau_{\rm BC}=0.5$, $\tau_{\rm diff}=0.6$, $\delta_{\rm dust}=-0.3$, $\log_{10}(Z_{\rm gas}/Z_\odot)=-0.2$, and $\log_{10}U=-2.2$. This gives the attenuation a clear imprint on the continuum.

We then fitted the galaxy with photometry alone, with four nested filter sets at fixed $\mathrm{S/N}=20$ per band: SDSS \textit{r}; SDSS \textit{ugriz}; SDSS plus GALEX \textit{FUV} and \textit{NUV}; and finally the addition of 2MASS \textit{JHK}$_{\rm s}$, for a total of one, five, seven, and ten bands. Each set contains all bands from the previous one. As in \citet[their Fig.~4]{Johnson:2021aa}, the posteriors narrow as the wavelength baseline increases, with little systematic shift in their centres.

Figure~\ref{fig:kl_phot_spec}b shows where the additional information enters. A single optical band constrains mainly the stellar mass ($1.0$\,nats), while the remaining parameters stay close to their priors. Adding the full SDSS set provides most of the photometric information on the SFH, increasing the summed information gain of the SFH ratios from $0.5$ to $1.7$\,nats ($2.3$\,nats with all ten bands). The GALEX bands mainly constrain $\tau_{\rm diff}$, whose information gain rises from $0.2$ to $0.7$\,nats. The 2MASS bands constrain the attenuation slope, from $0.3$ to $1.0$\,nats, and the stellar metallicity, from $0.25$ to $1.0$\,nats, through the red continuum shape. The stellar-mass information saturates at about $1.7$\,nats once seven bands are included. As expected, even with ten bands at $\mathrm{S/N}=20$, the gas-phase metallicity and the ionisation parameter gain only $0.3$ and $0.5$\,nats, far below the $3.2$ and $2.7$\,nats supplied by the spectrum in Fig.~\mbox{\ref{fig:kl_phot_spec}}a.

\begin{figure*}[!tb]
    \centering
    \includegraphics[width=\linewidth]{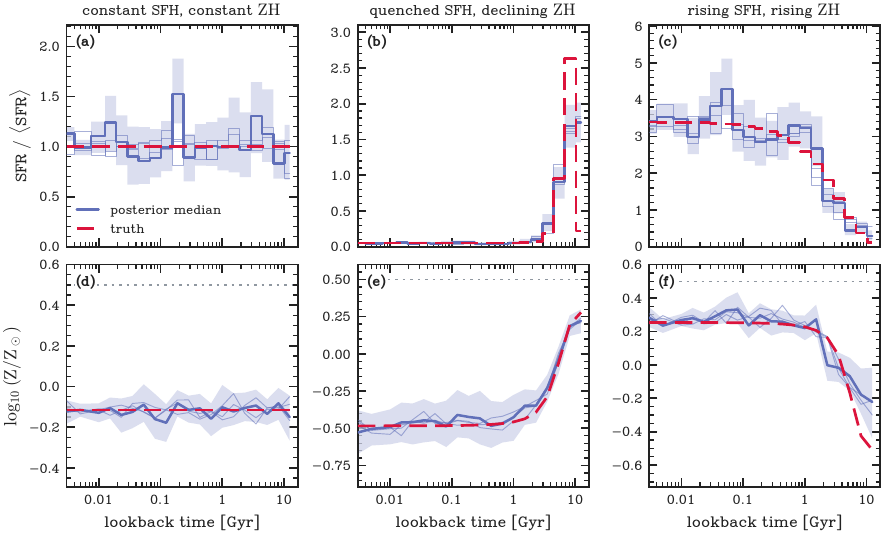}
    \caption{Joint recovery of the SFH and time-varying stellar metallicity from a noiseless $z=0.1$ ultraviolet-to-near-infrared spectrum with peak $\mathrm{S/N}=200$. Columns show the constant-SFH/constant-ZH, quenched-SFH/declining-ZH, and rising-SFH/rising-ZH mocks described in \S\ref{sec:demo_zh}. The top row shows $\mathrm{SFR}/\langle\mathrm{SFR}\rangle$ and the bottom row $\log_{10}(Z/Z_\odot)$, with posterior medians and $68\%$ credible intervals (blue) compared with the input histories (crimson dashed). Thin curves show the posterior medians from three additional nested-sampling seeds. The grey dotted line marks the upper metallicity limit of the SSP grid, $\log_{10}(Z/Z_\odot)=+0.5$.}
    \label{fig:zh_recovery}
\end{figure*}

\subsection{Joint star-formation and chemical-enrichment histories}
\label{sec:demo_zh}

A distinctive capability of \ceri\ is the joint inference of a time-varying stellar metallicity and SFH, rather than assigning a single metallicity to all stellar populations (\S\ref{sec:zh}). We test this with three increasingly structured input histories (Fig.~\ref{fig:zh_recovery}): a constant SFH with constant metallicity, a quenched SFH with declining enrichment, and a rising SFH with rising enrichment. In all cases, the fits used a non-parametric SFH and chemical-enrichment history with $N_T=20$ lookback-time nodes. Both histories are inferred non-parametrically: the SFH through adjacent-bin SFR ratios and the metallicity history through adjacent-node metallicity ratios, each with a Student-$t$ continuity prior. The SFH ratios use $\nu=2$ and $\sigma=0.3$. For the metallicity track we adopt $\nu=2$ and $\sigma_Z=0.2$. Thus, neither the qualitative form of the input SFH nor that of the enrichment history is imposed during the fit.

The setup otherwise differs from the fiducial configuration of \S\ref{sec:demo_mock} in several aspects. The data consist of a single spectrum with no photometry, spanning $1150$--$29975$~\AA\ observed ($1045$--$27250$~\AA\ rest-frame at $z=0.1$). The rest-optical region is sampled at $1$~\AA, while wavelengths outside the well-sampled MILES range retain the native SSP-grid sampling to avoid introducing artificial information through oversampling. The data are set equal to the model realisation, with per-pixel uncertainties corresponding to a peak $\mathrm{S/N}=200$. Birth-cloud attenuation is disabled, leaving only the diffuse \citet{kriek_dust_2013} component. The SSP grid uses the solar-scaled \textsc{MIST}$+$\textsc{MILES} models with $Z_\odot=0.0185$. The resulting posterior has 44 dimensions, and each fit used 600 live points, 120 replaced per iteration, and $3n_{\rm dim}=132$ slice steps per replacement.

Figure~\ref{fig:zh_recovery} shows the recovered SFHs and enrichment histories. The constant, quenched, and rising cases probe increasingly structured histories while keeping the data quality and inference setup fixed. For display, the SFHs are normalised to their time-averaged SFR and the metallicities are shown as $\log_{10}(Z/Z_\odot)$. We repeated each fit with four nested-sampling seeds; because the mock spectrum itself is unchanged, the variation between the recovered medians measures sampling variability alone.

We explore the metallicity-continuity prior by repeating the fits for $\sigma_Z=\{0.05,0.1,0.2,0.3\}$. Decreasing $\sigma_Z$ removes the pile-up of samples at the upper metallicity bound (at most $3.0$\%, in the rising model at $\sigma_Z=0.3$), while $\sigma_Z=0.05$ visibly over-smooths the recovered track. We adopted $\sigma_Z=0.2$, which lies between these two regimes and is ${\sim}2.4$ times the rms adjacent-node variation of the input tracks ($0.085$~dex). Across this range the evidence varies by up to $\Delta\ln Z\approx9$ in the constant model, and by $1.8$ and $5.0$ in the quenched and rising models. The evidence is highest at $\sigma_Z=0.05$ for the constant model and at $\sigma_Z=0.1$ for the quenched and rising models, higher than at $\sigma_Z=0.2$ by $\Delta\ln Z=4.5$, $1.4$ and $3.8$ (single seed). A narrow prior is favoured here because the input tracks are smooth, but it would suppress sharper enrichment episodes than those in our mocks. We therefore keep $\sigma_Z=0.2$ for the seed comparison of Fig.~\ref{fig:zh_recovery} and Table~\ref{tab:zh_seeds}, whose purpose is to test how stably the sampler explores the posterior, not which prior width the data prefer.
The four-seed comparison gives $R=0.2$--$0.5$ times the corresponding $68\%$ credible half-width for the recovered SFH and metallicity histories (Table~\ref{tab:zh_seeds}), implying an increase in the effective uncertainty of at most ${\sim}10$\%. The present-day SFR is recovered at $1.04\pm0.10$, $0.95\pm0.13$, and $0.97\pm0.04$ of the input value for the constant, quenched, and rising models, respectively. The evidence is less stable: its run-to-run scatter of $0.6$--$1.0$ exceeds the internal uncertainty of $0.3$--$0.4$ reported by individual runs. Differences of order unity in $\ln Z$ should therefore not be over-interpreted for fits of this complexity.

\begin{table}
    \centering
    \caption{Seed-to-seed stability of the recoveries in Fig.~\ref{fig:zh_recovery} for seeds $42$--$45$. $R_{\rm SFH}$ and $R_{\rm ZH}$ are the standard deviations of the posterior medians across the four seeds, averaged over the $20$ lookback-time nodes and expressed relative to the corresponding $68\%$ credible half-width. The final column gives the recovered present-day SFR relative to the input value, reported as the mean and standard deviation across seeds.}
    \label{tab:zh_seeds}
    \begin{tabular}{lccc}
        \hline
        Model & $R_{\rm SFH}$ & $R_{\rm ZH}$ & $\mathrm{SFR}_{0}/\mathrm{SFR}_{0,\rm true}$ \\
        \hline
        constant SFH, constant ZH  & $0.44$ & $0.32$ & $1.04\pm0.10$\\
        quenched SFH, declining ZH & $0.34$ & $0.20$ & $0.95\pm0.13$ \\
        rising SFH, rising ZH      & $0.46$ & $0.46$ & $0.97\pm0.04$\\
        \hline
    \end{tabular}
\end{table}

Two limitations are apparent. In the constant model, the recovered SFH shows node-to-node fluctuations of $11\pm4$\% rms around the flat input. Their persistence across seeds ($R=0.44$) indicates that they arise from degeneracies between neighbouring SFH bins rather than sampling noise. In the quenched model, the continuity prior smooths the single-node burst at $8$~Gyr across the two oldest bins: the recovered peak is $1.7~\langle\mathrm{SFR}\rangle$ ($68\%$: $1.4$--$2.0$), compared with an input of $2.6$, while the oldest bin is recovered at $1.7$ compared with an input of $0.2$. This reflects the prior penalty on the sharp node-to-node variation required by the input burst and is independent of $\sigma_Z$. 

Overall, these tests show that \ceri\ can jointly recover non-parametric SFHs and time-dependent chemical-enrichment histories, although their temporal resolution remains limited by the data and adopted priors. For real galaxies, the achievable resolution will additionally depend on wavelength coverage, signal-to-noise, and stellar-population systematics. We will explore these dependencies and the role of physically motivated priors in Stoffers et al.\ (in preparation).

\begin{figure*}[!tb]
    \centering
    \includegraphics[width=\linewidth]{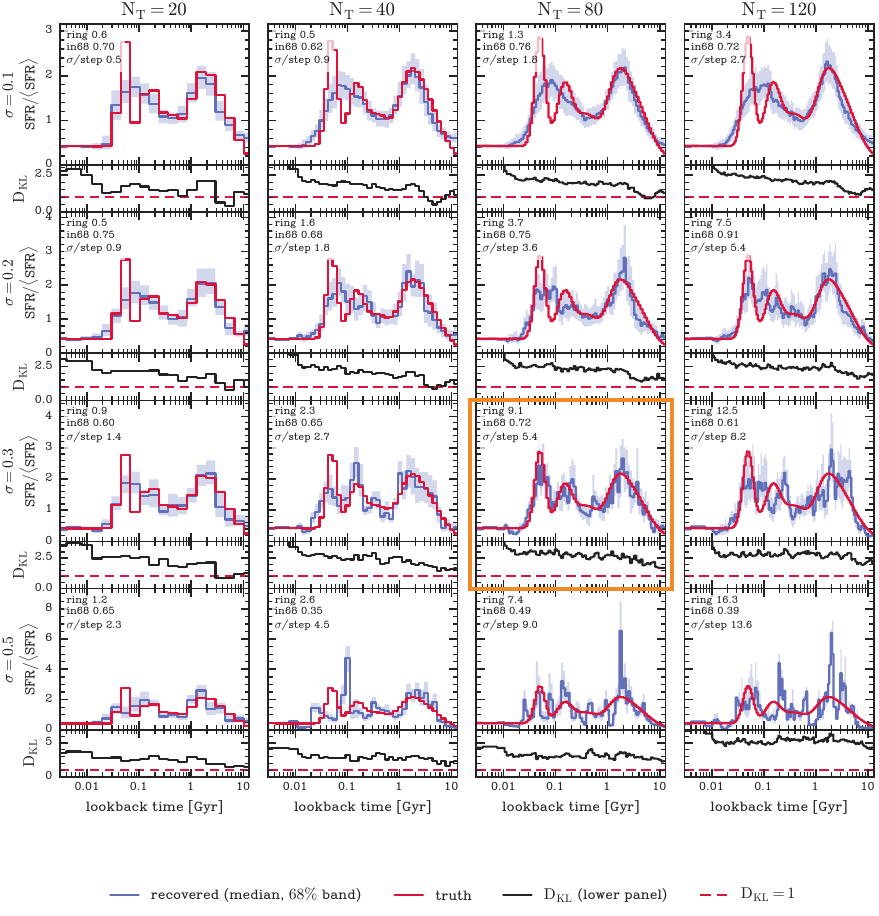}
    \caption{Recovery of a multi-burst SFH from the noiseless ultraviolet-to-near-infrared spectrum of \S\ref{sec:demo_zh}, as a function of continuity-prior width $\sigma$ (rows) and number of SFH nodes $N_T$ (columns). The input consists of six log-Gaussian bursts between $50$~Myr and $5$~Gyr lookback time on a slowly declining baseline. Each panel shows the input evaluated on that grid (crimson) and the posterior median with its $68\%$ credible interval (blue). The orange frame marks the $(\sigma,N_T)=(0.3,80)$ fit, the only one that recovers both the ${\sim}50$\,Myr burst and the dip between the ${\sim}50$ and ${\sim}150$\,Myr bursts (\mbox{\S\ref{sec:high_detail_sfh}}). Fits within a column use identical data; across columns the input is re-evaluated on each grid, changing the spectra by at most $1.5$\%. Panel annotations give the relative node-to-node curvature (\textit{ring}), $68\%$ coverage fraction (\textit{in68}), and $\sigma$ relative to the rms adjacent-node step of the input. The lower panels show the Kullback--Leibler divergence between the posterior normalised SFR and its propagated continuity prior; the dashed line marks $D_{\rm KL}=1$ nat.}
    \label{fig:sfh_prior_width}
\end{figure*}

\subsection{High-resolution star-formation histories}
\label{sec:high_detail_sfh}

The non-parametric SFH of \S\ref{sec:sfh} places no intrinsic limit on the number of time bins. In practice, however, the temporal resolution that can be recovered is set jointly by the information in the data and the prior imposed on variations between neighbouring bins.

We tested this with the same noiseless ultraviolet-to-near-infrared spectrum as in \S\ref{sec:demo_zh}: a single spectrum with no photometry, spanning $1150$--$29975$~\AA\ observed ($1045$--$27250$~\AA\ rest-frame at $z=0.1$) over $4876$ pixels, with per-pixel uncertainties corresponding to a peak $\mathrm{S/N}=200$. The SFH extends to an assumed formation redshift $z_{\rm f}=30$, while the stellar metallicity is fitted as a single constant value. Only diffuse \citet{kriek_dust_2013} attenuation is included. The model has $N_T+5$ free parameters, ranging from $25$ at $N_T=20$ to $125$ at $N_T=120$: the $N_T-1$ log-SFR ratios between adjacent nodes, the stellar mass, the constant stellar metallicity, the optical depth and power-law slope of the diffuse attenuation, and the gas-phase metallicity and ionisation parameter. The $N_T$ nodes lie at the present day, at $5$ and $10$\,Myr, and then log-uniformly in lookback time up to $z_{\rm f}=30$. The SFR is constant within each of the $N_T-1$ intervals between neighbouring nodes, at the mean of its two node values.

To separate the effects of temporal resolution and prior smoothness, we fitted all combinations $N_T\in\{20,40,80,120\},~~\sigma\in\{0.1,0.2,0.3,0.5\},$ for sixteen posteriors in total. Each fit used $400$ live points, $80$ replaced per iteration, and $3n_{\rm dim}$ slice steps per replacement, terminating at $\ln(Z_{\rm live}/Z)<-5$. Within a column of Fig.~\ref{fig:sfh_prior_width}, the data are identical and only $\sigma$ changes. Across a row, the input SFH is evaluated on each grid’s own nodes, changing the generated spectrum by at most $1.5\%$; evidences are therefore compared only within a fixed $N_T$ column. The input consists of six log-Gaussian bursts between $50$~Myr and $5$~Gyr lookback time on a slowly declining baseline. Figure~\ref{fig:sfh_prior_width} shows that increasing $N_T$ does not by itself improve the recovery. The $20$-node grid blends closely spaced young bursts, while the finer grids resolve more structure, but they also become increasingly sensitive to the adopted prior. At fixed $N_T$, broadening the prior eventually introduces strong node-to-node structure; at fixed $\sigma$, this effect becomes more pronounced as the grid is refined. The summary statistics in Fig.~\ref{fig:sfh_prior_width} try to quantify this behaviour. For $\sigma\le0.3$, the fraction of nodes whose input value lies within the $68\%$ credible interval is typically $0.60$--$0.76$, apart from the $(\sigma,N_T)=(0.2,120)$ fit, which over-covers at $0.91$. For $\sigma=0.5$ and $N_T\ge40$, the coverage falls to $0.35$--$0.49$. The relative node-to-node curvature of the recovered median (\textit{ring}) also increases with both $\sigma$ and $N_T$, reaching $3.4$, $7.5$, $12.5$, and $16.3$ at $N_T=120$ for $\sigma=0.1$, $0.2$, $0.3$, and $0.5$, respectively. The evidence likewise disfavours the broadest prior at high resolution: $\sigma=0.5$ is lower by $18.6$ and $20.9$ in $\ln Z$ at $N_T=80$ and $120$.

Individual SFH features show the same pattern. The older burst at $1.5$--$2$~Gyr has its integrated mass recovered to within $10\%$ in $13$ of the $16$ fits. The two youngest bursts are substantially harder to separate: the ${\sim}50$~Myr burst is recovered at only $0.5$--$0.7$ of its input amplitude in most fits, and the dip between the ${\sim}50$ and ${\sim}150$~Myr bursts is usually filled in. Only the $(\sigma,N_T)=(0.3,80)$ fit recovers both features well. Even at $N_T=120$, the evidence-preferred prior produces a single broad young component rather than two distinct bursts. The spectrum therefore contains much less information on these short timescales than on the older structure, while the continuity prior penalises the sharp adjacent-bin changes required to separate them.

The behaviour of the broad priors is not simply a widening of the posterior. With $\nu=2$, the Student-$t$ continuity prior has heavy tails, so a few large jumps between neighbouring bins can carry more prior probability than many moderate ones. At high $N_T$, the likelihood cannot distinguish many such sharply varying histories from smoother solutions, and the posterior can therefore become dominated by narrow bursts and strong ringing. This is a prior-volume effect instead of evidence that the data resolve this structure.

The origin of the problem is that a fixed $\sigma$ does not correspond to the same physical prior at different temporal resolutions. The continuity prior acts on changes in $\log\mathrm{SFR}$ between adjacent nodes, while the time interval between those nodes decreases as $N_T$ increases. For the input history, the rms adjacent-node step falls from $0.218$~dex at $N_T=20$ to $0.037$~dex at $N_T=120$, approximately as $1/(N_T-1)$. A value of $\sigma$ appropriate for the coarse grid therefore allows much larger physical fluctuations on the finer grid.

The evidence-preferred widths follow this scaling more closely: expressed relative to the rms step of the input history, they are $1.4$, $1.8$, $1.8$, and $2.7$ for $N_T=20$, $40$, $80$, and $120$, respectively. A practical choice is therefore to scale the prior width with the grid spacing, approximately $\sigma(N_T)\approx0.3~\frac{19}{N_T-1}$ for this experiment, rather than holding $\sigma$ fixed. More generally, a prior defined through physical amplitudes and characteristic timescales, such as the stochastic SFH prior of \citet{wan_stochastic_2024} based on the power-spectrum framework of \citet{Iyer:2020aa}, would provide a more consistent comparison across different temporal resolutions.

Overall, the long-timescale SFH is recovered across the tested grids: for $\sigma\le0.3$, the old burst and overall mass assembly are recovered to ${\sim}10\%$, with rms deviations of $0.09$--$0.15$~dex from the input. By contrast, features on timescales comparable to the node spacing remain strongly prior-dependent. Increasing $N_T$ therefore does not automatically increase the effective temporal resolution; a ${\sim}100$-bin posterior should be interpreted together with the prior that regularises it. The experiment demonstrates that \ceri\ can sample such high-dimensional SFHs and quantify their prior dependence, not that the data independently constrain ${\sim}100$ star-formation rates.

\begin{figure}[!tb]
    \centering
    \includegraphics[width=\linewidth]{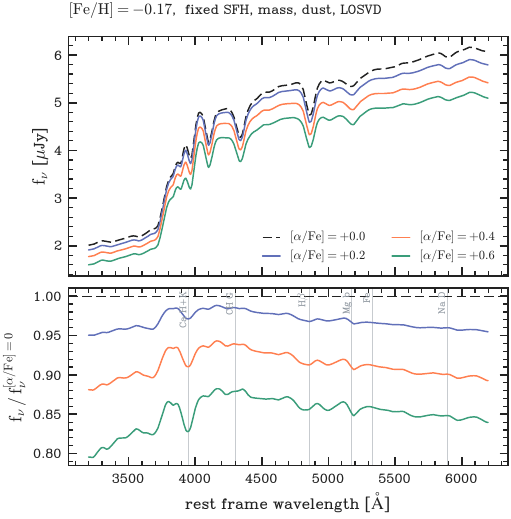}
    \caption{Effect of $[\alpha/\mathrm{Fe}]$ on the model spectrum, over the blue half of the wavelength range.  \emph{Top:} \ceri\ forward-model spectra at fixed $[\mathrm{Fe/H}]=-0.17$ with identical SFH, stellar mass, dust, and velocity dispersion, varying only $[\alpha/\mathrm{Fe}]=0.0$--$0.6$. \emph{Bottom:} ratio to the solar-scaled model (black dashed), with the $\alpha$-sensitive features marked. At fixed $[\mathrm{Fe/H}]$, raising $[\alpha/\mathrm{Fe}]$ adds $\alpha$-element electron donors and hence H$^{-}$ opacity: the optical continuum is depressed coherently by up to ${\sim}20$\% at $[\alpha/\mathrm{Fe}]=+0.6$, \ion{Ca}{ii}\,H+K deepens, and the iron blends weaken; the depression continues smoothly through the red half of the fitted range (not shown).  The signature is broad-band and coherent across many pixels, which is what makes $[\alpha/\mathrm{Fe}]$ measurable from a continuum fit.}
    \label{fig:afe_spectra}
\end{figure}

\subsection{Fitting \texorpdfstring{$[\alpha/\mathrm{Fe}]$}{[alpha/Fe]} enhancement}
\label{subsec:alpha}

All fits so far assume a scaled-solar abundance pattern, described by $\log_{10}(Z/Z_\odot)$ alone. This is a strong simplification: different elements follow different enrichment pathways and therefore carry independent information about a galaxy's formation history. We include the relative abundance of the $\alpha$ elements to the iron peak, $[\alpha/\mathrm{Fe}]$, as an additional fit parameter. The $\alpha$ elements are produced promptly by core-collapse supernovae, while much of the iron is released later by Type~Ia supernovae, making $[\alpha/\mathrm{Fe}]$ a chemical clock for the duration of star formation \citep[e.g.][]{Thomas:2003ab}. Rapidly formed, early-quenched systems are therefore expected to be $\alpha$-enhanced, whereas extended star formation drives the abundance ratio towards solar.

In \ceri, $[\alpha/\mathrm{Fe}]$ is fitted jointly with the SFH, stellar metallicity, mass, and dust. In the current implementation it is a single scalar parameter, $\theta[\texttt{afe}]$, shared by all stellar ages; a time-dependent $[\alpha/\mathrm{Fe}]$ history is left for future work (\S\ref{sec:future}). The SSP machinery is otherwise unchanged from \S\ref{sec:ssp}: the sampled abundance is interpolated across an additional library axis, here with five nodes spanning $[\alpha/\mathrm{Fe}]\in\{-0.2,0.0,+0.2,+0.4,+0.6\}$.

For these grids, the metallicity axis is $[\mathrm{Fe/H}]$ rather than total metallicity. Thus, $\theta[\texttt{logzsol}]$ represents $[\mathrm{Fe/H}]$, and increasing $[\alpha/\mathrm{Fe}]$ adds $\alpha$ elements at fixed iron abundance. The corresponding total metallicity can be recovered from
\begin{equation}
    [\mathrm{Z/H}] \;=\; [\mathrm{Fe/H}] + \log_{10}\!\left(1 - x_\alpha + x_\alpha\,10^{[\alpha/\mathrm{Fe}]}\right),
    \label{eq:zh_total}
\end{equation}

with $x_\alpha=0.687490$ for the \textsc{MIST} abundance pattern. As noted in \S\ref{sec:ssp}, the nebular component is disabled in these fits, which restricts them to continuum-dominated, quiescent systems. Lines that stem from other ionising sources can be marginalised with a flat prior (\S\ref{sec:likelihood}).

Figure~\ref{fig:afe_spectra} shows the effect of varying $[\alpha/\mathrm{Fe}]$ at fixed $[\mathrm{Fe/H}]$, SFH, mass, dust, and velocity broadening. Increasing $[\alpha/\mathrm{Fe}]$ depresses the continuum through enhanced H$^{-}$ opacity, strengthens $\alpha$-sensitive features such as \ion{Ca}{ii}\,H+K, Mg$b$, the TiO bands, and the \ion{Ca}{ii} triplet, and weakens many iron features. This is the signature of the \citet{Park:2025aa} models and what constrains $[\alpha/\mathrm{Fe}]$ in a continuum fit.

\begin{figure*}
    \centering
    \includegraphics[width=0.9\linewidth]{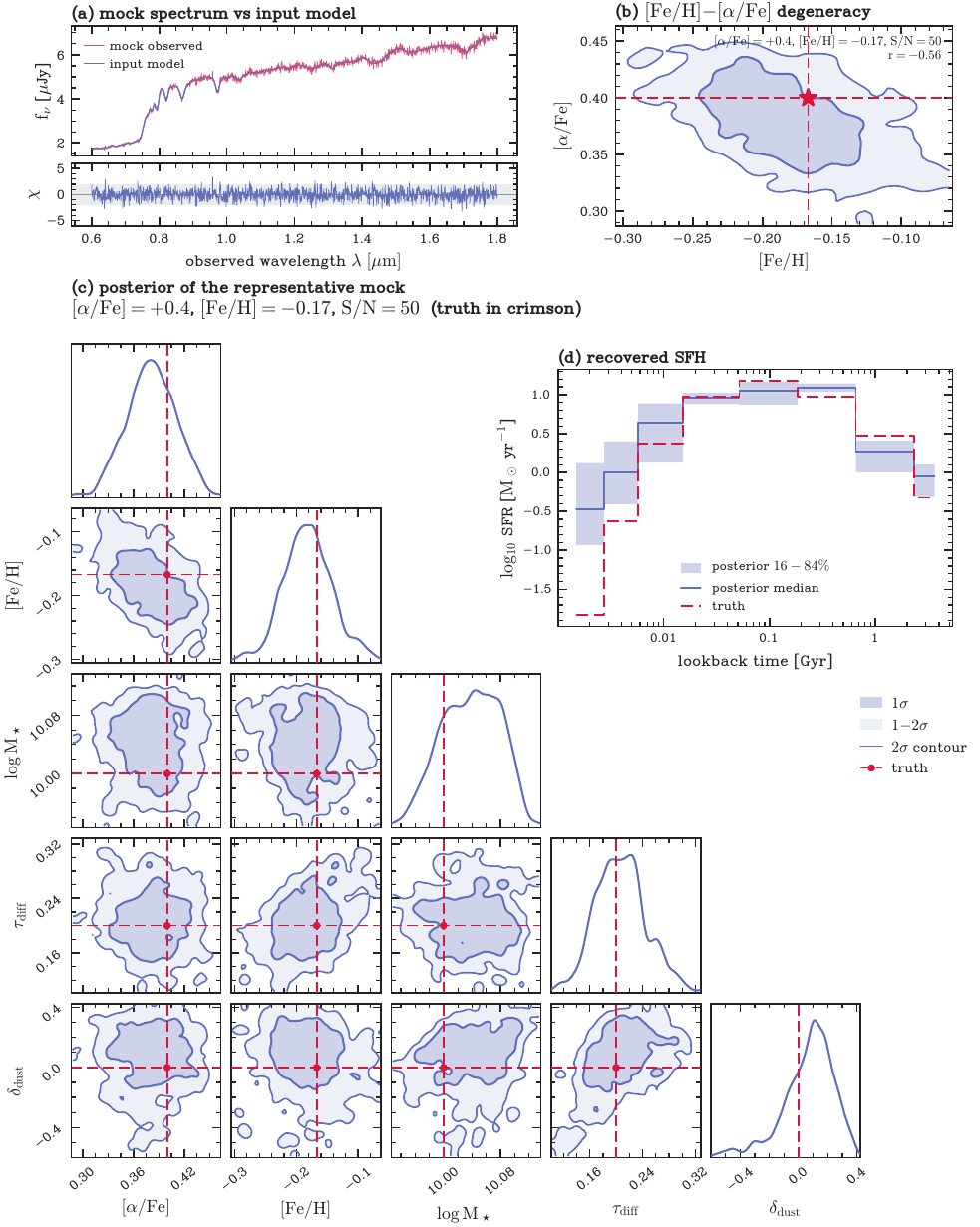}
    \caption{The $\alpha$-enhancement demonstration on a single representative continuum mock ($[\alpha/\mathrm{Fe}]=+0.4$, $[\mathrm{Fe/H}]=-0.17$, $\mathrm{S/N}=50$); every panel shows the same fit. \textbf{(a)} The mock spectrum (crimson) against the noiseless input model (blue), with the standardised residual $\chi=(\mathrm{obs}-\mathrm{model})/\sigma$ below: its scatter is consistent with unit variance and shows no wavelength-dependent structure. \textbf{(b)} The joint posterior of $[\mathrm{Fe/H}]$ and $[\alpha/\mathrm{Fe}]$ (filled $1\sigma$ and light $1$--$2\sigma$ regions, truth starred): the two are anticorrelated ($r=-0.56$) because raising either adds metal opacity and depresses the continuum, and the joint fit represents this degeneracy explicitly rather than tightening the metallicity under an assumed solar abundance pattern. \textbf{(c)} The full posterior of the sampled scalars (truth in crimson): all truths lie within the credible regions, and the $[\alpha/\mathrm{Fe}]$--$[\mathrm{Fe/H}]$ panel repeats the anticorrelation of panel~(b); the star-formation-history ratios are omitted for clarity. \textbf{(d)} The recovered SFH: posterior median and $16$--$84\%$ band per lookback bin against the input truth (dashed), sampled jointly with $[\alpha/\mathrm{Fe}]$, metallicity, mass, and dust under the eight-bin continuity prior. The older bins are recovered within their $16$--$84\%$ bands. The two youngest bins carry little light in this quiescent continuum mock; they are prior-dominated and overestimated.}
    \label{fig:afe_summary}
\end{figure*}

\begin{figure}[!tb]
    \centering
    \includegraphics[width=\linewidth]{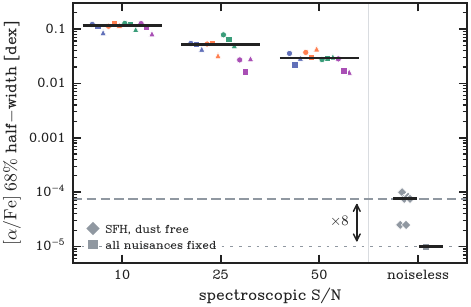}
    \caption{Identifiability of $[\alpha/\mathrm{Fe}]$: the $68\%$ half-width of the recovered $[\alpha/\mathrm{Fe}]$ posterior against the spectroscopic signal-to-noise. The realistic continuum mocks of Fig.~\ref{fig:afe_recovery} are grouped by S/N, with colour and symbol marking the true $[\alpha/\mathrm{Fe}]$ and $[\mathrm{Fe/H}]$ as there and black bars marking the medians. The right-hand group is the idealised limit of noiseless data at the native library resolution: with every nuisance parameter fixed at the truth (squares, dotted line) the posterior collapses to the interpolation floor, and freeing the SFH and dust (diamonds, dashed line) broadens it by a factor of ${\sim}8$, the cost of the star-formation--metallicity--$[\alpha/\mathrm{Fe}]$ covariances. Realistic noise dominates both.}
    \label{fig:afe_ideal}
\end{figure}

We tested the additional $\alpha$-abundance dimension with mock spectra (Fig.~\ref{fig:afe_summary}) based on the fiducial setup of \S\ref{sec:demo_mock}. The \textsc{BPASS} grid is replaced by the $\alpha$-variable \textsc{aMIST}$+$\textsc{C3K} grid, $[\alpha/\mathrm{Fe}]$ is sampled with a uniform prior over the grid support, and the nebular components are disabled. All remaining parameters are fitted as before, including a non-parametric SFH, stellar metallicity, mass, and dust. We first tested the interpolation itself in an idealised limit using noiseless spectra at the native library resolution, without velocity broadening, over $3000$--$10,000$~\AA. The injected $[\alpha/\mathrm{Fe}]$ is recovered without measurable bias both at and between grid nodes, with residual offsets $\lesssim0.004$~dex (Fig.~\ref{fig:afe_ideal}). Allowing the SFH and dust to vary rather than fixing them at their true values broadens the $[\alpha/\mathrm{Fe}]$ posterior by roughly a factor of eight, illustrating the importance of the covariance with the other stellar-population parameters.

Under more realistic conditions, these degeneracies become visible in the posterior. Figure~\ref{fig:afe_summary} shows a representative mock with $[\alpha/\mathrm{Fe}]=+0.4$ and $\mathrm{S/N}=50$. The SFH, $[\mathrm{Fe/H}]$, mass, dust, and $[\alpha/\mathrm{Fe}]$ are recovered jointly, while the $([\mathrm{Fe/H}],[\alpha/\mathrm{Fe}])$ posterior is anticorrelated ($r=-0.56$). This covariance arises because increasing either abundance increases the metal opacity and modifies the continuum in similar ways. Fitting both parameters jointly therefore captures a degeneracy that would otherwise be hidden by fixing $[\alpha/\mathrm{Fe}]$.

\begin{figure*}[!tb]
    \centering
    \includegraphics[width=\linewidth]{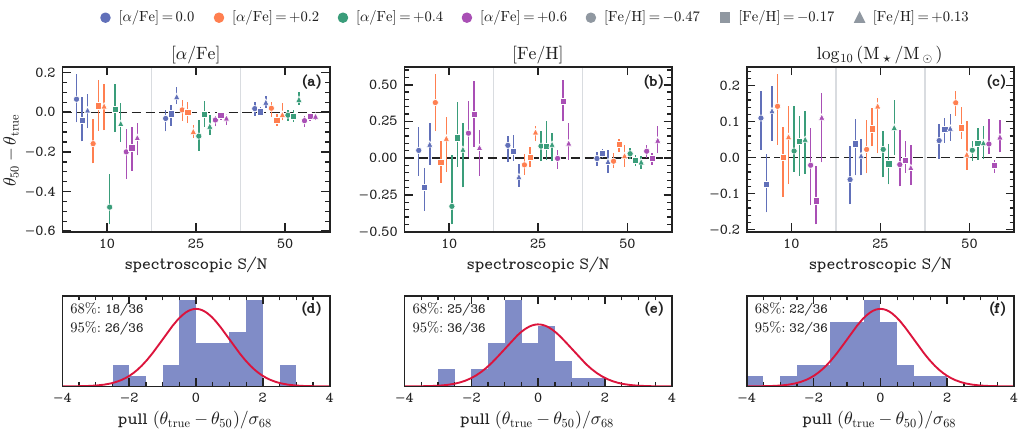}
    \caption{Recovery of $[\alpha/\mathrm{Fe}]$, $[\mathrm{Fe/H}]$ and stellar mass across the realistic continuum mock suite ($36$ mocks: four $[\alpha/\mathrm{Fe}]$ and three $[\mathrm{Fe/H}]$ truths at each of three spectroscopic signal-to-noise ratios, $\mathrm{S/N}=10$, $25$ and $50$; all mocks have $\log_{10}(M_\star/\mathrm{M}_\odot)=10$). \textbf{(a)--(c)} Offset of the posterior median from the truth with the $68\%$ credible interval, grouped by S/N; colour marks the true $[\alpha/\mathrm{Fe}]$ and symbol the true $[\mathrm{Fe/H}]$. \textbf{(d)--(f)} Distribution of the pull $(\theta_{\rm true}-\theta_{50})/\sigma_{68}$, with $\sigma_{68}$ the half-width of the $68\%$ interval on the side of the truth, against a unit Gaussian (red); the numbers give how many truths lie inside the $68\%$ and $95\%$ intervals. The credible intervals narrow with increasing S/N, and $[\alpha/\mathrm{Fe}]$ is recovered without pile-up at the grid edges (\S\ref{subsec:alpha}).}
    \label{fig:afe_recovery}
\end{figure*}

A broader recovery test uses observed-frame $6000$--$18,000$~\AA\ spectra with $\mathrm{S/N}=10$--$50$ (Fig.~\ref{fig:afe_recovery}). Across these mocks, the median $68\%$ half-width on $[\alpha/\mathrm{Fe}]$ is ${\sim}0.05$~dex and the mean absolute pull is $1.06$. At low signal-to-noise, the posterior shows a mild regression towards solar abundance along the $[\mathrm{Fe/H}]$--$[\alpha/\mathrm{Fe}]$ degeneracy. The median recovered $[\alpha/\mathrm{Fe}]$ lies ${\sim}0.05$\,dex below the truth at $\mathrm{S/N}=10$ and ${\sim}0.01$\,dex below it at $\mathrm{S/N}=50$. The recovery of the stellar metallicity, mass, and SFH remains comparable to the fiducial fits. These tests are intended to establish that the $[\alpha/\mathrm{Fe}]$ dimension is implemented correctly, identifiable in principle, and recoverable jointly with the other stellar-population parameters. A detailed study of the abundance-pattern science, the data quality required for robust measurements, and the extension to $\alpha$-enhanced nebular models is deferred to future work.

\subsection{Mock \jwst data}
\label{subsec:mock_jades_data}

One of the primary goals of \ceri\ is to fit high-redshift ($z>4$) \jwst galaxy observations with increasingly flexible physical models. This mock suite prepares the real-data comparison of \S\ref{sec:demo_real} and adopts the same model, including nebular emission. The $\alpha$-enhanced axis of \S\ref{subsec:alpha} is not included because it currently requires the nebular component to be disabled. We constructed mock datasets that reproduce the JADES \citep{eisenstein_overview_2023,eisenstein_jades_2023} NIRCam photometry and NIRSpec \citep{Jakobsen:2022aa} R1000 emission-line measurements \citep{curti_jades_2024,Curtis-Lake:2025aa}, using the band set, line list, and catalogue uncertainties of real JADES galaxies.

The forward model follows the fiducial setup of \S\ref{sec:demo_mock}, with several modifications tailored to these data. The SFH is represented by $12$ bins, with the three youngest fixed to $0$--$3$, $3$--$5$, and $5$--$10$~Myr and the remaining bins spaced uniformly in logarithmic lookback time. This finer sampling at young ages resolves the recent star formation that powers the nebular emission. The eleven adjacent-bin SFR ratios are assigned a Student-$t$ continuity prior with $\nu=2$ and $\sigma=0.5$.  Dust attenuation uses the \citet{charlot_simple_2000} birth-cloud component together with the diffuse \citet{kriek_dust_2013} law. Rather than sampling the birth-cloud optical depth directly, we sampled the ratio $\tau_{\rm BC}/\tau_{\rm diff}$, alongside $\tau_{\rm diff}$ and the diffuse-curve slope $\delta_{\rm dust}$. Nebular emission is controlled by the gas-phase metallicity $\log_{10}(Z_{\rm gas}/Z_\odot)$ and ionisation parameter $\log_{10}U$. To account for aperture differences and slit losses between the NIRSpec micro-shutter and NIRCam photometry, we additionally fitted an emission-line scaling parameter $s_{\rm el}$; its input value is drawn from the same prior used in the fit. The stellar mass $\log_{10}(M_\star/\mathrm{M}_\odot)$ is also free, while the redshift is fixed to its true value for each mock.  The sampled parameter vector therefore has $19$ dimensions. The adopted priors are summarised in Table~\ref{tab:mock_priors}.

\begin{table}
    \centering
    \caption{Priors of the \jwst-like mock recovery tests of \S\ref{subsec:mock_jades_data}.  All parameters are sampled jointly at fixed redshift; $\mathcal{N}_{[a,b]}$ denotes a normal distribution truncated to $[a,b]$, $\mathcal{U}_{[a,b]}$ a uniform distribution on the interval $[a,b]$. The ranges of $\log_{10}(Z_{\rm gas}/Z_\odot)$ and $\log_{10}U$ are those of the nebular grid.}
    \label{tab:mock_priors}
    \begin{tabular}{lll}
        \hline
        Parameter & Prior & Range/scale \\
        \hline
        $\log_{10} M_\star\,[{\rm M}_\odot]$ & $\mathcal{U}$ & $[6,\,12.5]$ \\
        $\log_{10} (Z/Z_\odot)$ & $\mathcal{U}$ & $[-2.30,\,+0.30]$ \\
        $\Delta\log_{10}{\rm SFR}$ (11 ratios) & Student-$t$ & $\nu=2$, $\sigma=0.5$ \\
        $\tau_{\rm diff}$ & $\mathcal{N}_{[0,4]}$ & $\mu=0.3$, $\sigma=1.0$ \\
        $\tau_{\rm BC}/\tau_{\rm diff}$ & $\mathcal{N}_{[0,2]}$ & $\mu=1.0$, $\sigma=0.3$ \\
        $\delta_{\rm dust}$ & $\mathcal{U}$ & $[-1,\,0.4]$ \\
        $\log_{10} Z_{\rm gas}/Z_\odot$ & $\mathcal{U}$ & $[-1.3,\,0.3]$ \\
        $\log_{10} U$ & $\mathcal{U}$ & $[-4,\,-1]$ \\
        $s_{\rm el}$ (line calibration) & LogNormal & $\mu_{\ln}=0$, $\sigma_{\ln}=0.2$ \\
        \hline
    \end{tabular}
\end{table}

\begin{figure*}[!tb]
    \centering
    \includegraphics[width=\linewidth]{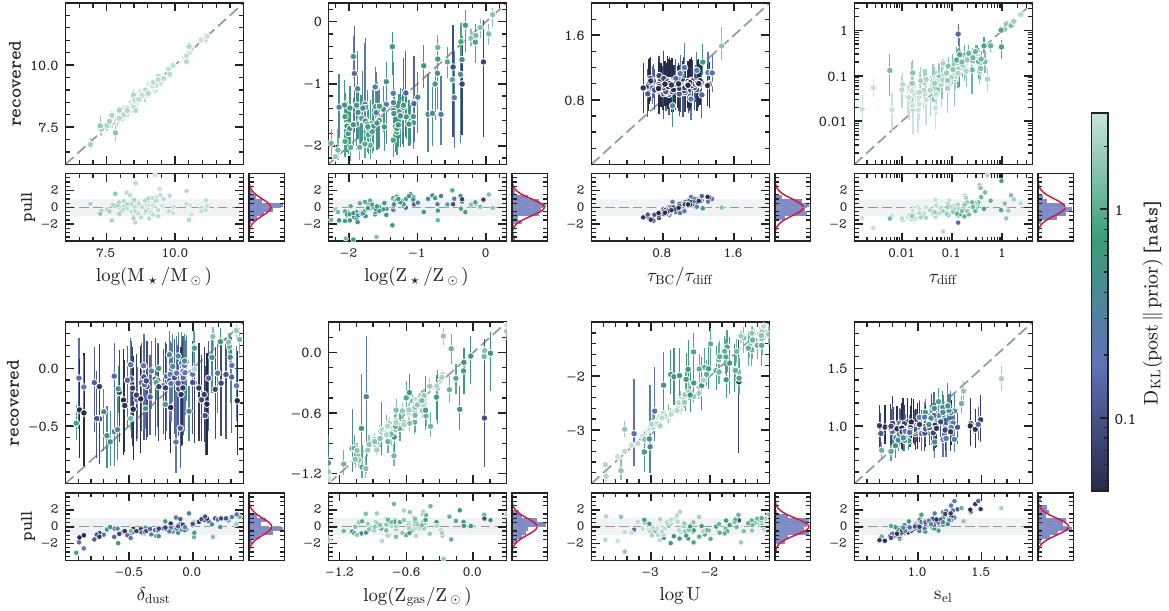}
    \caption{Parameter recovery for 100 mock galaxies matched to the JADES sample. Each mock adopts the NIRCam bands, NIRSpec R1000 line set, and catalogue uncertainties of a real galaxy, with the observations replaced by a known model realisation plus Gaussian noise (\S\ref{subsec:mock_jades_data}; Table~\ref{tab:mock_priors}). For each parameter, the upper panel compares the posterior median and $68\%$ credible interval with the input truth; the dashed line marks one-to-one recovery. The lower panel shows the pull, $(\theta_{\rm true}-\theta_{50})/\sigma_{68}$, with the grey band indicating $|\mathrm{pull}|<1$ and the corresponding pull distribution compared with a unit Gaussian (red). Points are coloured by the Kullback--Leibler divergence from the prior, $D_{\rm KL}$ (Fig.~\ref{fig:mock_kl}), indicating how strongly each parameter is constrained by the data. All mocks are fitted at their true, fixed redshift.}
    \label{fig:mock_recovery}
\end{figure*}

To generate the mock suite we selected $100$ sources at evenly spaced redshift ranks across the fitted JADES catalogue, spanning $z\simeq0.25$--$9.6$ with a median of $z\simeq3.1$. The truth vector of each mock is the posterior-predictive draw closest to the median of a preliminary fit to its source galaxy, so that every input lies in a region of parameter space the real data seem to support. The line-calibration factor $s_{\rm el}$ is instead drawn from its prior and the predicted line fluxes rescaled accordingly; its recovery is therefore a simulation-based-calibration (SBC) test. Each mock's data are the noiseless model prediction at the truth, perturbed with Gaussian noise at the source galaxy's catalogue uncertainties, and adopt the same NIRCam filters and NIRSpec emission lines observed for that source: between $8$ and $17$ bands and $10$ and $24$ lines, so the suite inherits the survey's per-datum signal-to-noise structure. Every mock was refitted at its true, fixed redshift, with $300$ live points, $60$ replaced per iteration and $40$ slice steps per replacement, terminating at $\ln(Z_{\rm live}/Z)<-5$ (\mbox{\S\ref{sec:sampling}}).

Figure~\ref{fig:mock_recovery} compares the recovered posterior medians and $68\%$ intervals with the truths. Every parameter is recovered without significant bias: the pull distributions, $(\theta_{\rm true}-\theta_{50})/\sigma_{68}$, are consistent with a standard normal, with median offsets no larger than $\sim\!0.3\sigma$, the largest being the diffuse optical depth ($-0.34\pm1.17$) and the dust slope ($-0.24\pm0.85$), while stellar mass ($+0.06\pm1.01$) and stellar metallicity ($+0.02\pm1.07$) are unbiased within the quoted uncertainties. The credible intervals are correspondingly well calibrated: across the eight sampled scalars the fraction of truths inside the $68\%$ ($95\%$) interval spans $0.69$--$0.87$ ($0.92$--$1.00$), the best-constrained parameters erring slightly conservative rather than overconfident. The line-calibration factor passes its SBC test ($+0.16\pm1.06$; coverage $0.69/0.92$).

\begin{figure}[!tb]
    \centering
    \includegraphics[width=\linewidth]{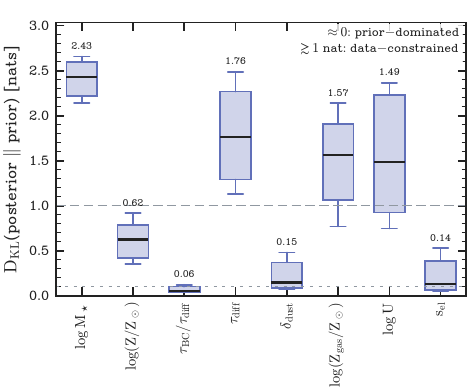}
    \caption{Information gained from the data in the mock recovery test. Boxes span the $16$--$84$th percentiles over the $100$ mocks, with the median marked and printed above each box; the dotted and dashed lines mark $0.1$ and $1$\,nat. The panels show the Kullback--Leibler divergence $D_{\rm KL}$ from prior to posterior, per sampled parameter. Divergences well above zero mark data-driven posteriors; the lowest $D_{\rm KL}$ distributions, like the birth-cloud-to-diffuse ratio $\tau_{\rm BC}/\tau_{\rm diff}$ and the dust slope $\delta_{\rm dust}$, identify the parameters the data constrain weakly. The line-calibration factor $s_{\rm el}$ has a small divergence by construction, its truths being drawn from the prior.}
    \label{fig:mock_kl}
\end{figure}

Coverage alone does not establish that the data are informative: a posterior identical to the prior could still contain the truth. Figure~\ref{fig:mock_kl} therefore shows the information gain, $D_{\rm KL}$, for every parameter and mock, with the same quantity used to colour Fig.~\ref{fig:mock_recovery}. Stellar mass is the best constrained, with a median gain of $2.4$~nats, while the gas-phase metallicity, diffuse optical depth, and ionisation parameter gain $1.5$--$1.8$~nats. By contrast, $\tau_{\rm BC}/\tau_{\rm diff}$ and the diffuse dust slope $\delta_{\rm dust}$ remain largely prior-dominated, with median gains of only $0.06$ and $0.15$~nats, respectively. Their posterior medians consequently remain close to the prior centre, producing the systematic pull trends seen in Fig.~\ref{fig:mock_recovery}.

The mock suite also illustrates how $D_{\rm KL}$ can be used to identify which parts of the data constrain a given parameter. For example, the information gain on $\tau_{\rm diff}$ correlates more strongly with median emission-line S/N ($\rho\simeq0.4$) than with photometric S/N ($\rho\simeq0.05$), and mocks in the upper half of the line-S/N distribution have ${\sim}37\%$ narrower posteriors. This is consistent with the Balmer decrement carrying much of the information on the dust attenuation. The $18$ mocks with $\tau_{\rm diff}>0.3$ are too few to confirm it on their own: for them the correlation with line S/N weakens to $\rho\simeq0.16$ (against $-0.27$ for photometric S/N), and their posterior widths show no trend with line S/N.

The same exercise can be applied to other parameters. As expected, gas-phase metallicity and ionisation parameter show stronger correlations with line S/N ($\rho\simeq0.84$ and $0.76$) than with photometric S/N, while stellar metallicity is more closely tied to the continuum. The information gain on the emission-line scaling parameter also increases where the lines carry more weight. These examples are not intended as a detailed analysis of the information content of JADES data, but to demonstrate how information gain can be combined with observational properties to diagnose which measurements drive each inferred parameter.

\section{Demonstrations on real data}
\label{sec:demo_real}
The mock tests of \S\ref{sec:demo_mock} establish that \ceri\ can recover known inputs under controlled conditions; we now apply it to real \jwst observations. We benchmark \ceri\ against the established SED-fitting code \pros\ on a sample of JADES galaxies \citep{eisenstein_overview_2023} with the same data and priors, and close with a posterior comparison on a single galaxy.

\subsection{The JADES sample and a comparison with \pros}
\label{sec:demo_jades}
\begin{figure*}[!tb]
    \centering
    \includegraphics[width=0.9\linewidth]{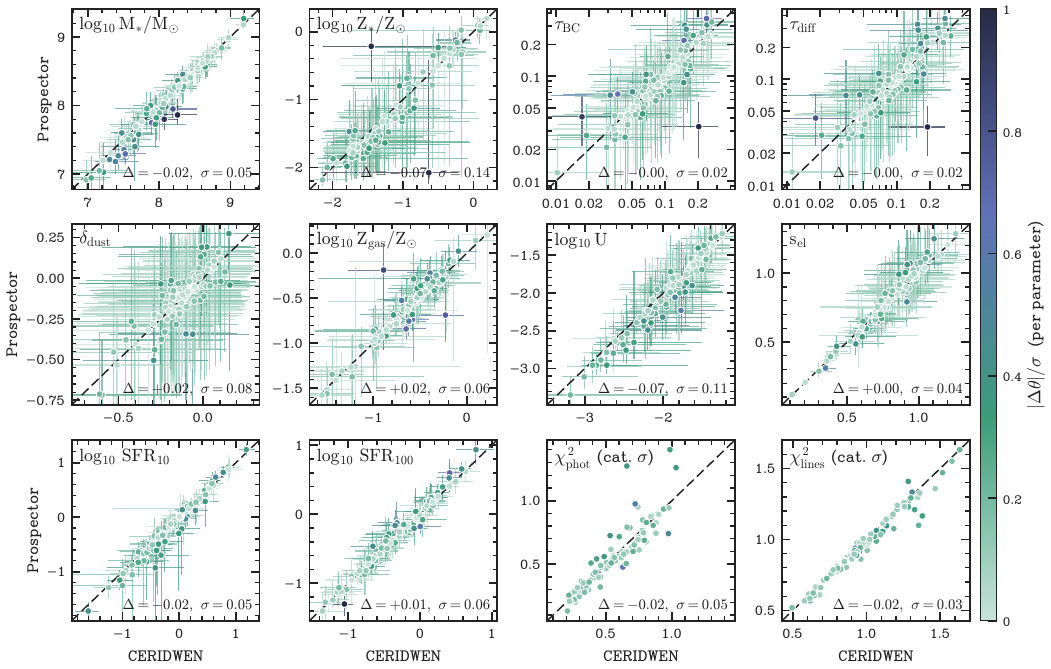}
    \caption{Controlled \ceri--\pros\ comparison for the $98$ matched JADES galaxies of \S\ref{sec:demo_jades}, selected for good \ceri\ fits. The panels show posterior medians (points) and $68\%$ intervals (bars) of \pros\ against \ceri\ for the eight sampled parameters, the $10$- and $100$-Myr star-formation rates, and the per-arm $\chi^{2}$ recomputed with the raw catalogue uncertainties, with the $1{:}1$ line (dashed) and the median offset $\Delta$ and robust scatter $\sigma$ annotated per panel.  Points are coloured by the per-parameter discrepancy $|\Delta\theta|/\sigma$, from exact agreement (pale) to one sigma (dark). The input data vectors, priors and eight-bin SFH are identical.}
    \label{fig:prospector_best100}
\end{figure*}

To benchmark \ceri\ against an established code on identical data, we refitted $100$ JADES galaxies with matched NIRCam photometry and R1000 emission-line fluxes, with the same data, priors and SFH in \ceri\ and \pros\ \citep{Johnson:2021aa}. The photometry is the JADES NIRCam Kron photometry of the GOODS-S and GOODS-N fields \citep[data release 5;][]{eisenstein_overview_2023,eisenstein_jades_2023}: $8$--$17$ bands per galaxy in this sample, from the wide filters F070W, F090W, F115W, F150W, F200W, F277W, F356W and F444W and the medium bands F162M, F182M, F210M, F250M, F300M, F335M, F410M, F430M, F460M and F480M, each entering a fit only where its catalogue flag is clean. The line fluxes and the spectroscopic redshifts, at which every fit is held fixed, are the NIRSpec R1000 measurements of data release 4 \citep{Curtis-Lake:2025aa,Scholtz:2025aa}; both arms carry per-datum error floors of $5$ and $10$\% of the measured flux on the photometry and lines respectively.  The input data vectors are identical, and both codes fit the model of \S\ref{subsec:mock_jades_data} restricted to eight star-formation-history bins: a \textsc{BPASS} SSP grid, the two-component dust model with free diffuse slope, nebular line and continuum emission with free $\log_{10}(Z_{\rm gas}/Z_\odot)$ and $\log_{10}U$, the line-calibration factor $s_{\rm el}$, and fixed \citet{madau_radiative_1995} IGM absorption.  The SFH has the three youngest bins fixed at $0$--$3$, $3$--$5$, and $5$--$10$\,Myr and five log-spaced bins beyond; both codes sample the same $15$ free parameters under the same priors (Table~\ref{tab:mock_priors}, here with seven bin ratios and $s_{\rm el}\sim\mathcal{N}_{[0,2]}(1.0,\,0.3)$). The \ceri\ fits used the nested-sampling settings that a dedicated tuning sweep selected for this model: $m=300$ live points, $k=64$ replaced per iteration and $40$ slice steps per replacement, terminating at $\ln(Z_{\rm live}/Z)=-5$ (\S\ref{sec:sampling}); the \pros\ fits used \texttt{nautilus} \citep{Lange:2023aa} through \pros's \texttt{fit\_model} interface with $3000$ live points and a target of $10^{4}$ effective posterior samples, the sampler's own default. Each \pros\ fit evaluates its likelihood serially in a single process, with \textsc{FSPS} and \textsc{NumPy} limited to two threads, and $64$ such fits share one $128$-core COSMA8 node (two $64$-core AMD EPYC processors and $1$\,TB of memory), so the \pros\ wall-clock times of \S\ref{sec:performance} are those of a fully packed node. Figure~\ref{fig:prospector_best100} compares the recovered posteriors parameter by parameter.

\begin{figure}
    \centering
    \includegraphics[width=0.95\linewidth]{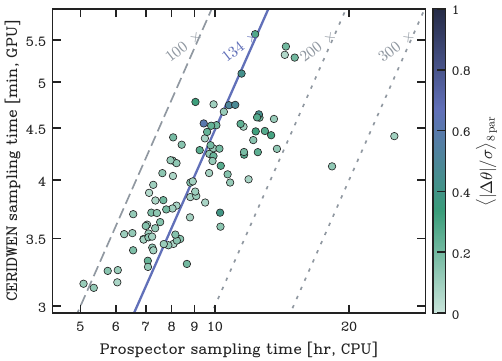}
    \caption{Per-galaxy sampling-time comparison of the nested-sampling fits between \ceri\ (\texttt{blackjax} nested slice sampling, single A100 GPU) and \pros\ (\texttt{nautilus}, CPU) on the matched refits.  Points are coloured by the per-galaxy mean parameter discrepancy $|\Delta\theta|/\sigma$, averaged over the eight sampled parameters as in Fig.~\ref{fig:prospector_best100}. The diagonal guides mark $100\times$, the median $134\times$, $200\times$ and $300\times$ speed-up.}
    \label{fig:jades_prospector}
\end{figure}

The colouring in Fig.~\ref{fig:prospector_best100} quantifies the agreement panel by panel: each point is shaded by the discrepancy in that parameter, $|\Delta\theta|/\sigma$ with $\sigma=(\sigma_{\ceri}^{2}+\sigma_{\rm Pro}^{2})^{1/2}$, on a common scale from exact agreement (pale) to one sigma (dark). The median discrepancy lies between $0.10$ and $0.16\,\sigma$ for every sampled parameter, and only 
$0.6$\% of all (galaxy, parameter) pairs exceed $1\,\sigma$.  The two right-hand panels show that this is not achieved at the expense of the fit: the raw $\chi^{2}$ values scatter tightly about the one-to-one relation, with median $\chi^{2}_{\nu}=1.56$ for \ceri\ against $1.55$ for \pros.

The median offsets between the two codes are consistent with zero for every parameter the likelihood constrains: $-0.018$~dex in $\log_{10} M_{\star}$, $-0.001$ in both dust optical depths, $+0.005$ in the emission-line scaling, and $-0.019$ and $+0.014$~dex in $\mathrm{SFR}_{10}$ and $\mathrm{SFR}_{100}$.  The two largest residuals, in stellar metallicity ($-0.071$~dex) and ionisation parameter ($-0.067$~dex), occur among the quantities the mock KL analysis identifies as weakly constrained and prior sensitive, and both remain small compared with their posterior widths.  The mock and real-data tests probe different things.  The mocks show that \ceri\ recovers known inputs with calibrated uncertainties when the assumed forward model generates the data (Figs.~\ref{fig:mock_recovery}--\ref{fig:mock_kl}); this is recovery \emph{conditional on that model}, and carries no guarantee for real galaxies whose physics it may not fully represent.  The \pros\ comparison instead asks whether two independent implementations, different languages, different samplers, and a Fortran population-synthesis library against a differentiable interpolation of the same stellar grids, infer the same parameters from the same observations. Since neither the real data nor the comparison supplies a ground truth, it should be read as evidence of consistency with an established modelling framework rather than as a test of absolute accuracy.

Appendix~\mbox{\ref{app:gal_comparison}} shows an object-level comparison for JADES~170891 at $z=4.43$ (Fig.~\mbox{\ref{fig:prospector_gal170891}}). The posterior-predictive SEDs, the emission-line residuals and the SFHs agree, and the joint posteriors occupy the same regions of parameter space.

\section{Computational performance}
\label{sec:performance}

The practical utility of \ceri\ depends on two computational properties beyond the convergence tests discussed in \S\ref{sec:sampling}: how the computational cost scales with model dimensionality, and how its wall-clock runtime compares with established SED-fitting frameworks. Here, we examine the cost per likelihood evaluation and the resulting wall-clock performance. 
The computational cost per likelihood evaluation changes only weakly with model dimensionality over the tested range of $25$--$125$ free parameters. Across this range the sampler performs $540$--$1440$ slice steps per second (median ${\sim}810$), with no systematic trend with the number of parameters (Appendix~\ref{app:benchmark}). Each slice step evaluates the likelihood at least once. At fixed $N_T$ the rate differs by up to a factor of two between fits that take the same number of steps. For the high-dimensional experiments of \S\ref{sec:high_detail_sfh}, run on a single NVIDIA A100 GPU, a converged fit takes $23$--$44$ minutes at $N_T=20$ ($25$ free parameters) and $2.7$--$5.0$ hours at $N_T=120$ ($125$ free parameters), including compilation. This increase in wall-clock time is driven primarily by the larger number of likelihood evaluations required to explore and converge the higher-dimensional posterior, and not by an increase in the cost of each evaluation.
The weak dependence of the per-evaluation cost on dimensionality arises from the compiled and vectorised structure of the forward model. Once compiled, additional model parameters add relatively little computational overhead. This scaling makes the flexible, high-dimensional SFHs explored in \S\ref{sec:demo_mock} computationally tractable.

Finally, Fig.~\ref{fig:jades_prospector} compares the per-galaxy sampling time of \ceri\ with \pros\ for the matched refits described in \S\ref{sec:demo_jades}, for which the data, physical model, and priors are controlled as closely as possible between the two frameworks. On a single A100 GPU, \ceri\ achieves a median sampling time of $4.0$ minutes per galaxy ($4.3$ minutes wall clock, including compilation), corresponding to a median speed-up of ${\sim}134\times$ relative to the CPU-based \pros\ fits. This factor should not be interpreted as an algorithm-independent benchmark: it is the median for a particular galaxy sample, a particular pair of matched model configurations, and a specific hardware comparison, and it therefore combines differences in implementation, sampling strategy, and GPU versus CPU execution. For JADES~170891 we also timed single likelihood evaluations after compilation. \mbox{\ceri} takes $0.045$\,ms per evaluation on one A100 when it evaluates $64$ parameter vectors at once, the batch the nested sampler uses, and $1.3$\,ms for a single vector. \mbox{\pros} takes $93$\,ms on one core of an AMD EPYC 7302 CPU. Per evaluation, \mbox{\ceri} is therefore ${\approx}2000$ times faster at the sampler's batch width and ${\approx}70$ times faster for a single vector. The \mbox{\ceri} fit used ${\approx}1.4\times10^{6}$ likelihood evaluations and the \mbox{\pros} fit $2.9\times10^{5}$. It is also a per-fit ratio of sampling times, not a throughput ratio. The \pros\ fits ran $64$ to a $128$-core node with two threads each (\S\ref{sec:demo_jades}) and took a median of $9.0$~h, so one fully packed node returned ${\approx}7$ posteriors per hour (${\approx}0.06$ per core-hour); one A100 running one fit at a time returns ${\approx}14$ per hour. Per unit of hardware time the gain is therefore a factor of ${\sim}2$ for this model, while the latency of an individual fit falls by two orders of magnitude, which is what makes interactive model exploration practical. First tests on a consumer-grade GPU (an NVIDIA GeForce RTX 5060) reached the same wall time as the A100 for these models, but a systematic benchmark across GPUs is still outstanding. The comparison further assumes one fit per GPU; whether several independent fits can share one device, and how far a single fit saturates its memory and compute, we have not measured.  Nevertheless, it provides a direct measure of the practical computational regime enabled by \ceri.  In the setting targeted by this work---full posterior inference for flexible, high-dimensional stellar-population models---fits that take hours per galaxy on a CPU finish in minutes on one GPU for the JADES model, and within a few hours for the $125$-parameter SFH fits.

\section{Current limitations and future extensions}
\label{sec:future}
\ceri\ is a working tool with known boundaries, and its modular, differentiable structure makes several of them straightforward to expand and evolve. We outline here the extensions under active development.

The first concerns the stellar ingredients. The \texttt{SSPData} interface already accepts any grid supplied in the expected format (\S\ref{sec:ssp}), and \S\ref{subsec:alpha} demonstrates the first higher-dimensional instance: a leading $[\alpha/\mathrm{Fe}]$ axis interpolated alongside age and $[\mathrm{Fe/H}]$. Because the CSP assembly is a single contraction over the grid axes (Eq.~\eqref{eq:csp_sum_main}), further axes, such as an IMF slope or more abundance ratios, follow the same pattern and leave the sampling machinery untouched. The cost falls on the offline grid generation and on memory. A next step is to promote $[\alpha/\mathrm{Fe}]$ from a single sampled scalar to a time-varying quantity, letting the enrichment mode of \S\ref{sec:zh} track abundance patterns and connecting the recovered enrichment histories directly to chemical-evolution models.

The second, and more pressing, concerns the ionised gas. \ceri\ currently interpolates the fixed \citet{Byler:2017aa} grids, which ties the nebular emission to the stellar models those grids assume (\S\ref{sec:nebular}) and represents each galaxy by what is effectively a single ionised cloud. Both assumptions are known to be limiting. \citet{marconi_homerun_2024} show that no single-cloud photoionisation model reproduces observed line ratios across ionisation stages, and that a data-constrained superposition of clouds recovers them to percent-level accuracy with direct consequences for inferred gas-phase abundances. Meanwhile, emulation has made flexible photoionisation tractable inside a fit: \citet{li_cue_2024} decouple the nebular calculation from any particular stellar ionising spectrum by emulating \textsc{Cloudy} over a flexible parametrisation of the ionising continuum. Both directions are within reach of a differentiable framework, and an emulator of this kind is an upgrade path for \ceri, making the nebular physics a sampled component rather than a fixed lookup table. Beyond a single ionised cloud, we plan to emulate full three-dimensional Monte Carlo radiative transfer with \textsc{colt} \citep{McClymont:2025ab}. An emulator trained on these calculations would predict the emergent nebular emission, including the escape of ionising photons, as a differentiable function of the fitted parameters, so that \ceri\ can sample it inside the likelihood like the current grid. The most immediate case is the $\alpha$-enhanced mode of \S\ref{subsec:alpha}, restricted to continuum-only fits because no $\alpha$-enhanced photoionisation tables exist: we plan \textsc{Cloudy} grids driven by the aMIST/C3K ionising continua with gas-phase abundance patterns matched to the stellar $[\alpha/\mathrm{Fe}]$, so that $\alpha$-enhancement can be fit self-consistently in star-forming galaxies. A related addition is an AGN component, both an accretion-disc continuum and its dusty-torus re-emission, entering as one more additive stream in the Tier-A/Tier-B structure of Fig.~\ref{fig:ceridwen_flow}.

On the inference side, we plan a broader treatment of the variational engines: the full-rank and flow-based guides already included in \ceri\ are candidates for amortised or preconditioned sampling schemes at survey scale, where millions of objects make even a $134\times$ speed-up per object only the starting point. The same architecture also makes \ceri\ a fast simulator for simulation-based inference. The forward model is vectorised over a batch dimension and compiled. It therefore generates parameter--observation pairs at a rate comparable to its likelihood throughput. Because it is differentiable, gradients of the simulator with respect to its parameters are also available to the training scheme. The objection raised in \S\ref{sec:introduction}, that amortised methods must be retrained whenever the physical model changes \citep{Hahn:2022aa,Khullar:2022aa}, is not thereby removed, but its cost is reduced: retraining requires a fast simulator over the new model, which is exactly what a modular, accelerator-native forward model provides.

Finally, a companion study is in preparation: a systematic exploration of physically motivated priors for the joint SFH–ZH inference of \S\ref{sec:demo_zh}, building on the prior-width experiment of \S\ref{sec:high_detail_sfh} and assessing the data required to recover the highest-resolution histories (Stoffers et al., in preparation).

\FloatBarrier

\section{Conclusions}
\label{sec:conclusions}
SED fitting faces a persistent tension between physical flexibility and computational tractability: models sufficiently flexible to capture the complexity of galaxy spectra can become prohibitively expensive to explore, while computational constraints often motivate restrictive assumptions about the underlying stellar populations. We presented \ceri, designed to reduce this tension. 

\ceri\ is a GPU-native framework for Bayesian SED fitting in which the stellar-population model, the observational projection and the likelihood form a single compiled, vectorised, differentiable computation, around which each sampler step is compiled. The forward model can be evaluated simultaneously for arbitrary batches of parameter vectors, allowing the sampler to replace a batch of live points in parallel on the GPU during nested sampling. At the same time, end-to-end differentiability provides exact gradients of the same log-posterior for gradient-based inference methods, including Hamiltonian Monte Carlo and variational inference. These samplers are included in the package but were not used for the results shown here. The physical model is constructed from interchangeable components, separating model specification from inference: new ingredients can inherit automatic differentiation, vectorisation, and accelerator execution without requiring changes to the inference machinery. The result is not only a faster implementation of a fixed SED model, but a framework designed to make increasingly flexible stellar-population models computationally accessible.

Under controlled mock conditions, \ceri\ accurately recovers parameters generated by its assumed forward model and reproduces the expected information-content trends with photometry, spectroscopy, and wavelength coverage. These experiments establish recovery and posterior calibration \emph{conditional on the adopted generative model}; they do not imply that the same parameters must be unbiased for real galaxies, whose stellar populations and spectra may differ from the assumptions encoded in the model. Within this controlled setting, the computational efficiency of \ceri\ makes substantially more flexible inference problems practical. We performed full posterior sampling of SFHs represented by up to ${\sim}120$ time nodes and jointly inferred time-dependent chemical-enrichment histories rather than reducing the stellar metallicity distribution to a single value (\S\ref{sec:demo_mock}). Increasing the number of SFH nodes does not by itself improve the recoverable temporal resolution: the inferred structure remains limited by the information in the data and the adopted prior (\S\ref{sec:high_detail_sfh}). Using $\alpha$-enhanced aMIST/C3K stellar libraries, we additionally treat $[\alpha/\mathrm{Fe}]$ as an independent stellar-population dimension in continuum-only fits and recover it jointly with metallicity, stellar mass, and SFH, and the joint posterior represents the resulting $[\mathrm{Fe/H}]$--$[\alpha/\mathrm{Fe}]$ degeneracy explicitly (\S\ref{subsec:alpha}).
A suite of 100 mock galaxies constructed using the observing configurations of the JADES sample provides a complementary test under realistic noise levels and wavelength coverage. For data generated within the assumed model family, the data-constrained parameters are recovered without significant bias and their credible intervals are well calibrated. Prior-dominated parameters regress towards the prior, which produces the pull trends of Fig.~\ref{fig:mock_recovery}. The corresponding Kullback--Leibler divergences quantify which parameters gain substantial information from the available observations and which remain predominantly prior constrained (\S\ref{subsec:mock_jades_data}). 

On observed galaxies, where the underlying physical parameters are not known, we instead test consistency against an independent and established inference framework. For matched fits to JADES galaxies, using the same data and closely controlled physical assumptions and priors, \ceri\ recovers posteriors consistent with those obtained using the established SED-fitting code \pros\ (\S\ref{sec:demo_real}). The median sampling time is $4.0$\,min per galaxy ($4.3$\,min including compilation) on a single A100 GPU, a median per-fit speed-up of ${\sim}134\times$ over the CPU-based \pros\ fits considered here (\S\ref{sec:performance}). Per unit of hardware time the gain is ${\sim}2\times$. This comparison does not establish the physical correctness of either model for real galaxies, nor does the speed-up isolate a single algorithmic contribution; rather, it demonstrates that \ceri\ can reproduce an established inference calculation while moving substantially more rapidly through the same statistical problem.

Many assumptions conventionally adopted in SED fitting are motivated at least partly by computational tractability, yet these assumptions can propagate directly into astrophysical conclusions. As the examples in \S\ref{sec:introduction} show, the adopted SFH and the physical components included can change inferred stellar masses substantially. Increasing computational flexibility does not remove model uncertainty: real galaxies may contain stellar populations, abundance patterns, dust geometries, or other physics absent from any adopted forward model. It does, however, remove computational cost as a reason to impose unnecessarily restrictive parameterisations.
This distinction is central to the purpose of \ceri. The goal is not to claim that a sufficiently flexible model guarantees the truth, but to make model flexibility itself testable. By making high-dimensional posterior inference computationally tractable, \ceri\ enables alternative assumptions about SFHs, chemical enrichment, abundance patterns, dust, and other stellar-population physics to be compared rather than fixed for computational convenience. The resulting framework shifts the practical limitation of SED fitting away from how much complexity can be afforded and towards the more scientifically meaningful question of which physical model is supported by the data. 
\ceri\ is publicly available: the source code at \url{https://github.com/Espe13/ceridwen}, the documentation at \url{https://www.amanda-stoffers.de/ceridwen/}, and the pre-computed stellar-population grids on Zenodo (see Data Availability).

\section*{Acknowledgements}

We thank Joel Leja, Emilie Burnham, and William McClymont for helpful discussions and comments. AS acknowledges support from the STFC Centre for Doctoral Training in Data Intensive Science at the University of Cambridge (ST/W006812/1), partly funded by the UKRI Frontier Research grant RISEandFALL. This work used the Tursa and COSMA high-performance computing facilities. Claude Opus~4.8 and Opus~5.0 was used to assist with drafting documentation and code comments of \ceri, generating some plotting scripts, and revising the manuscript for grammar and clarity and formatting the tikz figure; all outputs were reviewed and revised by the authors. This work made use of \textsc{JAX} \citep{jax2018github}, \textsc{BlackJAX} \citep{Cabezas:2024aa}, \textsc{TensorFlow Probability} \citep{Dillon:2017aa}, \textsc{FSPS} \citep{conroy_propagation_2009}, \texttt{sedpy} \citep{johnson_2021_4582723}, \textsc{NumPy} \citep{harris2020array}, \textsc{SciPy} \citep{Virtanen:2020aa}, \textsc{Matplotlib} \citep{Hunter:2007}, and \textsc{Astropy} \citep{Astropy-Collaboration:2022aa}.

\section*{Data Availability}
The \ceri\ source code and documentation are available at \url{https://github.com/Espe13/ceridwen} (v1.0.12) and \url{https://www.amanda-stoffers.de/ceridwen/}. The SSP grids used in this work are archived on Zenodo and can be downloaded with \texttt{ceridwen.ssps.fetch\_grid}. The JADES data used in \S\ref{sec:demo_real} are publicly available \citep{eisenstein_overview_2023}.

\bibliographystyle{mnras}
\bibliography{example}

@article{Tinsley:1980aa,
	author = {{Tinsley}, B.~M.},
	journal = {\fcp},
	month = jan,
	pages = {287-388},
	title = {{Evolution of the Stars and Gas in Galaxies}},
	volume = {5},
	year = 1980}

@article{Schaller:1992aa,
	author = {{Schaller}, G. and {Schaerer}, D. and {Meynet}, G. and {Maeder}, A.},
	journal = {\aaps},
	month = dec,
	pages = {269},
	title = {{New Grids of Stellar Models from 0.8-SOLAR-MASS to 120-SOLAR-MASSES at Z=0.020 and Z=0.001}},
	volume = {96},
	year = 1992}

@article{madau_radiative_1995,
	author = {Madau, Piero},
	journal = {\apj},
	month = {March},
	pages = {18},
	title = {Radiative {Transfer} in a {Clumpy} {Universe}: {The} {Colors} of {High}-{Redshift} {Galaxies}},
	volume = {441},
	year = {1995}}

@article{Miralda-Escude:1998aa,
	author = {{Miralda-Escud{\'e}}, Jordi},
	journal = {\apj},
	month = jul,
	number = {1},
	pages = {15-22},
	title = {{Reionization of the Intergalactic Medium and the Damping Wing of the Gunn-Peterson Trough}},
	volume = {501},
	year = 1998}

@article{calzetti_dust_2000,
	author = {Calzetti, Daniela and Armus, Lee and Bohlin, Ralph C. and Kinney, Anne L. and Koornneef, Jan and Storchi-Bergmann, Thaisa},
	journal = {\apj},
	month = {April},
	number = {2},
	pages = {682--695},
	title = {The {Dust} {Content} and {Opacity} of {Actively} {Star}-forming {Galaxies}},
	volume = {533},
	year = {2000}}

@article{charlot_simple_2000,
	author = {Charlot, St{\'e}phane and Fall, S. Michael},
	journal = {\apj},
	month = {August},
	pages = {718--731},
	title = {A {Simple} {Model} for the {Absorption} of {Starlight} by {Dust} in {Galaxies}},
	volume = {539},
	year = {2000}}

@article{Thomas:2003ab,
	author = {{Thomas}, D. and {Maraston}, C.},
	journal = {\aap},
	month = apr,
	pages = {429-432},
	title = {{The impact of alpha /Fe enhanced stellar evolutionary tracks on the ages of elliptical galaxies}},
	volume = {401},
	year = 2003}

@article{Gordon:2003aa,
	author = {{Gordon}, Karl D. and {Clayton}, Geoffrey C. and {Misselt}, K.~A. and {Landolt}, Arlo U. and {Wolff}, Michael J.},
	journal = {\apj},
	month = sep,
	number = {1},
	pages = {279-293},
	title = {{A Quantitative Comparison of the Small Magellanic Cloud, Large Magellanic Cloud, and Milky Way Ultraviolet to Near-Infrared Extinction Curves}},
	volume = {594},
	year = 2003}

@article{Bruzual:2003aa,
	author = {{Bruzual}, G. and {Charlot}, S.},
	journal = {\mnras},
	month = oct,
	number = {4},
	pages = {1000-1028},
	title = {{Stellar population synthesis at the resolution of 2003}},
	volume = {344},
	year = 2003}

@article{ocvirk_stecmap_2006,
	author = {Ocvirk, P. and Pichon, C. and Lan{\c c}on, A. and Thi{\'e}baut, E.},
	journal = {\mnras},
	month = {January},
	number = {1},
	pages = {46--73},
	title = {{STECMAP}: {STEllar} {Content} from high-resolution galactic spectra via {Maximum} {A} {Posteriori}},
	volume = {365},
	year = {2006}}

@article{Pietrinferni:2006aa,
	author = {{Pietrinferni}, Adriano and {Cassisi}, Santi and {Salaris}, Maurizio and {Castelli}, Fiorella},
	journal = {\apj},
	month = may,
	number = {2},
	pages = {797-812},
	title = {{A Large Stellar Evolution Database for Population Synthesis Studies. II. Stellar Models and Isochrones for an {\ensuremath{\alpha}}-enhanced Metal Distribution}},
	volume = {642},
	year = 2006}

@article{Totani:2006aa,
	author = {{Totani}, Tomonori and {Kawai}, Nobuyuki and {Kosugi}, George and {Aoki}, Kentaro and {Yamada}, Toru and {Iye}, Masanori and {Ohta}, Kouji and {Hattori}, Takashi},
	journal = {\pasj},
	month = jun,
	number = {3},
	pages = {485-498},
	title = {{Implications for Cosmic Reionization from the Optical Afterglow Spectrum of the Gamma-Ray Burst 050904 at z = 6.3$^{*}$}},
	volume = {58},
	year = 2006}

@article{Tepper-Garcia:2006aa,
	author = {{Tepper-Garc{\'\i}a}, Thorsten},
	journal = {\mnras},
	month = jul,
	number = {4},
	pages = {2025-2035},
	title = {{Voigt profile fitting to quasar absorption lines: an analytic approximation to the Voigt-Hjerting function}},
	volume = {369},
	year = 2006}

@article{Skilling2006,
	author = {Skilling, John},
	journal = {Bayesian Analysis},
	month = dec,
	number = {4},
	pages = {833--859},
	title = {Nested Sampling for General Bayesian Computation},
	volume = {1},
	year = {2006}}

@article{Eyles:2007aa,
	author = {{Eyles}, Laurence P. and {Bunker}, Andrew J. and {Ellis}, Richard S. and {Lacy}, Mark and {Stanway}, Elizabeth R. and {Stark}, Daniel P. and {Chiu}, Kuenley},
	journal = {\mnras},
	month = jan,
	number = {3},
	pages = {910-930},
	title = {{The stellar mass density at z \raisebox{-0.5ex}\textasciitilde 6 from Spitzer imaging of i'-drop galaxies}},
	volume = {374},
	year = 2007}

@article{Hunter:2007,
	author = {Hunter, J. D.},
	journal = {Computing in Science \& Engineering},
	number = {3},
	pages = {90--95},
	title = {Matplotlib: A 2D graphics environment},
	volume = {9},
	year = 2007}

@article{Draine:2007aa,
	author = {{Draine}, B.~T. and {Li}, Aigen},
	journal = {\apj},
	month = mar,
	number = {2},
	pages = {810-837},
	title = {{Infrared Emission from Interstellar Dust. IV. The Silicate-Graphite-PAH Model in the Post-Spitzer Era}},
	volume = {657},
	year = 2007}

@article{Marigo:2008aa,
	author = {{Marigo}, P. and {Girardi}, L. and {Bressan}, A. and {Groenewegen}, M.~A.~T. and {Silva}, L. and {Granato}, G.~L.},
	journal = {\aap},
	month = may,
	number = {3},
	pages = {883-905},
	title = {{Evolution of asymptotic giant branch stars. II. Optical to far-infrared isochrones with improved TP-AGB models}},
	volume = {482},
	year = 2008}

@article{da-Cunha:2008aa,
	author = {{\VAN{Da}{da}{da} Cunha}, Elisabete and {Charlot}, St{\'e}phane and {Elbaz}, David},
	journal = {\mnras},
	month = aug,
	number = {4},
	pages = {1595-1617},
	title = {{A simple model to interpret the ultraviolet, optical and infrared emission from galaxies}},
	volume = {388},
	year = 2008}

@article{Brammer:2008aa,
	author = {{Brammer}, Gabriel B. and {van Dokkum}, Pieter G. and {Coppi}, Paolo},
	journal = {\apj},
	month = oct,
	number = {2},
	pages = {1503-1513},
	title = {{EAZY: A Fast, Public Photometric Redshift Code}},
	volume = {686},
	year = 2008}

@article{conroy_propagation_2009,
	author = {Conroy, Charlie and Gunn, James E. and White, Martin},
	journal = {\apj},
	month = {June},
	number = {1},
	pages = {486},
	title = {{THE} {PROPAGATION} {OF} {UNCERTAINTIES} {IN} {S}℡{LAR} {POPULATION} {SYNTHESIS} {MODELING}. {I}. {THE} {RELEVANCE} {OF} {UNCERTAIN} {ASPECTS} {OF} {S}℡{LAR} {EVOLUTION} {AND} {THE} {INITIAL} {MASS} {FUNCTION} {TO} {THE} {DERIVED} {PHYSICAL} {PROPERTIES} {OF} {GALAXIES}},
	volume = {699},
	year = {2009}}

@article{Schaerer:2009aa,
	author = {{Schaerer}, D. and {de Barros}, S.},
	journal = {\aap},
	month = aug,
	number = {2},
	pages = {423-426},
	title = {{The impact of nebular emission on the ages of z{\ensuremath{\approx}} 6 galaxies}},
	volume = {502},
	year = 2009}

@article{Conroy:2010aa,
	author = {{Conroy}, Charlie and {Gunn}, James E.},
	journal = {\apj},
	month = apr,
	number = {2},
	pages = {833-857},
	title = {{The Propagation of Uncertainties in Stellar Population Synthesis Modeling. III. Model Calibration, Comparison, and Evaluation}},
	volume = {712},
	year = 2010}

@article{Hogg:2010aa,
	author = {{Hogg}, David W. and {Bovy}, Jo and {Lang}, Dustin},
	journal = {arXiv e-prints},
	month = aug,
	pages = {arXiv:1008.4686},
	title = {{Data analysis recipes: Fitting a model to data}},
	year = 2010}

@article{Cresci:2010aa,
	author = {{Cresci}, G. and {Mannucci}, F. and {Maiolino}, R. and {Marconi}, A. and {Gnerucci}, A. and {Magrini}, L.},
	journal = {\nat},
	month = oct,
	number = {7317},
	pages = {811-813},
	title = {{Gas accretion as the origin of chemical abundance gradients in distant galaxies}},
	volume = {467},
	year = 2010}

@article{walcher_fitting_2011,
	author = {Walcher, Jakob and Groves, Brent and Budav{\'a}ri, Tam{\'a}s and Dale, Daniel},
	journal = {Astrophysics and Space Science},
	month = {January},
	pages = {1--52},
	title = {Fitting the integrated spectral energy distributions of galaxies},
	volume = {331},
	year = {2011}}

@article{falcon-barroso_updated_2011,
	author = {Falc{\'o}n-Barroso, J. and S{\'a}nchez-Bl{\'a}zquez, P. and Vazdekis, A. and Ricciardelli, E. and Cardiel, N. and Cenarro, A. J. and Gorgas, J. and Peletier, R. F.},
	journal = {\aap},
	month = {August},
	pages = {A95},
	title = {An updated {MILES} stellar library and stellar population models},
	volume = {532},
	year = {2011}}

@article{Laureijs:2011aa,
	author = {{Laureijs}, R. and {Amiaux}, J. and {Arduini}, S. and {Augu{\`e}res}, J. -L. and {Brinchmann}, J. and {Cole}, R. and {Cropper}, M. and {Dabin}, C. and {Duvet}, L. and {Ealet}, A. and {Garilli}, B. and {Gondoin}, P. and {Guzzo}, L. and {Hoar}, J. and {Hoekstra}, H. and {Holmes}, R. and {Kitching}, T. and {Maciaszek}, T. and {Mellier}, Y. and {Pasian}, F. and {Percival}, W. and {Rhodes}, J. and {Saavedra Criado}, G. and {Sauvage}, M. and {Scaramella}, R. and {Valenziano}, L. and {Warren}, S. and {Bender}, R. and {Castander}, F. and {Cimatti}, A. and {Le F{\`e}vre}, O. and {Kurki-Suonio}, H. and {Levi}, M. and {Lilje}, P. and {Meylan}, G. and {Nichol}, R. and {Pedersen}, K. and {Popa}, V. and {Rebolo Lopez}, R. and {Rix}, H. -W. and {Rottgering}, H. and {Zeilinger}, W. and {Grupp}, F. and {Hudelot}, P. and {Massey}, R. and {Meneghetti}, M. and {Miller}, L. and {Paltani}, S. and {Paulin-Henriksson}, S. and {Pires}, S. and {Saxton}, C. and {Schrabback}, T. and {Seidel}, G. and {Walsh}, J. and {Aghanim}, N. and {Amendola}, L. and {Bartlett}, J. and {Baccigalupi}, C. and {Beaulieu}, J. -P. and {Benabed}, K. and {Cuby}, J. -G. and {Elbaz}, D. and {Fosalba}, P. and {Gavazzi}, G. and {Helmi}, A. and {Hook}, I. and {Irwin}, M. and {Kneib}, J. -P. and {Kunz}, M. and {Mannucci}, F. and {Moscardini}, L. and {Tao}, C. and {Teyssier}, R. and {Weller}, J. and {Zamorani}, G. and {Zapatero Osorio}, M.~R. and {Boulade}, O. and {Foumond}, J.~J. and {Di Giorgio}, A. and {Guttridge}, P. and {James}, A. and {Kemp}, M. and {Martignac}, J. and {Spencer}, A. and {Walton}, D. and {Bl{\"u}mchen}, T. and {Bonoli}, C. and {Bortoletto}, F. and {Cerna}, C. and {Corcione}, L. and {Fabron}, C. and {Jahnke}, K. and {Ligori}, S. and {Madrid}, F. and {Martin}, L. and {Morgante}, G. and {Pamplona}, T. and {Prieto}, E. and {Riva}, M. and {Toledo}, R. and {Trifoglio}, M. and {Zerbi}, F. and {Abdalla}, F. and {Douspis}, M. and {Grenet}, C. and {Borgani}, S. and {Bouwens}, R. and {Courbin}, F. and {Delouis}, J. -M. and {Dubath}, P. and {Fontana}, A. and {Frailis}, M. and {Grazian}, A. and {Koppenh{\"o}fer}, J. and {Mansutti}, O. and {Melchior}, M. and {Mignoli}, M. and {Mohr}, J. and {Neissner}, C. and {Noddle}, K. and {Poncet}, M. and {Scodeggio}, M. and {Serrano}, S. and {Shane}, N. and {Starck}, J. -L. and {Surace}, C. and {Taylor}, A. and {Verdoes-Kleijn}, G. and {Vuerli}, C. and {Williams}, O.~R. and {Zacchei}, A. and {Altieri}, B. and {Escudero Sanz}, I. and {Kohley}, R. and {Oosterbroek}, T. and {Astier}, P. and {Bacon}, D. and {Bardelli}, S. and {Baugh}, C. and {Bellagamba}, F. and {Benoist}, C. and {Bianchi}, D. and {Biviano}, A. and {Branchini}, E. and {Carbone}, C. and {Cardone}, V. and {Clements}, D. and {Colombi}, S. and {Conselice}, C. and {Cresci}, G. and {Deacon}, N. and {Dunlop}, J. and {Fedeli}, C. and {Fontanot}, F. and {Franzetti}, P. and {Giocoli}, C. and {Garcia-Bellido}, J. and {Gow}, J. and {Heavens}, A. and {Hewett}, P. and {Heymans}, C. and {Holland}, A. and {Huang}, Z. and {Ilbert}, O. and {Joachimi}, B. and {Jennins}, E. and {Kerins}, E. and {Kiessling}, A. and {Kirk}, D. and {Kotak}, R. and {Krause}, O. and {Lahav}, O. and {van Leeuwen}, F. and {Lesgourgues}, J. and {Lombardi}, M. and {Magliocchetti}, M. and {Maguire}, K. and {Majerotto}, E. and {Maoli}, R. and {Marulli}, F. and {Maurogordato}, S. and {McCracken}, H. and {McLure}, R. and {Melchiorri}, A. and {Merson}, A. and {Moresco}, M. and {Nonino}, M. and {Norberg}, P. and {Peacock}, J. and {Pello}, R. and {Penny}, M. and {Pettorino}, V. and {Di Porto}, C. and {Pozzetti}, L. and {Quercellini}, C. and {Radovich}, M. and {Rassat}, A. and {Roche}, N. and {Ronayette}, S. and {Rossetti}, E.},
	journal = {arXiv e-prints},
	month = oct,
	pages = {arXiv:1110.3193},
	title = {{Euclid Definition Study Report}},
	year = 2011}

@article{Hoffman:2011aa,
	author = {{Hoffman}, Matthew D. and {Gelman}, Andrew},
	journal = {arXiv e-prints},
	month = nov,
	pages = {arXiv:1111.4246},
	title = {{The No-U-Turn Sampler: Adaptively Setting Path Lengths in Hamiltonian Monte Carlo}},
	year = 2011}

@article{Bressan:2012aa,
	author = {{Bressan}, Alessandro and {Marigo}, Paola and {Girardi}, L{\'e}o. and {Salasnich}, Bernardo and {Dal Cero}, Claudia and {Rubele}, Stefano and {Nanni}, Ambra},
	journal = {\mnras},
	month = nov,
	number = {1},
	pages = {127-145},
	title = {{PARSEC: stellar tracks and isochrones with the PAdova and TRieste Stellar Evolution Code}},
	volume = {427},
	year = 2012}

@article{Sawicki:2012aa,
	author = {{Sawicki}, Marcin},
	journal = {\pasp},
	month = nov,
	number = {921},
	pages = {1208},
	title = {{SEDfit: Software for Spectral Energy Distribution Fitting of Photometric Data}},
	volume = {124},
	year = 2012}

@misc{ferland_2013_2013,
	author = {Ferland, G. J. and Porter, R. L. and van Hoof, P. A. M. and Williams, R. J. R. and Abel, N. P. and Lykins, M. L. and Shaw, Gargi and Henney, W. J. and Stancil, P. C.},
	month = {February},
	title = {The 2013 {Release} of {Cloudy}},
	year = {2013}}

@article{Chevallard:2013aa,
	author = {{Chevallard}, J. and {Charlot}, S. and {Wandelt}, B. and {Wild}, V.},
	journal = {\mnras},
	month = jul,
	number = {3},
	pages = {2061-2091},
	title = {{Insights into the content and spatial distribution of dust from the integrated spectral properties of galaxies}},
	volume = {432},
	year = 2013}

@article{conroy_modeling_2013,
	author = {Conroy, Charlie},
	journal = {\araa},
	month = {August},
	pages = {393--455},
	title = {Modeling the {Panchromatic} {Spectral} {Energy} {Distributions} of {Galaxies}},
	volume = {51},
	year = {2013}}

@article{Lilly:2013aa,
	author = {{Lilly}, Simon J. and {Carollo}, C. Marcella and {Pipino}, Antonio and {Renzini}, Alvio and {Peng}, Yingjie},
	journal = {\apj},
	month = aug,
	number = {2},
	pages = {119},
	title = {{Gas Regulation of Galaxies: The Evolution of the Cosmic Specific Star Formation Rate, the Metallicity-Mass-Star-formation Rate Relation, and the Stellar Content of Halos}},
	volume = {772},
	year = 2013}

@article{kriek_dust_2013,
	author = {Kriek, Mariska and Conroy, Charlie},
	journal = {\apj},
	month = {September},
	number = {1},
	pages = {L16},
	title = {The {Dust} {Attenuation} {Law} in {Distant} {Galaxies}: {Evidence} for {Variation} with {Spectral} {Type}},
	volume = {775},
	year = {2013}}

@article{price_direct_2014,
	author = {Price, Sedona H. and Kriek, Mariska and Brammer, Gabriel B. and Conroy, Charlie and F{\"o}rster Schreiber, Natascha M. and Franx, Marijn and Fumagalli, Mattia and Lundgren, Britt and Momcheva, Ivelina and Nelson, Erica J. and Skelton, Rosalind E. and van Dokkum, Pieter G. and Whitaker, Katherine E. and Wuyts, Stijn},
	journal = {\apj},
	month = {June},
	number = {1},
	pages = {86},
	title = {Direct {Measurements} of {Dust} {Attenuation} in z {\textasciitilde} 1.5 {Star}-forming {Galaxies} from {3D}-{HST}: {Implications} for {Dust} {Geometry} and {Star} {Formation} {Rates}},
	volume = {788},
	year = {2014}}

@article{Spergel:2015aa,
	author = {{Spergel}, D. and {Gehrels}, N. and {Baltay}, C. and {Bennett}, D. and {Breckinridge}, J. and {Donahue}, M. and {Dressler}, A. and {Gaudi}, B.~S. and {Greene}, T. and {Guyon}, O. and {Hirata}, C. and {Kalirai}, J. and {Kasdin}, N.~J. and {Macintosh}, B. and {Moos}, W. and {Perlmutter}, S. and {Postman}, M. and {Rauscher}, B. and {Rhodes}, J. and {Wang}, Y. and {Weinberg}, D. and {Benford}, D. and {Hudson}, M. and {Jeong}, W.-S. and {Mellier}, Y. and {Traub}, W. and {Yamada}, T. and {Capak}, P. and {Colbert}, J. and {Masters}, D. and {Penny}, M. and {Savransky}, D. and {Stern}, D. and {Zimmerman}, N. and {Barry}, R. and {Bartusek}, L. and {Carpenter}, K. and {Cheng}, E. and {Content}, D. and {Dekens}, F. and {Demers}, R. and {Grady}, K. and {Jackson}, C. and {Kuan}, G. and {Kruk}, J. and {Melton}, M. and {Nemati}, B. and {Parvin}, B. and {Poberezhskiy}, I. and {Peddie}, C. and {Ruffa}, J. and {Wallace}, J.~K. and {Whipple}, A. and {Wollack}, E. and {Zhao}, F.},
	journal = {arXiv e-prints},
	month = mar,
	pages = {arXiv:1503.03757},
	title = {{Wide-Field InfrarRed Survey Telescope-Astrophysics Focused Telescope Assets WFIRST-AFTA 2015 Report}},
	year = 2015}

@article{Peng:2015aa,
	author = {{Peng}, Y. and {Maiolino}, R. and {Cochrane}, R.},
	journal = {\nat},
	month = may,
	number = {7551},
	pages = {192-195},
	title = {{Strangulation as the primary mechanism for shutting down star formation in galaxies}},
	volume = {521},
	year = 2015}

@article{Reddy:2015aa,
	author = {{Reddy}, Naveen A. and {Kriek}, Mariska and {Shapley}, Alice E. and {Freeman}, William R. and {Siana}, Brian and {Coil}, Alison L. and {Mobasher}, Bahram and {Price}, Sedona H. and {Sanders}, Ryan L. and {Shivaei}, Irene},
	journal = {\apj},
	month = jun,
	number = {2},
	pages = {259},
	title = {{The MOSDEF Survey: Measurements of Balmer Decrements and the Dust Attenuation Curve at Redshifts z \raisebox{-0.5ex}\textasciitilde 1.4-2.6}},
	volume = {806},
	year = 2015}

@article{Dotter:2016aa,
	author = {{Dotter}, Aaron},
	journal = {\apjs},
	month = jan,
	number = {1},
	pages = {8},
	title = {{MESA Isochrones and Stellar Tracks (MIST) 0: Methods for the Construction of Stellar Isochrones}},
	volume = {222},
	year = 2016}

@article{reddy_connection_2016,
	author = {Reddy, Naveen A. and Steidel, Charles C. and Pettini, Max and Bogosavljevi{\'c}, Milan and Shapley, Alice E.},
	journal = {\apj},
	number = {2},
	pages = {108},
	title = {{THE} {CONNECTION} {BETWEEN} {REDDENING}, {GAS} {COVERING} {FRACTION}, {AND} {THE} {ESCAPE} {OF} {IONIZING} {RADIATION} {AT} {HIGH} {REDSHIFT}∗},
	volume = {828},
	year = {2016}}

@article{choi_mesa_2016,
	author = {Choi, Jieun and Dotter, Aaron and Conroy, Charlie and Cantiello, Matteo and Paxton, Bill and Johnson, Benjamin D.},
	journal = {\apj},
	month = {May},
	number = {2},
	pages = {102},
	title = {{MESA} {ISOCHRONES} {AND} {S}℡{LAR} {TRACKS} ({MIST}). {I}. {SOLAR}-{SCALED} {MODELS}},
	volume = {823},
	year = {2016}}

@article{chevallard_modelling_2016,
	author = {Chevallard, Jacopo and Charlot, St{\'e}phane},
	journal = {\mnras},
	month = {October},
	number = {2},
	pages = {1415--1443},
	title = {Modelling and interpreting spectral energy distributions of galaxies with beagle},
	volume = {462},
	year = {2016}}

@article{leja_deriving_2017,
	author = {Leja, Joel and Johnson, Benjamin D. and Conroy, Charlie and van Dokkum, Pieter G. and Byler, Nell},
	journal = {\apj},
	month = {March},
	pages = {170},
	title = {Deriving {Physical} {Properties} from {Broadband} {Photometry} with {Prospector}: {Description} of the {Model} and a {Demonstration} of its {Accuracy} {Using} 129 {Galaxies} in the {Local} {Universe}},
	volume = {837},
	year = {2017}}

@article{Byler:2017aa,
	author = {{Byler}, Nell and {Dalcanton}, Julianne J. and {Conroy}, Charlie and {Johnson}, Benjamin D.},
	journal = {\apj},
	month = may,
	number = {1},
	pages = {44},
	title = {{Nebular Continuum and Line Emission in Stellar Population Synthesis Models}},
	volume = {840},
	year = 2017}

@article{Jones:2017aa,
	author = {{Jones}, A.~P. and {K{\"o}hler}, M. and {Ysard}, N. and {Bocchio}, M. and {Verstraete}, L.},
	journal = {\aap},
	month = jun,
	pages = {A46},
	title = {{The global dust modelling framework THEMIS}},
	volume = {602},
	year = 2017}

@article{Dillon:2017aa,
	author = {{Dillon}, Joshua V. and {Langmore}, Ian and {Tran}, Dustin and {Brevdo}, Eugene and {Vasudevan}, Srinivas and {Moore}, Dave and {Patton}, Brian and {Alemi}, Alex and {Hoffman}, Matt and {Saurous}, Rif A.},
	journal = {arXiv e-prints},
	month = nov,
	pages = {arXiv:1711.10604},
	title = {{TensorFlow Distributions}},
	year = 2017}

@article{Conroy:2018aa,
	author = {{Conroy}, Charlie and {Villaume}, Alexa and {van Dokkum}, Pieter G. and {Lind}, Karin},
	journal = {\apj},
	month = feb,
	number = {2},
	pages = {139},
	title = {{Metal-rich, Metal-poor: Updated Stellar Population Models for Old Stellar Systems}},
	volume = {854},
	year = 2018}

@article{stanway_re-evaluating_2018,
	author = {Stanway, E. R. and Eldridge, J. J.},
	journal = {\mnras},
	month = {September},
	pages = {75--93},
	title = {Re-evaluating old stellar populations},
	volume = {479},
	year = {2018}}

@article{Carnall:2018aa,
	author = {{Carnall}, A.~C. and {McLure}, R.~J. and {Dunlop}, J.~S. and {Dav{\'e}}, R.},
	journal = {\mnras},
	month = nov,
	number = {4},
	pages = {4379-4401},
	title = {{Inferring the star formation histories of massive quiescent galaxies with BAGPIPES: evidence for multiple quenching mechanisms}},
	volume = {480},
	year = 2018}

@article{Boquien:2019aa,
	author = {{Boquien}, M. and {Burgarella}, D. and {Roehlly}, Y. and {Buat}, V. and {Ciesla}, L. and {Corre}, D. and {Inoue}, A.~K. and {Salas}, H.},
	journal = {\aap},
	month = feb,
	pages = {A103},
	title = {{CIGALE: a python Code Investigating GALaxy Emission}},
	volume = {622},
	year = 2019}

@article{carnall_how_2019,
	author = {Carnall, Adam C. and Leja, Joel and Johnson, Benjamin D. and McLure, Ross J. and Dunlop, James S. and Conroy, Charlie},
	journal = {\apj},
	month = {March},
	pages = {44},
	title = {How to {Measure} {Galaxy} {Star} {Formation} {Histories}. {I}. {Parametric} {Models}},
	volume = {873},
	year = {2019}}

@article{Hoffman:2019aa,
	author = {{Hoffman}, Matthew and {Sountsov}, Pavel and {Dillon}, Joshua V. and {Langmore}, Ian and {Tran}, Dustin and {Vasudevan}, Srinivas},
	journal = {arXiv e-prints},
	month = mar,
	pages = {arXiv:1903.03704},
	title = {{NeuTra-lizing Bad Geometry in Hamiltonian Monte Carlo Using Neural Transport}},
	year = 2019}

@article{Ivezic:2019aa,
	author = {{Ivezi{\'c}}, {\v{Z}}eljko and {Kahn}, Steven M. and {Tyson}, J. Anthony and {Abel}, Bob and {Acosta}, Emily and {Allsman}, Robyn and {Alonso}, David and {AlSayyad}, Yusra and {Anderson}, Scott F. and {Andrew}, John and {Angel}, James Roger P. and {Angeli}, George Z. and {Ansari}, Reza and {Antilogus}, Pierre and {Araujo}, Constanza and {Armstrong}, Robert and {Arndt}, Kirk T. and {Astier}, Pierre and {Aubourg}, {\'E}ric and {Auza}, Nicole and {Axelrod}, Tim S. and {Bard}, Deborah J. and {Barr}, Jeff D. and {Barrau}, Aurelian and {Bartlett}, James G. and {Bauer}, Amanda E. and {Bauman}, Brian J. and {Baumont}, Sylvain and {Bechtol}, Ellen and {Bechtol}, Keith and {Becker}, Andrew C. and {Becla}, Jacek and {Beldica}, Cristina and {Bellavia}, Steve and {Bianco}, Federica B. and {Biswas}, Rahul and {Blanc}, Guillaume and {Blazek}, Jonathan and {Blandford}, Roger D. and {Bloom}, Josh S. and {Bogart}, Joanne and {Bond}, Tim W. and {Booth}, Michael T. and {Borgland}, Anders W. and {Borne}, Kirk and {Bosch}, James F. and {Boutigny}, Dominique and {Brackett}, Craig A. and {Bradshaw}, Andrew and {Brandt}, William Nielsen and {Brown}, Michael E. and {Bullock}, James S. and {Burchat}, Patricia and {Burke}, David L. and {Cagnoli}, Gianpietro and {Calabrese}, Daniel and {Callahan}, Shawn and {Callen}, Alice L. and {Carlin}, Jeffrey L. and {Carlson}, Erin L. and {Chandrasekharan}, Srinivasan and {Charles-Emerson}, Glenaver and {Chesley}, Steve and {Cheu}, Elliott C. and {Chiang}, Hsin-Fang and {Chiang}, James and {Chirino}, Carol and {Chow}, Derek and {Ciardi}, David R. and {Claver}, Charles F. and {Cohen-Tanugi}, Johann and {Cockrum}, Joseph J. and {Coles}, Rebecca and {Connolly}, Andrew J. and {Cook}, Kem H. and {Cooray}, Asantha and {Covey}, Kevin R. and {Cribbs}, Chris and {Cui}, Wei and {Cutri}, Roc and {Daly}, Philip N. and {Daniel}, Scott F. and {Daruich}, Felipe and {Daubard}, Guillaume and {Daues}, Greg and {Dawson}, William and {Delgado}, Francisco and {Dellapenna}, Alfred and {de Peyster}, Robert and {de Val-Borro}, Miguel and {Digel}, Seth W. and {Doherty}, Peter and {Dubois}, Richard and {Dubois-Felsmann}, Gregory P. and {Durech}, Josef and {Economou}, Frossie and {Eifler}, Tim and {Eracleous}, Michael and {Emmons}, Benjamin L. and {Fausti Neto}, Angelo and {Ferguson}, Henry and {Figueroa}, Enrique and {Fisher-Levine}, Merlin and {Focke}, Warren and {Foss}, Michael D. and {Frank}, James and {Freemon}, Michael D. and {Gangler}, Emmanuel and {Gawiser}, Eric and {Geary}, John C. and {Gee}, Perry and {Geha}, Marla and {Gessner}, Charles J.~B. and {Gibson}, Robert R. and {Gilmore}, D. Kirk and {Glanzman}, Thomas and {Glick}, William and {Goldina}, Tatiana and {Goldstein}, Daniel A. and {Goodenow}, Iain and {Graham}, Melissa L. and {Gressler}, William J. and {Gris}, Philippe and {Guy}, Leanne P. and {Guyonnet}, Augustin and {Haller}, Gunther and {Harris}, Ron and {Hascall}, Patrick A. and {Haupt}, Justine and {Hernandez}, Fabio and {Herrmann}, Sven and {Hileman}, Edward and {Hoblitt}, Joshua and {Hodgson}, John A. and {Hogan}, Craig and {Howard}, James D. and {Huang}, Dajun and {Huffer}, Michael E. and {Ingraham}, Patrick and {Innes}, Walter R. and {Jacoby}, Suzanne H. and {Jain}, Bhuvnesh and {Jammes}, Fabrice and {Jee}, M. James and {Jenness}, Tim and {Jernigan}, Garrett and {Jevremovi{\'c}}, Darko and {Johns}, Kenneth and {Johnson}, Anthony S. and {Johnson}, Margaret W.~G. and {Jones}, R. Lynne and {Juramy-Gilles}, Claire and {Juri{\'c}}, Mario and {Kalirai}, Jason S. and {Kallivayalil}, Nitya J. and {Kalmbach}, Bryce and {Kantor}, Jeffrey P. and {Karst}, Pierre and {Kasliwal}, Mansi M. and {Kelly}, Heather and {Kessler}, Richard and {Kinnison}, Veronica and {Kirkby}, David and {Knox}, Lloyd and {Kotov}, Ivan V. and {Krabbendam}, Victor L. and {Krughoff}, K. Simon and {Kub{\'a}nek}, Petr and {Kuczewski}, John and {Kulkarni}, Shri and {Ku}, John and {Kurita}, Nadine R. and {Lage}, Craig S. and {Lambert}, Ron and {Lange}, Travis and {Langton}, J. Brian and {Le Guillou}, Laurent and {Levine}, Deborah and {Liang}, Ming and {Lim}, Kian-Tat and {Lintott}, Chris J. and {Long}, Kevin E. and {Lopez}, Margaux and {Lotz}, Paul J. and {Lupton}, Robert H. and {Lust}, Nate B. and {MacArthur}, Lauren A. and {Mahabal}, Ashish and {Mandelbaum}, Rachel and {Markiewicz}, Thomas W. and {Marsh}, Darren S. and {Marshall}, Philip J. and {Marshall}, Stuart and {May}, Morgan and {McKercher}, Robert and {McQueen}, Michelle and {Meyers}, Joshua and {Migliore}, Myriam and {Miller}, Michelle and {Mills}, David J.},
	journal = {\apj},
	month = mar,
	number = {2},
	pages = {111},
	title = {{LSST: From Science Drivers to Reference Design and Anticipated Data Products}},
	volume = {873},
	year = 2019}

@article{leja_how_2019,
	author = {Leja, Joel and Carnall, Adam C. and Johnson, Benjamin D. and Conroy, Charlie and Speagle, Joshua S.},
	journal = {\apj},
	month = {May},
	number = {1},
	pages = {3},
	title = {How to {Measure} {Galaxy} {Star} {Formation} {Histories}. {II}. {Nonparametric} {Models}},
	volume = {876},
	year = {2019}}

@article{anesthetic,
	author = {Will Handley},
	journal = {The Journal of Open Source Software},
	month = {Jun},
	number = {37},
	pages = {1414},
	title = {anesthetic: nested sampling visualisation},
	volume = {4},
	year = {2019}}

@article{Trussler:2020aa,
	author = {{Trussler}, James and {Maiolino}, Roberto and {Maraston}, Claudia and {Peng}, Yingjie and {Thomas}, Daniel and {Goddard}, Daniel and {Lian}, Jianhui},
	journal = {\mnras},
	month = feb,
	number = {4},
	pages = {5406-5434},
	title = {{Both starvation and outflows drive galaxy quenching}},
	volume = {491},
	year = 2020}

@article{Virtanen:2020aa,
	author = {{Virtanen}, Pauli and {Gommers}, Ralf and {Oliphant}, Travis E. and {Haberland}, Matt and {Reddy}, Tyler and {Cournapeau}, David and {Burovski}, Evgeni and {Peterson}, Pearu and {Weckesser}, Warren and {Bright}, Jonathan and {van der Walt}, St{\'e}fan J. and {Brett}, Matthew and {Wilson}, Joshua and {Millman}, K. Jarrod and {Mayorov}, Nikolay and {Nelson}, Andrew R.~J. and {Jones}, Eric and {Kern}, Robert and {Larson}, Eric and {Carey}, C.~J. and {Polat}, {\.I}lhan and {Feng}, Yu and {Moore}, Eric W. and {VanderPlas}, Jake and {Laxalde}, Denis and {Perktold}, Josef and {Cimrman}, Robert and {Henriksen}, Ian and {Quintero}, E.~A. and {Harris}, Charles R. and {Archibald}, Anne M. and {Ribeiro}, Ant{\^o}nio H. and {Pedregosa}, Fabian and {van Mulbregt}, Paul and {SciPy 1. 0 Contributors}},
	journal = {Nature Methods},
	month = feb,
	pages = {261-272},
	title = {{SciPy 1.0: fundamental algorithms for scientific computing in Python}},
	volume = {17},
	year = 2020}

@article{Alsing:2020aa,
	author = {{Alsing}, Justin and {Peiris}, Hiranya and {Leja}, Joel and {Hahn}, ChangHoon and {Tojeiro}, Rita and {Mortlock}, Daniel and {Leistedt}, Boris and {Johnson}, Benjamin D. and {Conroy}, Charlie},
	journal = {\apjs},
	month = jul,
	number = {1},
	pages = {5},
	title = {{SPECULATOR: Emulating Stellar Population Synthesis for Fast and Accurate Galaxy Spectra and Photometry}},
	volume = {249},
	year = 2020}

@article{salim_dust_2020,
	author = {Salim, Samir and Narayanan, Desika},
	journal = {\araa},
	month = {August},
	number = {1},
	pages = {529--575},
	title = {The {Dust} {Attenuation} {Law} in {Galaxies}},
	volume = {58},
	year = {2020}}

@article{harris2020array,
	author = {Charles R. Harris and K. Jarrod Millman and St{\'{e}}fan J. van der Walt and Ralf Gommers and Pauli Virtanen and David Cournapeau and Eric Wieser and Julian Taylor and Sebastian Berg and Nathaniel J. Smith and Robert Kern and Matti Picus and Stephan Hoyer and Marten H. van Kerkwijk and Matthew Brett and Allan Haldane and Jaime Fern{\'{a}}ndez del R{\'{i}}o and Mark Wiebe and Pearu Peterson and Pierre G{\'{e}}rard-Marchant and Kevin Sheppard and Tyler Reddy and Warren Weckesser and Hameer Abbasi and Christoph Gohlke and Travis E. Oliphant},
	journal = {\nat},
	month = sep,
	number = {7825},
	pages = {357--362},
	title = {Array programming with {NumPy}},
	volume = {585},
	year = {2020}}

@article{Fowlie:2020aa,
	author = {{Fowlie}, Andrew and {Handley}, Will and {Su}, Liangliang},
	journal = {\mnras},
	month = oct,
	number = {4},
	pages = {5256-5263},
	title = {{Nested sampling cross-checks using order statistics}},
	volume = {497},
	year = 2020}

@article{Iyer:2020aa,
	author = {{Iyer}, Kartheik G. and {Tacchella}, Sandro and {Genel}, Shy and {Hayward}, Christopher C. and {Hernquist}, Lars and {Brooks}, Alyson M. and {Caplar}, Neven and {Dav{\'e}}, Romeel and {Diemer}, Benedikt and {Forbes}, John C. and {Gawiser}, Eric and {Somerville}, Rachel S. and {Starkenburg}, Tjitske K.},
	journal = {\mnras},
	month = oct,
	number = {1},
	pages = {430-463},
	title = {{The diversity and variability of star formation histories in models of galaxy evolution}},
	volume = {498},
	year = 2020}

@article{Lower:2020aa,
	author = {{Lower}, Sidney and {Narayanan}, Desika and {Leja}, Joel and {Johnson}, Benjamin D. and {Conroy}, Charlie and {Dav{\'e}}, Romeel},
	journal = {\apj},
	month = nov,
	number = {1},
	pages = {33},
	title = {{How Well Can We Measure the Stellar Mass of a Galaxy: The Impact of the Assumed Star Formation History Model in SED Fitting}},
	volume = {904},
	year = 2020}

@software{johnson_2021_4582723,
	author = {Johnson, Benjamin D.},
	doi = {10.5281/zenodo.4582723},
	month = mar,
	publisher = {Zenodo},
	title = {bd-j/sedpy: sedpy v0.2.0},
	url = {https://doi.org/10.5281/zenodo.4582723},
	version = {v0.2.0},
	year = 2021}

@article{Johnson:2021aa,
	author = {{Johnson}, Benjamin D. and {Leja}, Joel and {Conroy}, Charlie and {Speagle}, Joshua S.},
	journal = {\apjs},
	month = jun,
	number = {2},
	pages = {22},
	title = {{Stellar Population Inference with Prospector}},
	volume = {254},
	year = 2021}

@article{Jakobsen:2022aa,
	author = {{Jakobsen}, P. and {Ferruit}, P. and {Alves de Oliveira}, C. and {Arribas}, S. and {Bagnasco}, G. and {Barho}, R. and {Beck}, T.~L. and {Birkmann}, S. and {B{\"o}ker}, T. and {Bunker}, A.~J. and {Charlot}, S. and {de Jong}, P. and {de Marchi}, G. and {Ehrenwinkler}, R. and {Falcolini}, M. and {Fels}, R. and {Franx}, M. and {Franz}, D. and {Funke}, M. and {Giardino}, G. and {Gnata}, X. and {Holota}, W. and {Honnen}, K. and {Jensen}, P.~L. and {Jentsch}, M. and {Johnson}, T. and {Jollet}, D. and {Karl}, H. and {Kling}, G. and {K{\"o}hler}, J. and {Kolm}, M.-G. and {Kumari}, N. and {Lander}, M.~E. and {Lemke}, R. and {L{\'o}pez-Caniego}, M. and {L{\"u}tzgendorf}, N. and {Maiolino}, R. and {Manjavacas}, E. and {Marston}, A. and {Maschmann}, M. and {Maurer}, R. and {Messerschmidt}, B. and {Moseley}, S.~H. and {Mosner}, P. and {Mott}, D.~B. and {Muzerolle}, J. and {Pirzkal}, N. and {Pittet}, J.-F. and {Plitzke}, A. and {Posselt}, W. and {Rapp}, B. and {Rauscher}, B.~J. and {Rawle}, T. and {Rix}, H.-W. and {R{\"o}del}, A. and {Rumler}, P. and {Sabbi}, E. and {Salvignol}, J.-C. and {Schmid}, T. and {Sirianni}, M. and {Smith}, C. and {Strada}, P. and {te Plate}, M. and {Valenti}, J. and {Wettemann}, T. and {Wiehe}, T. and {Wiesmayer}, M. and {Willott}, C.~J. and {Wright}, R. and {Zeidler}, P. and {Zincke}, C.},
	journal = {\aap},
	month = may,
	pages = {A80},
	title = {{The Near-Infrared Spectrograph (NIRSpec) on the James Webb Space Telescope. I. Overview of the instrument and its capabilities}},
	volume = {661},
	year = 2022}

@article{Astropy-Collaboration:2022aa,
	author = {{Astropy Collaboration} and {Price-Whelan}, Adrian M. and {Lim}, Pey Lian and {Earl}, Nicholas and {Starkman}, Nathaniel and {Bradley}, Larry and {Shupe}, David L. and {Patil}, Aarya A. and {Corrales}, Lia and {Brasseur}, C.~E. and {N{\"o}the}, Maximilian and {Donath}, Axel and {Tollerud}, Erik and {Morris}, Brett M. and {Ginsburg}, Adam and {Vaher}, Eero and {Weaver}, Benjamin A. and {Tocknell}, James and {Jamieson}, William and {van Kerkwijk}, Marten H. and {Robitaille}, Thomas P. and {Merry}, Bruce and {Bachetti}, Matteo and {G{\"u}nther}, H. Moritz and {Aldcroft}, Thomas L. and {Alvarado-Montes}, Jaime A. and {Archibald}, Anne M. and {B{\'o}di}, Attila and {Bapat}, Shreyas and {Barentsen}, Geert and {Baz{\'a}n}, Juanjo and {Biswas}, Manish and {Boquien}, M{\'e}d{\'e}ric and {Burke}, D.~J. and {Cara}, Daria and {Cara}, Mihai and {Conroy}, Kyle E. and {Conseil}, Simon and {Craig}, Matthew W. and {Cross}, Robert M. and {Cruz}, Kelle L. and {D'Eugenio}, Francesco and {Dencheva}, Nadia and {Devillepoix}, Hadrien A.~R. and {Dietrich}, J{\"o}rg P. and {Eigenbrot}, Arthur Davis and {Erben}, Thomas and {Ferreira}, Leonardo and {Foreman-Mackey}, Daniel and {Fox}, Ryan and {Freij}, Nabil and {Garg}, Suyog and {Geda}, Robel and {Glattly}, Lauren and {Gondhalekar}, Yash and {Gordon}, Karl D. and {Grant}, David and {Greenfield}, Perry and {Groener}, Austen M. and {Guest}, Steve and {Gurovich}, Sebastian and {Handberg}, Rasmus and {Hart}, Akeem and {Hatfield-Dodds}, Zac and {Homeier}, Derek and {Hosseinzadeh}, Griffin and {Jenness}, Tim and {Jones}, Craig K. and {Joseph}, Prajwel and {Kalmbach}, J. Bryce and {Karamehmetoglu}, Emir and {Ka{\l}uszy{\'n}ski}, Miko{\l}aj and {Kelley}, Michael S.~P. and {Kern}, Nicholas and {Kerzendorf}, Wolfgang E. and {Koch}, Eric W. and {Kulumani}, Shankar and {Lee}, Antony and {Ly}, Chun and {Ma}, Zhiyuan and {MacBride}, Conor and {Maljaars}, Jakob M. and {Muna}, Demitri and {Murphy}, N.~A. and {Norman}, Henrik and {O'Steen}, Richard and {Oman}, Kyle A. and {Pacifici}, Camilla and {Pascual}, Sergio and {Pascual-Granado}, J. and {Patil}, Rohit R. and {Perren}, Gabriel I. and {Pickering}, Timothy E. and {Rastogi}, Tanuj and {Roulston}, Benjamin R. and {Ryan}, Daniel F. and {Rykoff}, Eli S. and {Sabater}, Jose and {Sakurikar}, Parikshit and {Salgado}, Jes{\'u}s and {Sanghi}, Aniket and {Saunders}, Nicholas and {Savchenko}, Volodymyr and {Schwardt}, Ludwig and {Seifert-Eckert}, Michael and {Shih}, Albert Y. and {Jain}, Anany Shrey and {Shukla}, Gyanendra and {Sick}, Jonathan and {Simpson}, Chris and {Singanamalla}, Sudheesh and {Singer}, Leo P. and {Singhal}, Jaladh and {Sinha}, Manodeep and {Sip{\H{o}}cz}, Brigitta M. and {Spitler}, Lee R. and {Stansby}, David and {Streicher}, Ole and {{\v{S}}umak}, Jani and {Swinbank}, John D. and {Taranu}, Dan S. and {Tewary}, Nikita and {Tremblay}, Grant R. and {de Val-Borro}, Miguel and {Van Kooten}, Samuel J. and {Vasovi{\'c}}, Zlatan and {Verma}, Shresth and {de Miranda Cardoso}, Jos{\'e} Vin{\'\i}cius and {Williams}, Peter K.~G. and {Wilson}, Tom J. and {Winkel}, Benjamin and {Wood-Vasey}, W.~M. and {Xue}, Rui and {Yoachim}, Peter and {Zhang}, Chen and {Zonca}, Andrea and {Astropy Project Contributors}},
	journal = {\apj},
	month = aug,
	number = {2},
	pages = {167},
	title = {{The Astropy Project: Sustaining and Growing a Community-oriented Open-source Project and the Latest Major Release (v5.0) of the Core Package}},
	volume = {935},
	year = 2022}

@article{Hahn:2022aa,
	author = {{Hahn}, ChangHoon and {Melchior}, Peter},
	journal = {\apj},
	month = oct,
	number = {1},
	pages = {11},
	title = {{Accelerated Bayesian SED Modeling Using Amortized Neural Posterior Estimation}},
	volume = {938},
	year = 2022}

@article{Khullar:2022aa,
	author = {{Khullar}, Gourav and {Nord}, Brian and {{\'C}iprijanovi{\'c}}, Aleksandra and {Poh}, Jason and {Xu}, Fei},
	journal = {Machine Learning: Science and Technology},
	month = dec,
	number = {4},
	pages = {04LT04},
	title = {{DIGS: deep inference of galaxy spectra with neural posterior estimation}},
	volume = {3},
	year = 2022}

@article{Thorne:2022aa,
	author = {{Thorne}, Jessica E. and {Robotham}, Aaron S.~G. and {Bellstedt}, Sabine and {Davies}, Luke J.~M. and {Cook}, Robin H.~W. and {Cortese}, Luca and {Holwerda}, Benne and {Phillipps}, Steven and {Siudek}, Malgorzata},
	journal = {\mnras},
	month = dec,
	number = {4},
	pages = {6035-6059},
	title = {{DEVILS: cosmic evolution of SED-derived metallicities and their connection to star formation histories}},
	volume = {517},
	year = 2022}

@article{Labbe:2023aa,
	author = {{Labb{\'e}}, Ivo and {van Dokkum}, Pieter and {Nelson}, Erica and {Bezanson}, Rachel and {Suess}, Katherine A. and {Leja}, Joel and {Brammer}, Gabriel and {Whitaker}, Katherine and {Mathews}, Elijah and {Stefanon}, Mauro and {Wang}, Bingjie},
	journal = {\nat},
	month = apr,
	number = {7956},
	pages = {266-269},
	title = {{A population of red candidate massive galaxies 600 Myr after the Big Bang}},
	volume = {616},
	year = 2023}

@article{Hearin:2023aa,
	author = {{Hearin}, Andrew P. and {Chaves-Montero}, Jon{\'a}s and {Alarcon}, Alex and {Becker}, Matthew R. and {Benson}, Andrew},
	journal = {\mnras},
	month = may,
	number = {2},
	pages = {1741-1756},
	title = {{DSPS: Differentiable stellar population synthesis}},
	volume = {521},
	year = 2023}

@article{Boylan-Kolchin:2023aa,
	author = {{Boylan-Kolchin}, Michael},
	journal = {Nature Astronomy},
	month = jun,
	pages = {731-735},
	title = {{Stress testing {\ensuremath{\Lambda}}CDM with high-redshift galaxy candidates}},
	volume = {7},
	year = 2023}

@book{eisenstein_overview_2023,
	author = {Eisenstein, Daniel and Willott, Chris and Alberts, Stacey and Arribas, Santiago and Bonaventura, Nina and Bunker, Andrew and Cameron, Alex and Carniani, Stefano and Charlot, Stephane and Curtis-Lake, Emma and D'Eugenio, Francesco and Endsley, Ryan and Ferruit, Pierre and Giardino, Giovanna and Hainline, Kevin and Hausen, Ryan and Jakobsen, Peter and Johnson, Benjamin and Maiolino, Roberto and Woodrum, Charity},
	month = {June},
	title = {Overview of the {JWST} {Advanced} {Deep} {Extragalactic} {Survey} ({JADES})},
	year = {2023}}

@article{Kocevski:2023aa,
	author = {{Kocevski}, Dale D. and {Onoue}, Masafusa and {Inayoshi}, Kohei and {Trump}, Jonathan R. and {Arrabal Haro}, Pablo and {Grazian}, Andrea and {Dickinson}, Mark and {Finkelstein}, Steven L. and {Kartaltepe}, Jeyhan S. and {Hirschmann}, Michaela and {Aird}, James and {Holwerda}, Benne W. and {Fujimoto}, Seiji and {Juneau}, St{\'e}phanie and {Amor{\'\i}n}, Ricardo O. and {Backhaus}, Bren E. and {Bagley}, Micaela B. and {Barro}, Guillermo and {Bell}, Eric F. and {Bisigello}, Laura and {Calabr{\`o}}, Antonello and {Cleri}, Nikko J. and {Cooper}, M.~C. and {Ding}, Xuheng and {Grogin}, Norman A. and {Ho}, Luis C. and {Hutchison}, Taylor A. and {Inoue}, Akio K. and {Jiang}, Linhua and {Jones}, Brenda and {Koekemoer}, Anton M. and {Li}, Wenxiu and {Li}, Zhengrong and {McGrath}, Elizabeth J. and {Molina}, Juan and {Papovich}, Casey and {P{\'e}rez-Gonz{\'a}lez}, Pablo G. and {Pirzkal}, Nor and {Wilkins}, Stephen M. and {Yang}, Guang and {Yung}, L.~Y. Aaron},
	journal = {\apjl},
	month = sep,
	number = {1},
	pages = {L4},
	title = {{Hidden Little Monsters: Spectroscopic Identification of Low-mass, Broad-line AGNs at z > 5 with CEERS}},
	volume = {954},
	year = 2023}

@article{Mathews:2023aa,
	author = {{Mathews}, Elijah P. and {Leja}, Joel and {Speagle}, Joshua S. and {Johnson}, Benjamin D. and {Gibson}, Justus and {Nelson}, Erica J. and {Suess}, Katherine A. and {Tacchella}, Sandro and {Whitaker}, Katherine E. and {Wang}, Bingjie},
	journal = {\apj},
	month = sep,
	number = {2},
	pages = {132},
	title = {{As Simple as Possible but No Simpler: Optimizing the Performance of Neural Net Emulators for Galaxy SED Fitting}},
	volume = {954},
	year = 2023}

@misc{eisenstein_jades_2023,
	author = {Eisenstein, Daniel J. and Johnson, Benjamin D. and Robertson, Brant and Tacchella, Sandro and Hainline, Kevin and Jakobsen, Peter and Maiolino, Roberto and Bonaventura, Nina and Bunker, Andrew J. and Cameron, Alex J. and Cargile, Phillip A. and Curtis-Lake, Emma and Hausen, Ryan and Pusk{\'a}s, D{\'a}vid and Rieke, Marcia and Sun, Fengwu and Willmer, Christopher N. A. and Willott, Chris and Alberts, Stacey and Arribas, Santiago and Baker, William M. and Baum, Stefi and Bhatawdekar, Rachana and Carniani, Stefano and Charlot, Stephane and Chen, Zuyi and Chevallard, Jacopo and Curti, Mirko and DeCoursey, Christa and D'Eugenio, Francesco and de Graaff, Anna and Egami, Eiichi and Helton, Jakob M. and Ji, Zhiyuan and Jones, Gareth C. and Kumari, Nimisha and L{\"u}tzgendorf, Nora and Laseter, Isaac and Looser, Tobias J. and Lyu, Jianwei and Maseda, Michael V. and Nelson, Erica and Parlanti, Eleonora and Rauscher, Bernard J. and Rawle, Tim and Rieke, George and Rix, Hans-Walter and Rujopakarn, Wiphu and Sandles, Lester and Saxena, Aayush and Scholtz, Jan and Sharpe, Katherine and Shivaei, Irene and Simmonds, Charlotte and Smit, Renske and Topping, Michael W. and {\"U}bler, Hannah and Venturi, Giacomo and Williams, Christina C. and Witstok, Joris and Woodrum, Charity},
	month = {October},
	title = {The {JADES} {Origins} {Field}: {A} {New} {JWST} {Deep} {Field} in the {JADES} {Second} {NIRCam} {Data} {Release}},
	year = {2023}}

@article{Lange:2023aa,
	author = {{Lange}, Johannes U.},
	journal = {\mnras},
	month = oct,
	number = {2},
	pages = {3181-3194},
	title = {{NAUTILUS: boosting Bayesian importance nested sampling with deep learning}},
	volume = {525},
	year = 2023}

@article{Narayanan:2024aa,
	author = {{Narayanan}, Desika and {Lower}, Sidney and {Torrey}, Paul and {Brammer}, Gabriel and {Cui}, Weiguang and {Dav{\'e}}, Romeel and {Iyer}, Kartheik G. and {Li}, Qi and {Lovell}, Christopher C. and {Sales}, Laura V. and {Stark}, Daniel P. and {Marinacci}, Federico and {Vogelsberger}, Mark},
	journal = {\apj},
	month = jan,
	number = {1},
	pages = {73},
	title = {{Outshining by Recent Star Formation Prevents the Accurate Measurement of High-z Galaxy Stellar Masses}},
	volume = {961},
	year = 2024}

@article{Cabezas:2024aa,
	author = {{Cabezas}, Alberto and {Corenflos}, Adrien and {Lao}, Junpeng and {Louf}, R{\'e}mi and {Carnec}, Antoine and {Chaudhari}, Kaustubh and {Cohn-Gordon}, Reuben and {Coullon}, Jeremie and {Deng}, Wei and {Duffield}, Sam and {Dur{\'a}n-Mart{\'\i}n}, Gerardo and {Elantkowski}, Marcin and {Foreman-Mackey}, Dan and {Gregori}, Michele and {Iguaran}, Carlos and {Kumar}, Ravin and {Lysy}, Martin and {Murphy}, Kevin and {Orduz}, Juan Camilo and {Patel}, Karm and {Wang}, Xi and {Zinkov}, Rob},
	journal = {arXiv e-prints},
	month = feb,
	pages = {arXiv:2402.10797},
	title = {{BlackJAX: Composable Bayesian inference in JAX}},
	year = 2024}

@article{curti_jades_2024,
	author = {Curti, Mirko and Maiolino, Roberto and Curtis-Lake, Emma and Chevallard, Jacopo and Carniani, Stefano and D'Eugenio, Francesco and Looser, Tobias J. and Scholtz, Jan and Charlot, Stephane and Cameron, Alex and {\"U}bler, Hannah and Witstok, Joris and Boyett, Kristian and Laseter, Isaac and Sandles, Lester and Arribas, Santiago and Bunker, Andrew and Giardino, Giovanna and Maseda, Michael V. and Rawle, Tim and Pino, Bruno Rodr{\'\i}guez Del and Smit, Renske and Willott, Chris J. and Eisenstein, Daniel J. and Hausen, Ryan and Johnson, Benjamin and Rieke, Marcia and Robertson, Brant and Tacchella, Sandro and Williams, Christina C. and Willmer, Christopher and Baker, William M. and Bhatawdekar, Rachana and Egami, Eiichi and Helton, Jakob M. and Ji, Zhiyuan and Kumari, Nimisha and Perna, Michele and Shivaei, Irene and Sun, Fengwu},
	journal = {\aap},
	month = {April},
	pages = {A75},
	title = {{JADES}: {Insights} into the low-mass end of the mass--metallicity--{SFR} relation at 3 {\textless} z {\textless} 10 from deep {JWST}/{NIRSpec} spectroscopy},
	volume = {684},
	year = {2024}}

@article{Glazebrook:2024aa,
	author = {{Glazebrook}, Karl and {Nanayakkara}, Themiya and {Schreiber}, Corentin and {Lagos}, Claudia and {Kawinwanichakij}, Lalitwadee and {Jacobs}, Colin and {Chittenden}, Harry and {Brammer}, Gabriel and {Kacprzak}, Glenn G. and {Labbe}, Ivo and {Marchesini}, Danilo and {Marsan}, Z. Cemile and {Oesch}, Pascal A. and {Papovich}, Casey and {Remus}, Rhea-Silvia and {Tran}, Kim-Vy H. and {Esdaile}, James and {Chandro-Gomez}, Angel},
	journal = {\nat},
	month = apr,
	number = {8007},
	pages = {277-281},
	title = {{A massive galaxy that formed its stars at z {\ensuremath{\approx}} 11}},
	volume = {628},
	year = 2024}

@misc{wan_stochastic_2024,
	author = {Wan, Jenny T. and Tacchella, Sandro and Johnson, Benjamin D. and Iyer, Kartheik G. and Speagle, Joshua S. and Maiolino, Roberto},
	month = {April},
	title = {Stochastic prior for non-parametric star-formation histories},
	year = {2024}}

@misc{li_cue_2024,
	author = {Li, Yijia and Leja, Joel and Johnson, Benjamin D. and Tacchella, Sandro and Davies, Rebecca and Belli, Sirio and Park, Minjung and Emami, Razieh},
	month = {May},
	title = {Cue: {A} {Fast} and {Flexible} {Photoionization} {Emulator} for {Modeling} {Nebular} {Emission} {Powered} {By} {Almost} {Any} {Ionizing} {Source}},
	year = {2024}}

@article{Wang:2024aa,
	author = {{Wang}, Bingjie and {Leja}, Joel and {de Graaff}, Anna and {Brammer}, Gabriel B. and {Weibel}, Andrea and {van Dokkum}, Pieter and {Baggen}, Josephine F.~W. and {Suess}, Katherine A. and {Greene}, Jenny E. and {Bezanson}, Rachel and {Cleri}, Nikko J. and {Hirschmann}, Michaela and {Labb{\'e}}, Ivo and {Matthee}, Jorryt and {McConachie}, Ian and {Naidu}, Rohan P. and {Nelson}, Erica and {Oesch}, Pascal A. and {Setton}, David J. and {Williams}, Christina C.},
	journal = {\apjl},
	month = jul,
	number = {1},
	pages = {L13},
	title = {{RUBIES: Evolved Stellar Populations with Extended Formation Histories at z {\ensuremath{\sim}} 7─8 in Candidate Massive Galaxies Identified with JWST/NIRSpec}},
	volume = {969},
	year = 2024}

@misc{narayanan_ultraviolet_2024,
	author = {Narayanan, Desika and Stark, Daniel P. and Finkelstein, Steven L. and Torrey, Paul and Li, Qi and Cullen, Fergus and Topping, Micheal W. and Marinacci, Federico and Sales, Laura V. and Shen, Xuejian and Vogelsberger, Mark},
	month = {August},
	title = {The {Ultraviolet} {Slopes} of {Early} {Universe} {Galaxies}: {The} {Impact} of {Bursty} {Star} {Formation}, {Dust}, and {Nebular} {Continuum} {Emission}},
	year = {2024}}

@article{marconi_homerun_2024,
	author = {Marconi, A. and Amiri, A. and Feltre, A. and Belfiore, F. and Cresci, G. and Curti, M. and Mannucci, F. and Bertola, E. and Brazzini, M. and Carniani, S. and Cataldi, E. and D'Amato, Q. and Rosa, G. de and Teodoro, E. Di and Ginolfi, M. and Kumari, N. and Marconcini, C. and Maiolino, R. and Magrini, L. and Marasco, A. and Mingozzi, M. and Moreschini, B. and Nagao, T. and Oliva, E. and Scialpi, M. and Tomicic, N. and Tozzi, G. and Ulivi, L. and Venturi, G.},
	journal = {\aap},
	month = {September},
	pages = {A78},
	title = {{HOMERUN} a new approach to photoionization modelling. {I} -- reproducing observed emission lines with percent accuracy and obtaining accurate physical properties of the ionized gas},
	volume = {689},
	year = {2024}}

@article{Xiao:2024aa,
	author = {{Xiao}, Mengyuan and {Oesch}, Pascal A. and {Elbaz}, David and {Bing}, Longji and {Nelson}, Erica J. and {Weibel}, Andrea and {Illingworth}, Garth D. and {van Dokkum}, Pieter and {Naidu}, Rohan P. and {Daddi}, Emanuele and {Bouwens}, Rychard J. and {Matthee}, Jorryt and {Wuyts}, Stijn and {Chisholm}, John and {Brammer}, Gabriel and {Dickinson}, Mark and {Magnelli}, Benjamin and {Leroy}, Lucas and {Schaerer}, Daniel and {Herard-Demanche}, Thomas and {Lim}, Seunghwan and {Barrufet}, Laia and {Endsley}, Ryan and {Fudamoto}, Yoshinobu and {G{\'o}mez-Guijarro}, Carlos and {Gottumukkala}, Rashmi and {Labb{\'e}}, Ivo and {Magee}, Dan and {Marchesini}, Danilo and {Maseda}, Michael and {Qin}, Yuxiang and {Reddy}, Naveen A. and {Shapley}, Alice and {Shivaei}, Irene and {Shuntov}, Marko and {Stefanon}, Mauro and {Whitaker}, Katherine E. and {Wyithe}, J. Stuart B.},
	journal = {\nat},
	month = nov,
	number = {8038},
	pages = {311-315},
	title = {{Accelerated formation of ultra-massive galaxies in the first billion years}},
	volume = {635},
	year = 2024}

@article{Turner:2025aa,
	author = {{Turner}, Crispin and {Tacchella}, Sandro and {D'Eugenio}, Francesco and {Carniani}, Stefano and {Curti}, Mirko and {Glazebrook}, Karl and {Johnson}, Benjamin D. and {Lim}, Seunghwan and {Looser}, Tobias and {Maiolino}, Roberto and {Nanayakkara}, Themiya and {Wan}, Jenny},
	journal = {\mnras},
	month = feb,
	number = {2},
	pages = {1826-1848},
	title = {{Age-dating early quiescent galaxies: high star formation efficiency, but consistent with direct, higher-redshift observations}},
	volume = {537},
	year = 2025}

@article{Yallup:2025aa,
	author = {{Yallup}, David and {Prathaban}, Metha and {Alvey}, James and {Handley}, Will},
	journal = {arXiv e-prints},
	month = sep,
	pages = {arXiv:2509.24949},
	title = {{Parallel Nested Slice Sampling for Gravitational Wave Parameter Estimation}},
	year = 2025}

@article{Curtis-Lake:2025aa,
	author = {{Curtis-Lake}, Emma and {Cameron}, Alex J. and {Bunker}, Andrew J. and {Scholtz}, Jan and {Carniani}, Stefano and {Parlanti}, Eleonora and {D'Eugenio}, Francesco and {Jakobsen}, Peter and {Willmer}, Christopher N.~A. and {Arribas}, Santiago and {Baker}, William M. and {Charlot}, St{\'e}phane and {Chevallard}, Jacopo and {Circosta}, Chiara and {Curti}, Mirko and {Eisenstein}, Daniel J. and {Hainline}, Kevin and {Ji}, Zhiyuan and {Johnson}, Benjamin D. and {Jones}, Gareth C. and {Maiolino}, Roberto and {Maseda}, Michael V. and {P{\'e}rez-Gonz{\'a}lez}, Pablo G. and {Rawle}, Tim and {Rieke}, Marcia and {Rinaldi}, Pierluigi and {Robertson}, Brant and {Rodr{\'\i}gez Del Pino}, Bruno and {Saxena}, Aayush and {Shivaei}, Irene and {Smit}, Renske and {Tacchella}, Sandro and {{\"U}bler}, Hannah and {Venturi}, Giacomo and {Williams}, Christina C. and {Willott}, Chris and {Duan}, Qiao},
	journal = {arXiv e-prints},
	month = oct,
	pages = {arXiv:2510.01033},
	title = {{JADES Data Release 4 Paper I: Sample Selection, Observing Strategy and Redshifts of the complete spectroscopic sample}},
	year = 2025}

@article{McClymont:2025ab,
	author = {{McClymont}, William and {Smith}, Aaron and {Tacchella}, Sandro},
	journal = {arXiv e-prints},
	month = oct,
	pages = {arXiv:2510.13952},
	title = {{Modelling the nebular emission of galaxies across cosmic time with COLT}},
	year = 2025}

@article{Scholtz:2025aa,
	author = {{Scholtz}, J. and {Carniani}, S. and {Parlanti}, E. and {D'Eugenio}, F. and {Curtis-Lake}, E. and {Jakobsen}, P. and {Bunker}, A.~J. and {Cameron}, A.~J. and {Arribas}, S. and {Baker}, W.~M. and {Charlot}, S. and {Chevellard}, J. and {Circosta}, C. and {Curti}, M. and {Duan}, Q. and {Eisenstein}, D.~J. and {Hainline}, K. and {Ji}, Z. and {Johnson}, B.~D. and {Jones}, G.~C. and {Kumari}, N. and {Maiolino}, R. and {Maseda}, M.~V. and {Perna}, M. and {P{\'e}rez-Gonz{\'a}lez}, P.~G. and {Rawle}, T. and {Rieke}, M. and {Rinaldi}, P. and {Robertson}, B. and {Saxena}, A. and {Shivaei}, I. and {Silcock}, M.~S. and {Sun}, Y. and {Rodr{\'\i}guez Del Pino}, B. and {Tacchella}, S. and {{\"U}bler}, H. and {Venturi}, G. and {Williams}, C.~C. and {Willmer}, C.~N.~A. and {Willott}, C. and {Witstok}, J.},
	journal = {arXiv e-prints},
	month = oct,
	pages = {arXiv:2510.01034},
	title = {{JADES Data Release 4 -- Paper II: Data reduction, analysis and emission-line fluxes of the complete spectroscopic sample}},
	year = 2025}

@article{Park:2025aa,
	author = {{Park}, Minjung and {Conroy}, Charlie and {Johnson}, Benjamin D. and {Leja}, Joel and {Dotter}, Aaron and {Cargile}, Phillip A.},
	journal = {\apj},
	month = dec,
	number = {2},
	pages = {165},
	title = {{{\ensuremath{\alpha}}-MC: Self-consistent {\ensuremath{\alpha}}-enhanced Stellar Population Models Covering a Wide Range of Age, Metallicity, and Wavelength}},
	volume = {994},
	year = 2025}

@article{Lapasia:2026aa,
	author = {{Lapasia}, Yash and {Tacchella}, Sandro and {D'Eugenio}, Francesco and {Pusk{\'a}s}, D{\'a}vid and {Bunker}, Andrew J. and {Danhaive}, A. Lola and {Johnson}, Benjamin D. and {Maiolino}, Roberto and {Robertson}, Brant and {Simmonds}, Charlotte and {Shivaei}, Irene and {Williams}, Christina C. and {Willmer}, Christopher and {Xiao}, Mengyuan},
	journal = {arXiv e-prints},
	month = jan,
	pages = {arXiv:2601.08693},
	title = {{Stellar masses of optically dark galaxies: uncertainty introduced by the attenuation law and star-formation histories}},
	year = 2026}

@article{Stoffers:2026aa,
	author = {{Stoffers}, Amanda and {Tacchella}, Sandro and {Simmonds}, Charlotte and {Johnson}, Benjamin D. and {Maiolino}, Roberto},
	journal = {\mnras},
	month = may,
	title = {{The Challenge in Illuminating the Invisible: Constraining LyC Escape with Bayesian Modelling and Symbolic Regression}},
	year = 2026}

@software{jax2018github,
	author = {James Bradbury and Roy Frostig and Peter Hawkins and Matthew James Johnson and Yash Katariya and Chris Leary and Dougal Maclaurin and George Necula and Adam Paszke and Jake Vander{P}las and Skye Wanderman-{M}ilne and Qiao Zhang},
	title = {{JAX}: composable transformations of {P}ython+{N}um{P}y programs},
	url = {http://github.com/jax-ml/jax},
	version = {0.3.13},
	year = {2018}}

\appendix

\section{The full forward-model data flow}
\label{app:flow}

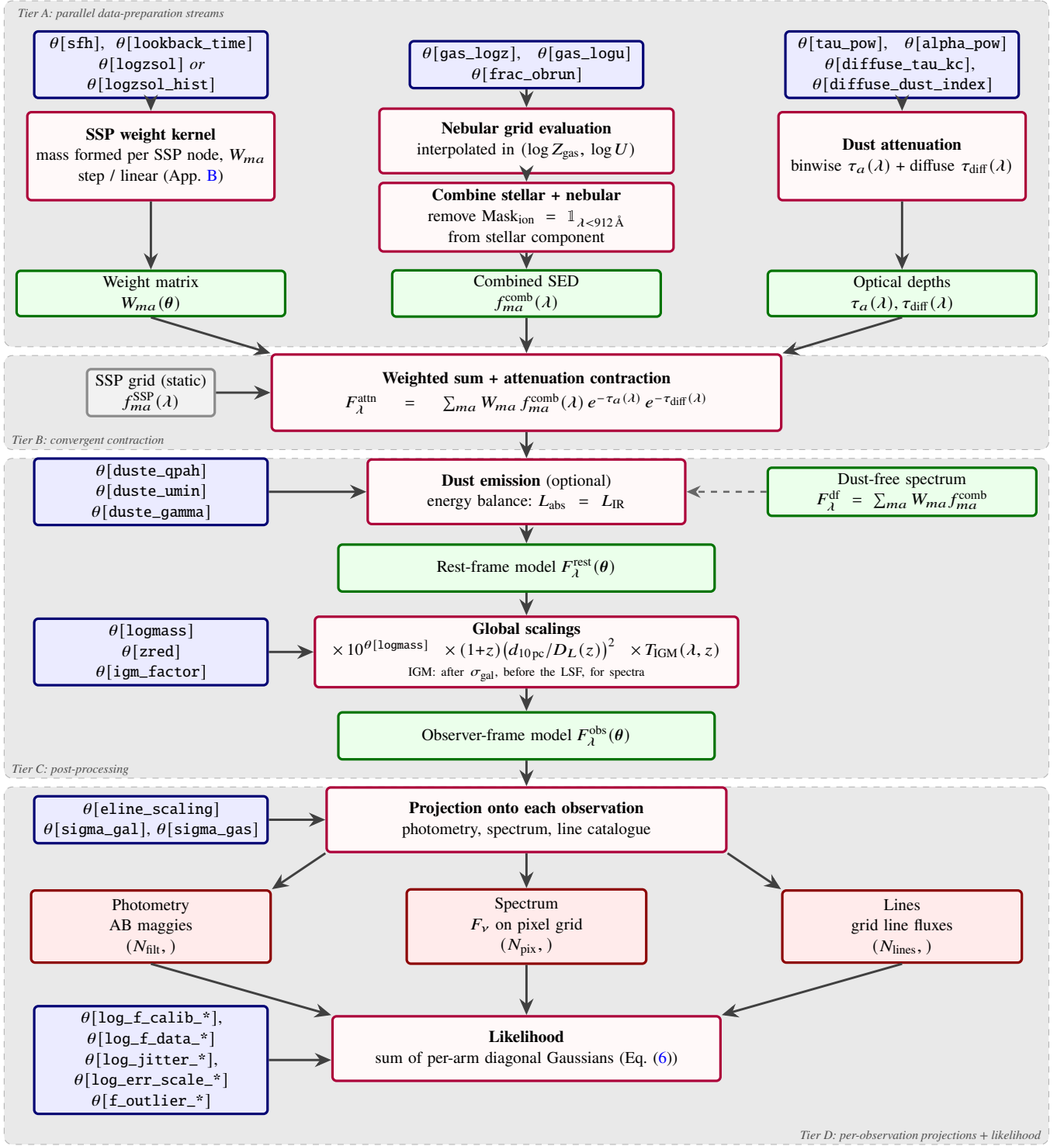
\begin{figure*}
\centering
\begin{tikzpicture}[
    >={Stealth[length=2.2mm,width=2.6mm]},
    font=\footnotesize,
    ceridwenParam/.style ={draw=blue!45!black,   fill=blue!8},
    ceridwenStatic/.style={draw=black!45,        fill=black!5},
    ceridwenOp/.style    ={draw=purple!90!black, fill=pink!10},
    ceridwenBigOp/.style ={draw=purple!90!black, fill=pink!10},
    ceridwenQty/.style   ={draw=green!45!black,  fill=green!8},
    ceridwenQtyB/.style  ={draw=green!45!black,  fill=green!8},
    ceridwenObs/.style   ={draw=red!55!black,    fill=red!8},
    param/.style ={rectangle, rounded corners=2pt, ceridwenParam,very thick,
                   align=center, inner sep=2.5pt,
                   minimum height=8mm, text width=38mm},
    static/.style={rectangle, rounded corners=2pt, ceridwenStatic,very thick,
                   align=center, inner sep=2.5pt,
                   minimum height=8mm, text width=20mm},
    op/.style    ={rectangle, rounded corners=2pt, ceridwenOp,very thick,
                   align=center, inner sep=3pt,
                   minimum height=11mm, text width=48mm},
    bigop/.style ={rectangle, rounded corners=2pt, ceridwenBigOp, very thick,
                   align=center, inner sep=4pt,
                   minimum height=13mm, text width=84mm},
    qty/.style   ={rectangle, rounded corners=2pt, ceridwenQty,very thick,
                   align=center, inner sep=2.5pt,
                   minimum height=7mm, text width=44mm},
    Qty/.style   ={rectangle, rounded corners=2pt, ceridwenQtyB,very thick,
                   align=center, inner sep=3pt,
                   minimum height=8mm, text width=62mm},
    obs/.style   ={rectangle, rounded corners=2pt, ceridwenObs, very thick,
                   align=center, inner sep=4pt,
                   minimum height=12mm, text width=38mm},
    flow/.style  ={-{Stealth[length=2.8mm,width=2.4mm]}, line width=1.1pt, draw=black!75},
    sidein/.style={-{Stealth[length=2.4mm,width=2.1mm]}, line width=0.95pt, draw=black!65},
    branch/.style={-{Stealth[length=2.4mm,width=2.1mm]}, line width=0.95pt, dashed, draw=black!60},
    glabel/.style={font=\scriptsize\itshape, text=black!70, inner sep=1pt},
]

\def\xL{-64mm}
\def\xC{0mm}
\def\xR{64mm}
\def\R{13.0mm}

\node[param ] (sfhpar)  at (\xL, 0)
        {$\theta[\texttt{sfh}]$,\;
         $\theta[\texttt{lookback\_time}]$\\
         $\theta[\texttt{logzsol}]$ \emph{or} $\theta[\texttt{logzsol\_hist}]$};
\node[param ] (dustpar) at (\xR, 0)
        {$\theta[\texttt{tau\_pow}]$, \;
        $\theta[\texttt{alpha\_pow}]$\\
         $\theta[\texttt{diffuse\_tau\_kc}]$, \;
         $\theta[\texttt{diffuse\_dust\_index}]$};

\node[param ] (nebpar) at (\xC, 0)
        {$\theta[\texttt{gas\_logz}]$, \;
         $\theta[\texttt{gas\_logu}]$\\
         $\theta[\texttt{frac\_obrun}]$};

\node[op, text width=40mm, minimum height=15mm] (wcalc)  at (\xL, -1.2*\R)
        {\textbf{SSP weight kernel}\\
         mass formed per SSP node, $W_{ma}$\\
         step / linear (App.~\ref{app:sfh_interp})};
\node[op] (nebop)  at (\xC, -1*\R)
        {\textbf{Nebular grid evaluation}\\
         interpolated in $(\log Z_{\rm gas},\,\log U)$};

\node[op, text width=40mm, minimum height=15mm] (dustop) at (\xR, -1.2*\R)
        {\textbf{Dust attenuation}\\
         binwise $\tau_{a}(\lambda)$\;+\;diffuse $\tau_{\rm diff}(\lambda)$};

\node[op] (combine) at (\xC, -2*\R)
        {\textbf{Combine stellar + nebular}\\
        remove ${\rm Mask}_{\rm ion} = \mathbbm{1}_{\lambda < 912\,\text{\AA}}$ from stellar component};

\node[qty] (W)     at (\xL, -3*\R)
        {Weight matrix\\$W_{ma}(\boldsymbol\theta)$};
\node[qty] (fcomb) at (\xC, -3*\R)
        {Combined SED\\$f^{\rm comb}_{ma}(\lambda)$};
\node[qty] (tau)   at (\xR, -3*\R)
        {Optical depths\\$\tau_{a}(\lambda)$,\,$\tau_{\rm diff}(\lambda)$};

\node[static] (sspgrid) at (\xL, -4.3*\R)
        {SSP grid (static)\\
         $f^{\rm SSP}_{ma}(\lambda)$};

\node[bigop] (sum) at (\xC, -4.3*\R)
        {\textbf{Weighted sum + attenuation contraction}\\[2pt]
         $F^{\rm attn}_{\lambda}\;=\;
            \sum_{ma}W_{ma}\,f^{\rm comb}_{ma}(\lambda)\,
            e^{-\tau_{a}(\lambda)}\,e^{-\tau_{\rm diff}(\lambda)}$};

\node[qty] (fdfree) at (\xR, -5.6*\R)
        {Dust-free spectrum\\$F^{\rm df}_{\lambda}=\sum_{ma}W_{ma}f^{\rm comb}_{ma}$};

\node[op, text width=5.2cm, align=center] (demop) at (\xC, -5.6*\R)
        {\textbf{Dust emission} (optional)\\
         energy balance: $L_{\rm abs}=L_{\rm IR}$};

\node[param] (dempar) at (\xL, -5.6*\R)
        {$\theta[\texttt{duste\_qpah}]$\\
         $\theta[\texttt{duste\_umin}]$\\
         $\theta[\texttt{duste\_gamma}]$};

\node[Qty] (Frest) at (\xC, -6.6*\R)
        {Rest-frame model\;$F^{\rm rest}_{\lambda}(\boldsymbol\theta)$};

\node[op, text width=70mm] (scale) at (\xC, -7.7*\R)
        {\textbf{Global scalings}\\
         $\times\,10^{\theta[\texttt{logmass}]}$\;\;
         $\times\,(1{+}z)\bigl(d_{10\,\rm pc}/D_{L}(z)\bigr)^{2}$\;\;
         $\times\,T_{\rm IGM}(\lambda,z)$\\
         \scriptsize IGM: after $\sigma_{\rm gal}$, before the LSF, for spectra};
\node[param] (scalepar) at (\xL, -7.7*\R)
        {$\theta[\texttt{logmass}]$\\
         $\theta[\texttt{zred}]$\\
         $\theta[\texttt{igm\_factor}]$};

\node[Qty] (Fobs) at (\xC, -8.8*\R)
        {Observer-frame model\;$F^{\rm obs}_{\lambda}(\boldsymbol\theta)$};

\node[op, text width=66mm]  (proj) at (\xC, -9.9*\R)
        {\textbf{Projection onto each observation}\\
         photometry, spectrum, line catalogue};

\node[param] (elinescaling) at (\xL,  -9.9*\R)
        {$\theta[\texttt{eline\_scaling}]$\\
        $\theta[\texttt{sigma\_gal}]$,\;$\theta[\texttt{sigma\_gas}]$};

\node[obs] (phot)  at (\xL, -11.3*\R)
        {Photometry\\AB maggies\\$(N_{\rm filt},)$};
\node[obs] (spec)  at (\xC, -11.3*\R)
        {Spectrum\\$F_{\nu}$ on pixel grid\\$(N_{\rm pix},)$};
\node[obs] (lines) at (\xR, -11.3*\R)
        {Lines\\ grid line fluxes\\$(N_{\rm lines},)$};
\node[op, text width=64mm] (like) at (\xC, -12.9*\R)
        {\textbf{Likelihood}\\
         sum of per-arm diagonal Gaussians (Eq.~\eqref{eq:total_lnlike})};
\node[param] (floorpar) at (\xL, -13.05*\R)
        {$\theta[\texttt{log\_f\_calib\_*}]$,\;
         $\theta[\texttt{log\_f\_data\_*}]$\\
         $\theta[\texttt{log\_jitter\_*}]$,\; $\theta[\texttt{log\_err\_scale\_*}]$\\
         $\theta[\texttt{f\_outlier\_*}]$};

\draw[flow] (sfhpar)  -- (wcalc);
\draw[flow] (sspgrid) -- (sum);
\draw[flow] (dustpar) -- (dustop);
\draw[flow] (nebpar)  -- (nebop);

\draw[flow] (wcalc)   -- (W);
\draw[flow] (nebop)   -- (combine);
\draw[flow] (combine) -- (fcomb);
\draw[flow] (dustop)  -- (tau);

\draw[flow] (W.south)     -- (sum.north west);
\draw[flow] (fcomb.south) -- (sum.north);
\draw[flow] (tau.south)   -- (sum.north east);

\draw[flow] (sum.south) -- (demop.north);
\draw[branch] (fdfree.west) -- (demop.east)
              node[glabel, pos=0.5, above] {};

\draw[flow] (demop) -- (Frest);
\draw[flow] (Frest) -- (scale);
\draw[flow] (scale) -- (Fobs);
\draw[flow] (Fobs)  -- (proj);

\draw[flow] (dempar)   -- (demop);
\draw[flow] (scalepar) -- (scale);

\draw[flow] (elinescaling.east) -- (proj.west);
\draw[flow] (proj.south) -- (spec.north);
\draw[flow] (proj.south west) -- (phot.north east);
\draw[flow] (proj.south east) -- (lines.north west);
\draw[flow] (phot.south)  -- (like.north west);
\draw[flow] (spec.south)  -- (like.north);
\draw[flow] (lines.south) -- (like.north east);
\draw[flow] (floorpar.east) -- (floorpar.east -| like.west);

\coordinate (LeftRef)  at (-89mm, 0);
\coordinate (RightRef) at ( 89mm, 0);

\def\PadA{5mm}
\def\PadB{3mm}
\def\PadC{3mm}
\def\PadD{5mm}

\def\GapAB{1.5mm}
\def\GapBC{1.5mm}
\def\GapCD{1.5mm}

\coordinate (TierA-nw) at ([yshift=\PadA]LeftRef  |- sfhpar.north);
\coordinate (TierA-se) at ([yshift=-\PadA]RightRef |- W.south);

\coordinate (TierB-topref) at ([yshift=-\GapAB]TierA-se);
\coordinate (TierB-nw)     at (LeftRef |- TierB-topref);
\coordinate (TierB-se)     at ([yshift=-\PadB]RightRef |- sum.south);

\coordinate (TierC-topref) at ([yshift=-\GapBC]TierB-se);
\coordinate (TierC-nw)     at (LeftRef |- TierC-topref);
\coordinate (TierC-se)     at ([yshift=-\PadC]RightRef |- Fobs.south);

\coordinate (TierD-topref) at ([yshift=-\GapCD]TierC-se);
\coordinate (TierD-nw)     at (LeftRef |- TierD-topref);
\coordinate (TierD-se)     at ([yshift=-\PadD]RightRef |- floorpar.south);

\begin{scope}[on background layer]

\node[
    draw=black!30, fill=black!30, fill opacity=0.28,
    dashed, rounded corners,
    fit=(TierA-nw)(TierA-se),
    inner sep=0pt,
    label={[glabel, anchor=north west, xshift=2mm, yshift=-1mm]north west:
        Tier A: parallel data-preparation streams}
] {};

\node[
    draw=black!30, fill=black!30, fill opacity=0.28,
    dashed, rounded corners,
    fit=(TierB-nw)(TierB-se),
    inner sep=0pt,
    label={[glabel, anchor=west, yshift=1.5mm, xshift=1mm]south west:
        Tier B: convergent contraction}
] {};

\node[
    draw=black!30, fill=black!30, fill opacity=0.28,
    dashed, rounded corners,
    fit=(TierC-nw)(TierC-se),
    inner sep=0pt,
    label={[glabel, anchor=west, yshift=1.5mm, xshift=1mm]south west:
        Tier C: post-processing}
] {};

\node[
    draw=black!30, fill=black!30, fill opacity=0.28,
    dashed, rounded corners,
    fit=(TierD-nw)(TierD-se),
    inner sep=0pt,
    label={[glabel, anchor=east, yshift=1.5mm, xshift=-1mm]south east:
        Tier D: per-observation projections + likelihood}
] {};

\end{scope}

\end{tikzpicture}
\caption{Forward-model data flow in \ceri.  The model is organised as three parallel data-preparation streams (Tier~A) that converge at the weighted-sum contraction (Tier~B), followed by a sequential post-processing column (Tier~C) and the per-observation projection fan-out (Tier~D).  Free parameters (\textcolor{blue!60!black}{blue}) feed each stream; the static SSP grid (\textcolor{black!60}{grey}) enters the contraction, and its ionising-photon rates, computed once from the grid, scale the nebular grids.  The three streams supply the three operands of the single weighted-sum contraction shown in Tier~B, whose dust-free side product $F^{\rm df}_{\lambda}$ supplies the energy-balance constraint of the optional dust-emission step.  The rest-frame spectrum is scaled to the observer frame (mass, cosmological flux factor, IGM transmission) before being projected onto every observation.  Optional components (nebular emission, dust, IGM, the ionising escape fraction) are switched on or off per model, and the whole graph is differentiable end to end.  The projected observables share one likelihood, the sum of the per-arm diagonal Gaussians; the optional noise nuisances $\theta[\texttt{log\_f\_calib\_*}]$, $\theta[\texttt{log\_f\_data\_*}]$, $\theta[\texttt{log\_jitter\_*}]$ and $\theta[\texttt{log\_err\_scale\_*}]$ and the outlier fraction $\theta[\texttt{f\_outlier\_*}]$, one per observation (Appendix~\ref{app:sys_floor}) and the galaxy's velocity dispersions $\theta[\texttt{sigma\_gal}]$ and $\theta[\texttt{sigma\_gas}]$ when they are sampled (\S\ref{sec:spectra}) enter at this final tier.}
\label{fig:ceridwen_flow}
\end{figure*}

Figure~\ref{fig:ceridwen_flow} gives the complete data-flow diagram of the \ceri\ forward model summarised by the cartoon of Fig.~\ref{fig:flow_cartoon}.

\section{From the SFH to SSP weights}
\label{app:sfh_interp}

The composite stellar population (CSP) spectrum is the mass-weighted sum over the SSP grid,
\begin{equation}
    F_{\lambda}(\boldsymbol{\theta}) \;=\;
    \sum_{m}\sum_{a}
    W_{ma}(\boldsymbol{\theta})\,f^{\rm SSP}_{ma}(\lambda),
    \label{eq:csp_sum}
\end{equation}
the discrete approximation to $F_{\lambda}=\int\!\mathrm{d}t\!\int\!\mathrm{d}Z\,\psi(t,Z)\,s_{\lambda}(t,Z)$, where $f^{\rm SSP}_{ma}$ is the SSP spectrum at metallicity node $m$ and log-age node $a$ and $W_{ma}$ is the stellar mass formed there. This appendix sets out how $W_{ma}$ is obtained from a tabulated star-formation and chemical-enrichment history, expanding on the treatments of \citet{Johnson:2021aa} and \citet{Hearin:2023aa}; $i$ indexes an SFH time bin, $a$ an SSP age node, and $m$ an SSP metallicity node.
The SFH is specified by its star-formation rate $\psi_{i}$ on a lookback-time grid $T_{i}$ ordered from the present day ($T_{0}=0$) to the oldest node, with bin $i$ spanning $[T_{i},T_{i+1}]$ of width $\Delta T_{i}$ and forming mass $m_{i}=\bar\psi_{i}\,\Delta T_{i}$ (the trapezoidal mass $\tfrac{1}{2}(\psi_{i}+\psi_{i+1})\Delta T_{i}$ for the linear scheme below). \ceri\ provides two schemes for turning this into weights on the SSP age axis, both mass-conserving by construction; either may be combined with either metallicity mode below.

\begin{figure}[!tb]
    \centering
    \includegraphics[width=\linewidth]{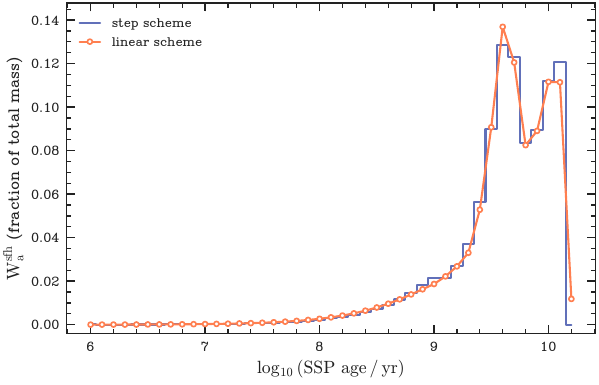}
    \caption{From a tabulated star-formation history to weights on the SSP age axis, at constant metallicity. The step scheme (\S\ref{sec:sfh}, Appendix~\ref{app:sfh_interp}) attributes mass by linear-time Voronoi overlap; the linear scheme (Appendix~\ref{app:sfh_interp}) integrates a piecewise-linear SFR analytically against the log-age tent basis. The two agree for smooth histories and differ where the SFH varies on a timescale comparable to the SSP age spacing. Both schemes form exactly the same mass in every SFH bin, $m_i=\tfrac{1}{2}(\psi_i+\psi_{i+1})\Delta T_i$ for the per-node input used here, and the same total mass; they differ only in how that mass is distributed among the SSP age nodes that a bin overlaps. The curves are the weights per SSP age node, which are per logarithmic age cell, so they need not be flat even for a constant star-formation rate; both are divided by the common total mass, so the two curves sum to the same value.}
    \label{fig:sfh_interp}
\end{figure}
The default \emph{step} scheme treats the SFR as piecewise-constant across each bin and apportions its mass to the SSP age nodes in proportion to the overlap, in linear time, between the SFH bin and the age cell each node owns (bounded by the midpoints to its neighbours). The resulting weights are non-negative, which makes the scheme numerically robust for steep histories, and it reproduces the \texttt{FastStepBasis} convention of \pros.
The \emph{linear} scheme \citep{Johnson:2021aa} instead treats the SFR as piecewise-linear in time and integrates it against the linear-in-$\log t$ interpolation \ceri\ uses to evaluate the SSP grid (SSP fluxes vary approximately linearly in $\log t$; \citealt{ocvirk_stecmap_2006}). Writing the spectrum between adjacent age nodes as $s_{\lambda}(t)=a_{j}(t)\,s_{\lambda}(t_{j})+b_{j+1}(t)\,s_{\lambda}(t_{j+1})$ with $a_{j},b_{j+1}$ the log-age interpolation weights, the mass assigned to node $j$ is
\begin{equation}
    m_{j}\;=\;
    \int_{t_{j}}^{t_{j+1}}\!\!\mathrm{d}t\;\psi(t)\,a_{j}(t)
    \;+\;
    \int_{t_{j-1}}^{t_{j}}\!\!\mathrm{d}t\;\psi(t)\,b_{j}(t),
    \label{eq:tent_integral}
\end{equation}
which \ceri\ evaluates analytically for the piecewise-linear $\psi$ and renormalises so that each bin contributes exactly its mass $m_{i}$. The linear scheme captures rapidly varying histories more accurately, at the cost of a small clip that enforces non-negativity where a steep gradient would otherwise drive a weight marginally below zero.
The age weights are then spread across the SSP metallicity grid. The metallicity parameters $\theta[\texttt{logzsol}]$ and $\theta[\texttt{logzsol\_hist}]$ are $\log_{10}(Z/Z_\odot)$, relative to the solar metallicity of the SSP grid (\S\ref{sec:zh}); the grid axis is converted once, at construction, by subtracting $\log_{10}Z_\odot$. In the constant-metallicity mode a single value $\theta[\texttt{logzsol}]$ is applied to every bin and the mass at each age node is split between the two bracketing SSP metallicity nodes by linear interpolation in $\log_{10}(Z/Z_\odot)$. In the time-varying mode each SFH bin is instead assigned its own metallicity from the enrichment track $\theta[\texttt{logzsol\_hist}]$ and split in the same way, so the chemical history evolves with the SFH at the same temporal resolution. The resulting $W_{ma}$ enters Eq.~\eqref{eq:csp_sum}.

\section{Nebular emission interpolation and broadening}
\label{app:nebular}
\ceri's nebular model wraps the photoionisation grids of \citet{Byler:2017aa}, which provide $\log_{10}$-tabulated continuum and line luminosities for an idealised, ionisation-bounded \ion{H}{ii} region as a function of three parameters: gas-phase metallicity $\log_{10}(Z_{\rm g}/Z_\odot)\equiv\theta[\texttt{gas\_logz}]$, ionising-source age $\tau$, and ionisation parameter $\log_{10}U\equiv\theta[\texttt{gas\_logu}]$. The continuum cube $\mathcal{C}_{i,j,k}(\lambda)=\log_{10}\bigl[L_{\nu}/Q\bigr]$ is tabulated on a $(n_{Z}\times n_{\tau}\times n_{U})$ grid, and the line cube $\mathcal{L}_{i,j,k}^{(\ell)}=\log_{10}\bigl[L_{\ell}/Q\bigr]$ in the same shape for each of the $n_{\rm line}$ emission lines.
The nebular luminosity scales with the rate of hydrogen-ionising photons, which \ceri\ computes directly from each unattenuated SSP spectrum,
\begin{equation}
    Q_{ma}
    \;=\;
    \frac{L_{\odot}}{h}\,
    \int_{0}^{912\,\textup{\AA}}\!\frac{f^{\rm SSP}_{ma}(\lambda)}{\lambda}\,d\lambda,
    \label{eq:app_q}
\end{equation}
matching the run-time formula of \textsc{FSPS}. For a given $(Z_{\rm g},U)$ the continuum and line cubes are trilinearly interpolated in gas-phase metallicity, ionising-source age, and ionisation parameter (each cube against the axes it was generated with), and the result is scaled by $Q_{ma}$ to give the nebular continuum and per-line luminosities at each SSP node,
\begin{equation}
    L^{\rm neb,c}_{ma}(\lambda)=Q_{ma}\,10^{\widehat{\mathcal{C}}},
    \qquad
    L^{\rm neb,\ell}_{ma}=Q_{ma}\,10^{\widehat{\mathcal{L}}^{(\ell)}},
\end{equation}
the cube being queried at the SSP age $\tau_{a}$ but at the free gas-phase parameters. Only SSP ages that fall inside the overlap of the continuum and line grids' tabulated age ranges contribute; older nodes are set to zero.
On the model wavelength grid, which the photometry integrates, each line is placed as a Gaussian of width $\Delta\lambda_{\ell}=2\,\Delta\lambda^{\rm pix}_{\ell}$, the grid's two-pixel floor, so that its flux is conserved on the grid; it carries no intrinsic width there. The profile a line acquires in a spectrum is set at the projection stage from the gas dispersion and the instrumental LSF, Eq.~\eqref{eq:one_kernel} and Appendix~\ref{app:spec_smoothing}.
The escape parameter $\fob$ scales the nebular emission by $(1-\fob)$ and restores the fraction $\fob$ of the ionising continuum (\S\ref{sec:frac_obrun}).

\section{Dust attenuation and emission}
\label{app:dust}
\subsection{Per-bin attenuation contraction}
\label{app:dust_attn}
\ceri\ partitions the SSP age axis into $n_{\rm bin}$ user-supplied bins, each with its own attenuation law $\kappa_{b}(\lambda;\theta_{b})$ from the library of twelve curves carried with the code. The bin assignment is encoded in an age-to-bin matrix $\mathbf{M}\in[0,1]^{N_{\rm age}\times n_{\rm bin}}$ whose populated rows are normalised to sum to unity: an age node lying in a single bin gives a one-hot row, a node shared between overlapping bins gives the corresponding convex combination, and a node outside every bin gives a zero row (no birth-cloud attenuation). The optical depth seen by stars at SSP age node $a$ is then
\begin{equation}
    \tau_{a}(\lambda)
    \;=\;
    \sum_{b=1}^{n_{\rm bin}} M_{a,b}\,\kappa_{b}(\lambda;\theta_{b}),
    \label{eq:app_tau_bin}
\end{equation}
and the diffuse component contributes a single age-independent
$\tau_{\rm diff}(\lambda)$. The full attenuated CSP spectrum is
\begin{equation}
    F^{\rm attn}_{\lambda}(\boldsymbol{\theta})
    \;=\;
    e^{-\tau_{\rm diff}(\lambda)}\!
    \sum_{ma} W_{ma}\,f^{\rm comb}_{ma}(\lambda)\,
    e^{-\tau_{a}(\lambda)},
    \label{eq:app_attn}
\end{equation}
where $f^{\rm comb}_{ma}=f^{\rm SSP}_{ma}+L^{\rm neb}_{ma}$ is the combined stellar-plus-nebular SED, so the two attenuation factors enter the same contraction as the SSP-grid weighted sum (Tier~B of Fig.~\ref{fig:ceridwen_flow}).
When $\fob\equiv\theta[\texttt{frac\_obrun}]$ is present, a fraction $\fob$ of the young-star light bypasses the birth-cloud attenuation, i.e.\ $e^{-\tau_{a}(\lambda)}$ is replaced by $(1-\fob)\,e^{-\tau_{a}(\lambda)}+\fob$ at the attenuated (birth-cloud) nodes. The escaping ionising continuum ($\lambda<912\,$\AA\ of every population inside the age range of the photoionisation grid) is treated as a special case: because it is exactly the flux restored to the nebular channel in Appendix~\ref{app:nebular}, it bypasses the birth cloud completely (transmission set to unity) rather than entering the partial-escape mixture above, and so is not double-counted. The operation is a no-op for old nodes, which carry no birth-cloud law; when \fob\ is absent or $0$, the model reduces to Eq.~\eqref{eq:app_attn} exactly.

\subsection{Dust emission and energy balance}
\label{app:dust_emission}
When dust emission is requested, the bolometric absorbed stellar
luminosity is
\begin{equation}
    L_{\rm abs}
    \;=\;
    \int_{0}^{\infty}\!\bigl[F^{\rm df}_{\lambda}-F^{\rm attn}_{\lambda}\bigr]\,d\nu,
    \label{eq:app_labs}
\end{equation} 
where $F^{\rm df}_{\lambda}=\sum_{ma}W_{ma}f^{\rm comb}_{ma}(\lambda)$ is the dust-free side product of the same contraction (Eq.~\eqref{eq:app_attn}). This luminosity is re-emitted in the infrared with a template $m_{\rm duste}(\lambda;\theta_{\rm duste})$ obtained from a bilinear interpolation of the \citet{Draine:2007aa} (DL07) or \citet{Jones:2017aa} (THEMIS) grids in $(q_{\rm PAH},U_{\rm min})\equiv(\theta[\texttt{duste\_qpah}],\,\theta[\texttt{duste\_umin}])$, linearly mixed between the minimum-$U$ and warm-$U$ template components with fraction $\gamma\equiv\theta[\texttt{duste\_gamma}]\in[0,1]$. Writing $\mathcal{N}=\int m_{\rm duste}\,d\nu$ for its bolometric integral, the unit-luminosity template is $\hat{m}_{\rm duste}=m_{\rm duste}/\mathcal{N}$. The initial emission is $\hat{m}_{\rm duste}\,L_{\rm abs}$; the luminosity it loses to the diffuse component, $L_{\rm abs}^{(1)}=\int\hat{m}_{\rm duste}\,L_{\rm abs}\, \bigl(1-e^{-\tau_{\rm diff}}\bigr)\,d\nu$, is re-emitted once with the same template, and both the initial and the reprocessed emission are then attenuated by the diffuse curve. The final spectrum is
\begin{equation}
    F^{\rm full}_{\lambda}
    \;=\;
    F^{\rm attn}_{\lambda}
    + e^{-\tau_{\rm diff}(\lambda)}\,\hat{m}_{\rm duste}(\lambda)
        \bigl(L_{\rm abs}+L_{\rm abs}^{(1)}\bigr),
\end{equation}
and the dust mass follows from the standard \citet{Draine:2007aa} normalisation
\begin{equation}
    M_{\rm dust}\;=\;\frac{3.21\times10^{-3}}{4\pi}\,\frac{L_{\rm abs}}{\mathcal{N}}.
\end{equation}

\section{Photometric projection}
\label{app:phot}
This appendix gives the discrete form of the band projection of \S\ref{sec:photometry}. The filter set is resampled onto a shared wavelength grid $\{\lambda_{j}\}$ and stored as a transmission matrix whose row for band $b$ is
\begin{equation}
    R_{bj}
    \;=\;
    \frac{R_{b}(\lambda_{j})\,\lambda_{j}\,\Delta\lambda_{j}}
         {\int \lambda\,R_{b}(\lambda)\,f_{\lambda}^{\rm AB}(\lambda)\,\mathrm{d}\lambda},
    \label{eq:app_phot_row}
\end{equation}
where $\Delta\lambda_{j}$ is the local grid spacing, the numerator is the photon-counting transmission weight, and the denominator is the AB zero-point flux of a $3631\,$Jy source, $f_{\nu}^{\rm AB}=3.631\times10^{-20}\,\mathrm{erg\,s^{-1}\,cm^{-2}\,Hz^{-1}}$, integrated through the same band. The numerator uses a Riemann sum with $\Delta\lambda_{j}$ the local grid spacing (one-sided at the grid edges) while the zero-point integral uses the trapezoidal rule; on the finely sampled filter grids the two quadratures agree to well below the photometric precision.
The observer-frame model arrives as $F_{\nu}^{\rm obs}$ on the rest-frame model grid $\lambda_{k}^{\rm rest}$, already carrying the mass, flux-factor, and IGM scalings of \S\ref{sec:global}. The predicted maggie in band $b$ is then
\begin{equation}
    m_{b}
    \;=\;
    \sum_{k} T_{bk}\,F_{\nu,k}^{\rm obs},
    \qquad
    \mathbf{T}
    \;=\;
    \bigl(\mathbf{R}\,\mathbf{H}\bigr)\,\mathrm{diag}\!\left(\frac{c}{\lambda_{k}^{\rm obs\,2}}\right),
    \label{eq:app_phot_T}
\end{equation}
where $\lambda_{k}^{\rm obs}=(1+z)\,\lambda_{k}^{\rm rest}$, the diagonal factor performs the $F_{\nu}\!\to\!f_{\lambda}$ conversion, and $\mathbf{H}$ is the sparse linear-interpolation matrix gathering the model grid onto the filter grid. When the redshift is fixed the projection $\mathbf{T}$ reduces to a single matrix product, and the gradients of the band fluxes with respect to every model parameter follow from the chain rule through that linear map: the gradient spectra are projected through the filters exactly as the spectrum is; when $\theta[\texttt{zred}]$ is sampled, the observer-frame grid $\lambda_{k}^{\rm obs}$ and the interpolation onto the filter grid are rebuilt at each evaluation and the gradient with respect to $z$ flows through that interpolation. The interpolation weights of $\mathbf{H}$ are continuous, piecewise-linear functions of $z$, so the band flux is continuous in $z$ and differentiable everywhere except on the measure-zero set where a filter node crosses a model pixel. The kinks occur at the discrete redshifts at which a filter node coincides with a model pixel. A continuous proposal lands on one with probability zero, and even there the gradient is finite (one-sided). No smoothing of the transmission curves is involved. The velocity broadening applied to the spectrum before projection changes broad-band fluxes by less than $5\times10^{-4}$~mag at $300\,$km\,s$^{-1}$ even for extreme equivalent widths. The band maggies feed the diagonal Gaussian likelihood of \S\ref{sec:likelihood} directly.

\section{Spectral broadening}
\label{app:spec_smoothing}
This appendix collects the broadening operators referenced in \S\ref{sec:spectra}. The projection of the rest-frame model onto the detector grid applies, at every observed pixel, the continuum kernel of Eq.~\eqref{eq:one_kernel}: the free galaxy dispersion, the  instrumental width and the fixed library width are one kernel because all three are Gaussians in $\ln\lambda$. The emission lines do not pass through this kernel. They are added afterwards with the line width of Eq.~\eqref{eq:one_kernel}, at the end of this appendix. The operator is split into the broadening parts that are static and the ones that depend on sampled parameters.

Throughout, widths are Gaussian dispersions; a full width at half
maximum converts as
\begin{equation}
    \mathrm{FWHM} \;=\; 2\sqrt{2\ln 2}\;\sigma \;\simeq\; 2.3548\,\sigma .
    \label{eq:app_fwhm}
\end{equation}
An \texttt{Instrument} converts whatever unit the LSF is quoted in into a velocity dispersion at each observed pixel,

\begin{equation}
    \sigma_{\rm inst}(\lambda_{\rm obs}) =
    \begin{cases}
        c/(2.3548\,R),                         & \texttt{R\_fwhm},\\[2pt]
        c/R,                                   & \texttt{R\_sigma},\\[2pt]
        c\,\mathrm{FWHM}_{\lambda}/(2.3548\,\lambda_{\rm obs}), & \texttt{fwhm\_aa},\\[2pt]
        c\,\sigma_{\lambda}/\lambda_{\rm obs}, & \texttt{sigma\_aa},\\[2pt]
        \sigma_{\rm v},                        & \texttt{sigma\_kms},\\[2pt]
        \mathrm{FWHM}_{\rm v}/2.3548,          & \texttt{fwhm\_kms},
    \end{cases}
    \label{eq:app_modes}
\end{equation}

with $R=\lambda/\mathrm{FWHM}_{\lambda}$ for \texttt{R\_fwhm} and $R=\lambda/\sigma_{\lambda}$ for \texttt{R\_sigma}, each either a constant or a curve tabulated on the observed grid. Resolving powers are quoted in two conventions, $R=\lambda/\mathrm{FWHM}_{\lambda}$ (instrument datasheets) and $R=\lambda/\sigma_{\lambda}$ (\pros\ and \texttt{sedpy}), which differ by a factor of $2.3548$. \ceri\ therefore requires the convention to be stated and rejects a bare $R$. A width quoted in \AA\ is applied as a Gaussian in $\ln\lambda$ of the same dispersion at that pixel; the difference from a Gaussian in $\lambda$ is of order $\sigma_{\lambda}/\lambda$.

The stellar templates carry a finite intrinsic resolution, which each \ceri\ SSP grid stores as a per-pixel velocity dispersion on its own wavelength array,
\begin{equation}
    \sigma_{\rm lib}(\lambda)
    \;=\;
    \max\!\left[
        \frac{c}{2\sqrt{2\ln 2}}\,\frac{2\,\Delta\lambda}{\lambda},\;
        \sigma_{\rm LSF,lib}(\lambda)
    \right],
    \label{eq:app_siglib}
\end{equation}
the first term being the local two-pixel sampling width of the stored grid treated as a Gaussian FWHM, which is the resolution floor of the tabulated spectra whatever the parent library's native LSF, and the second the documented library LSF where one exists (for MILES, a FWHM of $2.54$~\AA; \citealt{falcon-barroso_updated_2011}). The curve is computed from the wavelength array of the stored grid itself, so it is valid regardless of the parent library's native resolution, and at fit time it is interpolated to the fitted pixels at their rest-frame wavelengths. Where the instrumental resolution is finer than the library curve the subtraction below floors at zero: no deconvolution is attempted, and the user is warned that the grid limits the deliverable resolution. Because Gaussians add in quadrature, the fixed part of the continuum kernel has per-pixel width
\begin{equation}
    \sigma_{\rm fix}(\lambda_{i})
    \;=\;
    \sqrt{\max\!\bigl[\sigma_{\rm inst}^{2}(\lambda_{i})-\sigma_{\rm lib}^{2}(\lambda_{i}/(1+z)),\,0\bigr]},
    \label{eq:app_sigeff}
\end{equation}
velocity widths being frame-invariant, so the rest-frame curve applies to observed pixels at $\lambda_{\rm obs}/(1+z)$ without modification. No width is converted between frames. With a sampled redshift, $z$ in Eq.~\eqref{eq:app_sigeff} is the reference redshift of the projector, set at model construction. \mbox{\ceri} warns if this changes the fixed kernel by more than $10$ per cent at any pixel over the redshift prior.  The floor at zero keeps the operator well defined when the instrumental resolution approaches or exceeds that of the library. When the resolution shifts into that regime, \ceri\ warns, stating the number and wavelength range of the affected pixels, and continues with the continuum delivered at the library resolution there.

The continuum is projected in two stages. First, the model is interpolated onto a grid uniform in $\ln\lambda$ over the observed window plus a margin, with spacing equal to the finest model pixel in the window. The free width $\sigma_{\rm gal}$ is applied there as a Gaussian in Fourier space, $\exp(-2\pi^{2}\sigma_{\rm gal}^{2}\nu^{2})$, with $\nu$ conjugate to $c\ln\lambda$. The transform buffer is padded with the edge values, so the window edges are not darkened. Second, a banded response matrix carries the result onto the observed pixels,

\begin{align}
    F_{\nu}(\lambda_{i}) &\;=\; \sum_{j} W_{ij}\,\tilde F_{\nu}(\lambda_{j}),\qquad \sum_{j}W_{ij}=1,\nonumber\\
    W_{ij} &\;\propto\; \exp\!\left[-\frac{1}{2}\left(\frac{\ln\lambda_{j}-\ln\lambda_{i}+\ln(1+z)}{\sigma_{\rm fix}(\lambda_{i})/c}\right)^{\!2}\right],
    \label{eq:app_response}
\end{align}

whose row $i$ is the Gaussian of the fixed width centred on observed pixel $i$, evaluated at the nodes $j$ of the log grid and truncated at five dispersions; rows whose fixed width falls below half a log pixel reduce to linear interpolation, since the model cannot deliver a feature sharper than its own grid. The matrix performs the resampling as well, so the continuum path is one interpolation, one fast Fourier transform and one banded product. Splitting the kernel this way is exact, since the two Gaussians compose to Eq.~\eqref{eq:one_kernel}. A single operator for the total width, such as the resolution-normalised coordinate of \citet{Johnson:2021aa}, would have to be rebuilt at every sample.

The emission lines inside the window are added to the continuum on the observed pixels, each from its observer-frame integrated flux $F_{\ell}$ (Appendix~\ref{app:line_proj}):

\begin{align}
    f^{\rm lines}_{\nu}(\lambda_{i}) &\;=\; \sum_{\ell} F_{\ell}\,\frac{\lambda_{i}}{c}\,
    \frac{1}{\sqrt{2\pi}\,s_{\ell}}\exp\!\left[-\frac{1}{2}\left(\frac{\ln\lambda_{i}-\ln\lambda_{\ell}^{\rm obs}}{s_{\ell}}\right)^{\!2}\right],\nonumber\\
    s_{\ell} &\;=\; \frac{1}{c}\sqrt{\sigma_{\rm gas}^{2}+\sigma_{\rm inst}^{2}(\lambda_{\ell}^{\rm obs})},
    \label{eq:app_line_profile}
\end{align}

a unit-area Gaussian in $\ln\lambda$ turned into a flux density per unit frequency by $|\mathrm{d}\ln\lambda/\mathrm{d}\nu|=\lambda/c$; the same $\lambda^{2}/c$ factor as the \textsc{FSPS} line profiles, written per unit $\ln\lambda$. The profile is not renormalised on the grid, so a line partly outside the observed range is truncated and not compressed. The LSF is understood as measured on the detector (from arc lines), so it already contains the pixel width and the profiles are sampled at pixel centres rather than integrated over the pixel a second time. For photometry only the Fourier stage is applied, with no instrumental or library term: $\sigma_{\rm gal}$ on the continuum, on the rest-frame range covered by the filters, and $\sigma_{\rm gas}$ on the line profiles of the model grid, and the broadened window is scattered back into the full spectrum before the filter projection of Appendix~\ref{app:phot}.

\section{The systematic noise floor}
\label{app:sys_floor}
Every arm of the likelihood (\S\ref{sec:likelihood}) may optionally inflate its per-datum uncertainty by a common rescaling of the formal measurement error and by one or more nuisance terms added in quadrature to it, to account for the limits of the model itself:
\begin{equation}
    \sigma_{{\rm eff},i}^{2}
    \;=\;
     s^{2}\,\sigma_{i}^{2}
    \;+\;\bigl(f_{\rm calib}\,|m_{i}|\bigr)^{2}
    \;+\;\bigl(f_{\rm data}\,|d_{i}|\bigr)^{2}
    \;+\;j^{2},
    \label{eq:app_noise_floor}
\end{equation}
where $m_{i}$ is the model prediction and $d_{i}$ the datum. Each term can be switched on independently, per arm, and each is sampled in log space ($\theta[\texttt{log\_err\_scale}]$, $\theta[\texttt{log\_f\_calib}]$, $\theta[\texttt{log\_f\_data}]$, $\theta[\texttt{log\_jitter}]$, each suffixed by the kind of the observation and, with several of a kind, by its name, e.g.\ $\theta[\texttt{log\_jitter\_phot}]$), which guarantees positivity without a constrained sampler. The scale factor $s$ keeps the relative weights of the data and only rescales their overall level. It suits a pipeline whose errors are uniformly too small or too large, for example when the reduced $\chi^{2}$ of a good fit is far from unity. The additive terms instead up-weight the faint data. The two may be combined. No term is shared between observations: a jitter in maggies and one in erg\,s$^{-1}$\,cm$^{-2}$\,Hz$^{-1}$ could not be the same number. Without the log-normalisation the likelihood would be maximised by $s\to\infty$; with it, $\ln\mathcal{L}$ contains $-N\ln s-\chi^{2}/(2s^{2})$ and is maximised at $s^{2}=\chi^{2}/N$, so the data themselves set the level. The choice of coordinate is also a choice of prior: a uniform prior on $\ln f$ is a $1/f$ prior on $f$ itself, which favours small values of the term and lets the data switch it off cleanly when it is not needed; this is usually the desired behaviour, but a user who wants the floor to be present at a definite level should place the prior on $f$ accordingly. The model-anchored fraction can alternatively be fixed at a chosen value rather than sampled.

The data uncertainty describes only the measurement; the model itself is also uncertain. For high-S/N data, the dominant error can be the fidelity of the templates: for example, the residual line-ratio error of the \textsc{Cloudy} grid can exceed the photon noise for line fluxes with $\mathrm{S/N}\gtrsim10$. Without a floor, a handful of such high-significance data acquire enormous statistical weight and drive the fit through model imperfections, pulling the entire posterior to accommodate lines the model cannot reproduce at the quoted precision.

The two fractional terms differ in what they are anchored to. The \emph{model-anchored} floor $f_{\rm calib}|m_{i}|$ scales with the predicted flux; because the resulting variance then depends on $\boldsymbol{\theta}$, the likelihood can be increased by shrinking the prediction, and the posterior is biased weakly towards smaller fluxes (an Eddington-type effect). For the few-per-cent floors relevant here the resulting shift in inferred quantities such as $\log M_{\star}$ is $\lesssim0.02\,\mathrm{dex}$, but the bias grows with the floor and is not removed by more data. The \emph{data-anchored} floor $f_{\rm data}|d_{i}|$ scales with the measured flux instead, so the variance is independent of $\boldsymbol{\theta}$ and no such bias arises; it is the appropriate choice for zero-point and flux-calibration uncertainties, and more generally for any systematic that should be pinned to the measurement rather than to the current model. Equivalently, users can inflate the supplied uncertainties before the fit. The additive jitter $j$ captures flux-independent excess noise, such as sky-subtraction residuals or read noise beyond the formal photon estimate.

The model-anchored floor makes $\sigma_{{\rm eff},i}$ a function of the parameters. This is why \ceri\ keeps the normalisation term of the log-likelihood, which is usually dropped. The per-datum log-normalisation $\tfrac{1}{2}\ln(2\pi\sigma_{{\rm eff},i}^{2})$ in Eq.~\eqref{eq:total_lnlike} is no longer a constant offset and needs to be retained. 

All floors are disabled by default, recovering the standard diagonal Gaussian, and should be enabled per arm only where the model is known to be mismatched at a quantifiable level.

\section{Spectrum likelihood and noise model}
\label{app:spec_likelihood}
This appendix collects the data-level operators alluded to in \S\ref{sec:spectra}. Let $\boldsymbol{d}\in\mathbb{R}^{n_{\rm pix}}$ be the observed flux on the detector grid $\boldsymbol{\lambda}_{\rm obs}$, with per-pixel $1\sigma$ uncertainty $\boldsymbol{\sigma}$ and boolean mask $\mathcal{M}$. Let $m(\boldsymbol{\lambda}_{\rm obs})=\mathcal{S}\,F^{\rm full}_{\lambda}$ be the model flux produced by the smoothing-and-resampling closure of Appendix~\ref{app:spec_smoothing}, and $\boldsymbol{s}$ the optional unsubtracted sky vector.

\subsection{Sky subtraction and masking}
\label{app:spec_skynoise}
When a sky vector is provided the data are corrected before the residual is formed, $\boldsymbol{d}\!\to\!\boldsymbol{d}-\boldsymbol{s}$. The per-pixel variance $\sigma_{{\rm eff},i}^{2}$ is the formal uncertainty optionally inflated by the systematic floor of Eq.~\eqref{eq:app_noise_floor}, with $i$ running over detector pixels. The mask $\mathcal{M}$ combines the user's boolean input, finite-data checks on both flux and uncertainty, the requirement $\sigma_{i}>0$, and any line-by-line exclusions placed around user-supplied rest-frame wavelengths.

\subsection{Diagonal Gaussian likelihood}
\label{app:spec_lndiag}
The standard pixel-independent contribution to the log-likelihood is the full diagonal Gaussian,
\begin{equation}
    \ln\mathcal{L}_{\rm diag}
    \;=\;
    -\tfrac{1}{2}\!\sum_{i\in\mathcal{M}}\!
    \left\{
        \left[\frac{(d_{i}-s_{i})-m_{i}}{\sigma_{{\rm eff},i}}\right]^{2}
        +\ln \bigl(2\pi\,\sigma_{{\rm eff},i}^{2}\bigr)
    \right\},
    \label{eq:app_spec_lndiag}
\end{equation}
matching the per-arm form of Eq.~\eqref{eq:total_lnlike}. The normalisation term is kept (Appendix~\ref{app:sys_floor}). The likelihood a \texttt{Spectrum} can evaluate after a fit, for a given model spectrum and outside the compiled code (\S\ref{sec:likelihood}), can alternatively form the residuals in $\ln$-flux units, the appropriate linearisation when multiplicative residuals dominate over additive ones; this option is not available in the compiled likelihood used by the samplers.

\subsection{Marginalised calibration polynomial}
\label{app:spec_polymarg}
The calibrated model is $\mu_i\,(1+\sum_{m=0}^{M}c_m T_m(x_i))$, with $T_m$ the Chebyshev polynomials on the wavelengths mapped to $x\in[-1,1]$. The residual $\boldsymbol{r}=\boldsymbol{d}-\boldsymbol{s}-\boldsymbol{\mu}=\mathbf{D}\boldsymbol{c}+\boldsymbol{n}$ is therefore linear in the coefficients, with $\mathbf{D}=\mathrm{diag}(\boldsymbol{\mu})\,\mathbf{A}$, $A_{im}=T_m(x_i)$ and $\boldsymbol{n}\sim\mathcal{N}(0,\mathbf{C})$, $\mathbf{C}=\mathrm{diag}(\sigma_{{\rm eff},i}^{2})$ evaluated at the uncalibrated model. With the prior $\boldsymbol{c}\sim\mathcal{N}(0,\boldsymbol{\Lambda})$, $\boldsymbol{\Lambda}=\mathrm{diag}(s_m^{2})$, the coefficients integrate out in closed form:

\begin{equation}
    \ln\mathcal{L} \;=\; \ln\mathcal{L}_{\rm diag}(\boldsymbol{r}) \;+\; \tfrac{1}{2}\,\boldsymbol{b}^{\top}\mathbf{M}^{-1}\boldsymbol{b} \;-\; \tfrac{1}{2}\ln|\mathbf{M}| \;-\; \tfrac{1}{2}\ln|\boldsymbol{\Lambda}|,
    \label{eq:app_polymarg}
\end{equation}
with $\mathbf{M}=\mathbf{D}^{\top}\mathbf{C}^{-1}\mathbf{D}+\boldsymbol{\Lambda}^{-1}$, $\boldsymbol{b}=\mathbf{D}^{\top}\mathbf{C}^{-1}\boldsymbol{r}$ and $\ln\mathcal{L}_{\rm diag}$ of Eq.~\eqref{eq:app_spec_lndiag}. The second term is the gain from the best-fitting calibration; the last two form an Occam penalty, $-\tfrac{1}{2}\ln|\mathbf{M}\boldsymbol{\Lambda}|\le0$, for calibration freedom the data do not need. For $s_m\to0$ the expression reduces to $\ln\mathcal{L}_{\rm diag}$. Given $\boldsymbol{\theta}$, the coefficients are Gaussian with mean $\mathbf{M}^{-1}\boldsymbol{b}$ and covariance $\mathbf{M}^{-1}$; the mean equals the optimised polynomial of \S\ref{sec:spectra} with regularisation $1/s_m$.

\subsection{Gaussian-process correction for correlated residuals}
\label{app:spec_GP}
Correlated residuals between neighbouring pixels, such as those induced by the resampling and combination of spectral pixels from separate dithers or by imperfect sky subtraction (a calibration error, by contrast, would be multiplicative in the data), can be modelled by an optional squared-exponential Gaussian process (GP). The GP is part of the likelihood used by all samplers. For each spectrum, its amplitude $a$ and correlation length $\ell$ are fixed or sampled, as $\ln a$ and $\ln\ell$ ($\theta[\texttt{log\_gp\_amp\_spec}]$ and $\theta[\texttt{log\_gp\_length\_spec}]$). The residuals are first whitened by the diagonal noise model, so every noise term of Eq.~\eqref{eq:app_noise_floor} enters $\sigma_{\rm eff}$. On the same spectrum, the GP cannot currently be combined with the outlier mixture, upper limits, the analytic line marginalisation or a profiled or marginalised calibration polynomial; the sampled calibration can be combined with it. Rather than adding a second likelihood term to the diagonal Gaussian of Eq.~\eqref{eq:app_spec_lndiag}, the GP promotes the pixel-independent covariance to a full covariance on the normalised residuals: the white noise carried by each whitened pixel enters as the identity and the correlated structure is added to it,
\begin{equation}
    \Sigma_{ij}
    \;=\;
    \delta_{ij}
    +a^{2}\exp\Bigl[-\tfrac{1}{2}\bigl(\tfrac{\lambda_{i}-\lambda_{j}}{\ell}\bigr)^{2}\Bigr]
    +\varepsilon\,\delta_{ij},
    \label{eq:app_spec_GP}
\end{equation}
where the leading $\delta_{ij}$ is the (unit) whitened per-pixel variance, $a$ is the kernel amplitude in units of the per-pixel $\sigma_{\rm eff}$, $\ell$ is the observer-frame correlation length in \AA, and $\varepsilon$ is a small diagonal jitter for numerical stability. The full log-likelihood is the single Gaussian
\begin{equation}
    \ln\mathcal{L}
    \;=\;
    -\tfrac{1}{2}\,\boldsymbol{r}^{\top}\Sigma^{-1}\boldsymbol{r}
    \;-\;\tfrac{1}{2}\ln\!|\Sigma|
    \;-\;\tfrac{1}{2}\sum_{i\in\mathcal{M}}\ln\!\bigl(2\pi\,\sigma_{{\rm eff},i}^{2}\bigr),
    \label{eq:app_spec_lnL}
\end{equation}
with $\boldsymbol{r}$ the vector of normalised residuals $r_i=(d_{i}-s_{i}-m_{i})/\sigma_{{\rm eff},i}$ over the unmasked subset of length $n$ ($d_i$ the observed flux, $s_i$ the sky vector, $m_i$ the model flux on the detector grid and $\sigma_{{\rm eff},i}$ the effective uncertainty of Eq.~\eqref{eq:app_noise_floor}, all at pixel $i$), and the final term the per-pixel normalisation carried over from the whitening of Eq.~\eqref{eq:app_spec_lndiag}. When the GP is disabled ($a\!\to\!0$, $\varepsilon\!\to\!0$) $\Sigma\!\to\!I$ and Eq.~\eqref{eq:app_spec_lnL} reduces exactly to the diagonal likelihood $\ln\mathcal{L}_{\rm diag}$; the white noise is never counted twice.

\section{Emission-line fluxes and likelihood}
\label{app:lines}

This appendix describes how \ceri\ predicts the catalogue line fluxes of a \texttt{Lines} observation (\S\ref{sec:lines}) from the nebular grid, and the likelihood of catalogues that mix detections and upper limits.

\subsection{Line fluxes from the nebular grid}
\label{app:line_proj}

With a nebular component, the model flux of catalogued line $k$ is the line luminosity carried through the same operations as the continuum. The per-node line luminosities $L^{\rm neb,\ell}_{ma}$ (Appendix~\ref{app:nebular}) are contracted with $W_{ma}$ and with the birth-cloud and diffuse attenuation at the line wavelength (Appendix~\ref{app:dust_attn}). The result is scaled by $10^{\theta[\texttt{logmass}]}$, by the flux factor of \S\ref{sec:global} divided by $(1+z)$ (an integrated flux carries no bandwidth factor), by the IGM transmission at the line, and by the optional aperture correction $\theta[\texttt{eline\_scaling}]$. An unresolved blend is the sum of its component grid lines. No line width enters.

\subsection{Detection and upper-limit likelihood}
\label{app:line_lik}
The contribution of detected lines to the chi-squared is the standard pixel-independent Gaussian form
\begin{equation}
    \chi^{2}_{\rm det}
    \;=\;\sum_{k\in\mathcal{M}_{\rm det}}\!
    \!\left(\frac{d_{k}-m_{k}}{\sigma_{k}}\right)^{2},
    \label{eq:app_line_chi2_det}
\end{equation}
with $\mathcal{M}_{\rm det}$ the set of unmasked, positively detected
lines. For lines that are catalogued as non-detections, flagged as
upper limits, the chi-squared takes a one-sided form,
\begin{equation}
    \chi^{2}_{\rm UL}
    \;=\;\sum_{k\in\mathcal{M}_{\rm UL}}\!
    \begin{cases}
        \bigl((d_{k}-m_{k})/\sigma_{k}\bigr)^{2}, & m_{k}>d_{k},\\
        0,                                       & m_{k}\le d_{k},
    \end{cases}
    \label{eq:app_line_chi2_UL}
\end{equation}
where $d_{k}$ now denotes the catalogued upper-limit value rather than a measured flux. This is a simplified one-sided treatment, in which any model that respects the threshold incurs no penalty and any model that exceeds it is penalised by the standard Gaussian residual against the upper-limit value. A more rigorous alternative is the survival form $\chi^{2}_{{\rm UL},k}=-2\ln\Phi\bigl[(d_{k}-m_{k})/\sigma_{k}\bigr]$, with $\Phi$ the standard-normal cumulative distribution function \citep{Sawicki:2012aa}. \ceri\ uses the step form because it makes no assumption about the catalogue's reported upper-limit convention (e.g.\ $1\sigma$ vs $3\sigma$ thresholds) and because for any model that is well clear of the limit the two forms differ by an additive constant that does not affect the posterior. The full line-side chi-squared is $\chi^{2}_{\rm lines}=\chi^{2}_{\rm det}+\chi^{2}_{\rm UL}$. Its log-likelihood contribution, $-\tfrac{1}{2}\chi^{2}_{\rm lines}$, is added to those of the other observations.

\subsection{Analytic marginalisation of line fluxes}
\label{app:line_marg}
When the line fluxes $\boldsymbol{\alpha}$ are marginalised (\S\ref{sec:likelihood}), every observation $k$ that sees the lines is linear in them, $\boldsymbol{d}_k=\boldsymbol{\mu}_k+\mathbf{A}_k\boldsymbol{\alpha}+\boldsymbol{n}_k$. Here $\boldsymbol{\mu}_k$ is the model without the fitted lines, and $\mathbf{A}_k$ holds the unit-flux line profiles of the spectrum, the line-to-band projection of the photometry, or the blend matrix of a line catalogue. With weights $\mathbf{W}_k=\mathrm{diag}(\sigma_{{\rm eff},i}^{-2})$, the prior $\alpha_j\sim\mathcal{N}(\bar\alpha_j,s_j^{2})$ around the photoionisation-grid flux ($s_j$ a fixed fraction of it), or a flat prior, the fluxes integrate out exactly:

\begin{equation}
    \begin{split}\ln Z_{\alpha} \;=\;& \sum_{k}\ln\mathcal{L}_{{\rm diag},k}(\boldsymbol{r}_k) \;+\; \tfrac{1}{2}\,\boldsymbol{b}^{\top}\mathbf{P}^{-1}\boldsymbol{b}\\ &-\; \tfrac{1}{2}\ln|\mathbf{P}| \;-\; \sum_{j\in{\rm G}}\ln s_{j} \;+\; \tfrac{m_{\rm flat}}{2}\ln 2\pi,\end{split}
    \label{eq:app_line_marg}
\end{equation}
with $\boldsymbol{r}_k=\boldsymbol{d}_k-\boldsymbol{\mu}_k-\mathbf{A}_k\bar{\boldsymbol{\alpha}}$ ($\bar\alpha_j=0$ for flat lines), $\boldsymbol{b}=\sum_k\mathbf{A}_k^{\top}\mathbf{W}_k\boldsymbol{r}_k$ and $\mathbf{P}=\sum_k\mathbf{A}_k^{\top}\mathbf{W}_k\mathbf{A}_k+\mathrm{diag}(s_j^{-2})$, the last term only for the lines with a Gaussian prior (set G). $m_{\rm flat}$ is the number of lines with a flat prior. Given $\boldsymbol{\theta}$, the fluxes are Gaussian with mean $\bar{\boldsymbol{\alpha}}+\mathbf{P}^{-1}\boldsymbol{b}$ and covariance $\mathbf{P}^{-1}$.

\section{Supplementary figures for the mock demonstrations}
\label{app:extra_figs}

\subsection{Computational cost and scaling}
\label{app:benchmark}
\begin{figure}[ht]
    \centering
    \includegraphics[width=\linewidth]{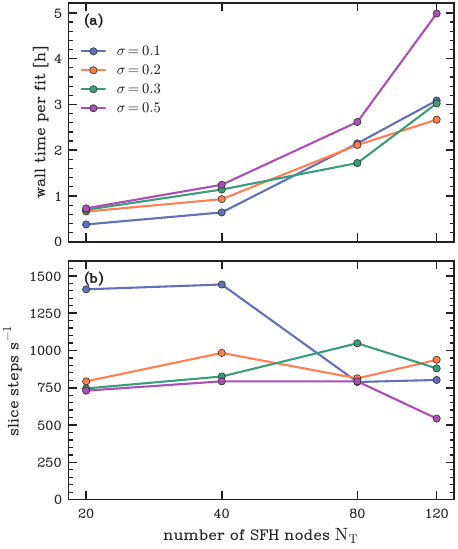}
    \caption{Computational cost of \ceri\ for the sixteen nested-sampling fits of \S\ref{sec:high_detail_sfh} (noiseless ultraviolet-to-near-infrared spectrum; continuity-prior width $\sigma$ in colour; $N_T=20$--$120$ SFH nodes, i.e.\ $25$--$125$ free parameters; $400$ live points, $80$ replaced per iteration, $3n_{\rm dim}$ slice steps per replacement), each run on one A100 GPU. \textbf{(a)} Wall-clock time per fit, including compilation. \textbf{(b)} Slice-sampling steps per second. Every slice step evaluates the likelihood at least once, so this is a lower bound on the likelihood-evaluation rate.}
    \label{fig:benchmark}
\end{figure}

Figure~\ref{fig:benchmark} quantifies the cost of the high-dimensional fits of \S\ref{sec:high_detail_sfh}, underlying the scaling quoted in \S\ref{sec:performance}.

\subsection{Single-galaxy comparison with \pros}
\label{app:gal_comparison}
\begin{figure*}[p]
    \centering
    \includegraphics[width=0.84\linewidth]{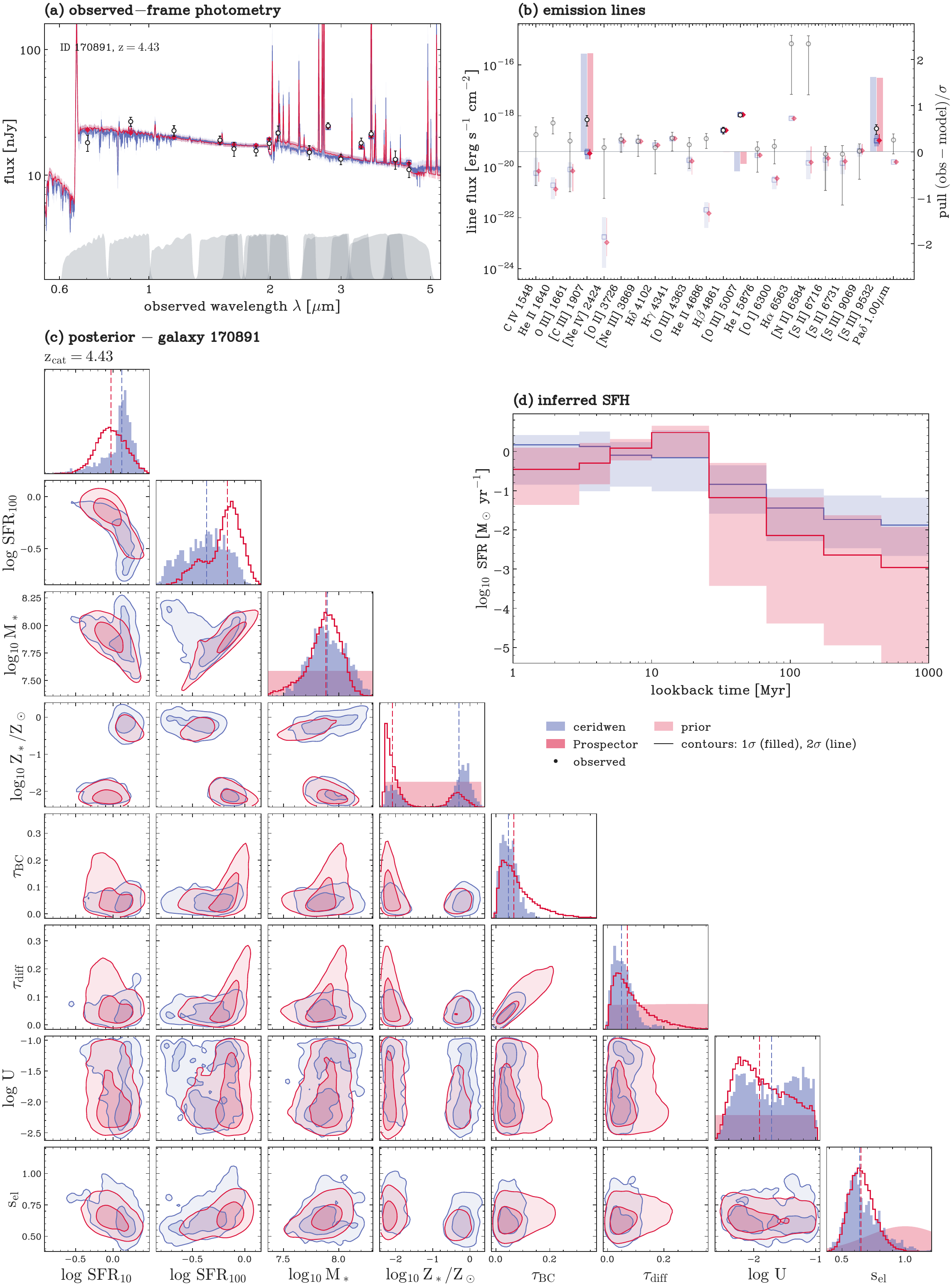}
    \caption{The controlled comparison for a single galaxy, JADES 170891 at $z=4.43$, with the same data and priors in both codes. \ceri\ is shown in blue, \pros\ in crimson and the observations as open circles. \textbf{(a)} The observed NIRCam photometry ($15$ bands) with both codes' posterior-predictive SEDs ($16$--$84\%$ bands) and the filter transmission curves (grey). \textbf{(b)} All $22$ fitted R1000 emission-line fluxes against both codes' posterior predictions; lines detected above $\mathrm{S/N}=2$ are drawn in full colour with background bars (right axis) giving each code's per-line pull $(\mathrm{obs}-\mathrm{model})/\sigma$; undetected catalogue entries, which carry no flux information, are faded. \textbf{(c)} The joint posterior ($1\sigma$ filled, $2\sigma$ contours; dashed lines mark each code's median, pink shading the shared priors). \textbf{(d)} The inferred SFH (median and $68\%$ band): both codes fit the same eight-bin history, and the two overlap within their $68\%$ bands in every bin.}
    \label{fig:prospector_gal170891}
\end{figure*}

Figure~\mbox{\ref{fig:prospector_gal170891}} makes the code-to-code comparison of \mbox{\S\ref{sec:demo_jades}} concrete for one galaxy, JADES 170891 at $z=4.43$ ($15$ bands, $22$ lines), at the level of the full joint posterior. The two codes evaluate the same model. At six parameter vectors drawn from the \mbox{\pros} posterior, the log-priors of the two codes agree to $10^{-4}$. The posterior-predictive SEDs of the two codes track each other across the full wavelength range, and the line fluxes are reproduced with per-line pulls that agree in sign and magnitude. The joint posteriors occupy the same regions. The stellar mass, the dust optical depths and the line-calibration factor agree within their uncertainties. Both codes find the same two stellar-metallicity modes, near $\log_{10}(Z/Z_\odot)\approx-2.1$ and $-0.3$, but weight them differently. The recovered SFHs overlap within their $68\%$ bands in every bin.

\bsp
\label{lastpage}
\end{document}